\documentstyle[epsf]{ptptex}

\renewcommand{\theequation}{{\thesection.\arabic{equation}}}
\title{Non-Abelian Stokes Theorem and Quark Confinement 
 in SU(N) Yang-Mills Gauge Theory  
}
\author{Kei-Ichi Kondo$^{1,2}{}^{\dagger}$
 and Yutaro Taira$^2{}^{\ddagger}$}
\inst{$^1$ Department of Physics, Faculty of Science, 
Chiba University,  Chiba 263-8522, Japan
$^2$ Graduate School of Science and Technology,
  Chiba University, Chiba 263-8522, Japan
}
\abst{We derive a new version of the non-Abelian Stokes theorem for the
Wilson loop in the $SU(N)$ case by making use of the coherent state
representation on the coset space $SU(N)/U(1)^{N-1}=F_{N-1}$, the
flag space. 
We consider the $SU(N)$ Yang-Mills theory in the maximal Abelian
gauge in which $SU(N)$ is broken down to $U(1)^{N-1}$. 
First, we show that the Abelian dominance in the string tension
follows from this theorem and the Abelian-projected effective gauge
theory that was derived by one of the authors. 
Next (but independently), combining the non-Abelian Stokes theorem
with a novel reformulation of the Yang-Mills theory recently
proposed by one of the authors, we proceed to derive the area law of
the Wilson loop in four-dimensional
$SU(N)$ Yang-Mills theory in the maximal Abelian gauge. 
Owing to dimensional reduction, the planar Wilson loop at least for the
fundamental representation in four-dimensional $SU(N)$ Yang-Mills
theory can be estimated by the diagonal (Abelian) Wilson loop defined
in the two-dimensional $CP^{N-1}$ model.
This derivation shows that the fundamental quarks are confined by a
single species of magnetic monopole.  The origin of the area law is
related to the geometric phase of the Wilczek-Zee holonomy for
$U(N-1)$.  
The calculations are performed using the instanton calculus (in the
dilute instanton-gas approximation) and using the large $N$ expansion
(in the leading order).
}
\recdate{}
\begin{document} 
\maketitle
\section{Introduction}

Gell-Mann and Zweig \cite{GZ64} predicted in the mid-1960s that all hadrons (i.e., baryons and mesons) are
composed of the fundamental constituents having fractional charges, $\pm {1\over 3}e$ or $\mp{2\over 3}e$, with $e$
being the elementary charge.  
Now, the fundamental constituent is called the quark and the proposed theory is called the quark model.
The predictions of this model are consistent with the results of
experiments performed over the past thirty years.  The strong interaction among quarks and anti-quarks is
mediated by the gluon, which is described by the quantized Yang-Mills
gauge field theory.
\cite{YM54}  
The present fundamental theory describing the
quark and the gluon is provided by quantum chromodynamics (QCD),
which is a non-Abelian gauge theory with the gauge group  $G=SU(3)$
corresponding to three colors.
  However, neither an isolated quark nor an isolated anti-quark has ever
been observed experimentally.  
In the present understanding, they are believed to be
confined in hadrons.  This is the hypothesis of quark confinement. Quark confinement could be explained theoretically within
the framework of QCD, although no one has achieved a rigorous proof
of quark confinement.  This is one of the most important
problems to be solved in theoretical physics.
\par
QCD has a remarkable property, called asymptotic freedom, which was discovered 
by Gross, Wilczek and Politzer and independently by 't Hooft. \cite{AF}  Asymptotic freedom does not appear in the most successful quantized field theory, quantum
electrodynamics (QED).  As is well known, QED is the Abelian gauge theory for the
electron and the photon in which the electromagnetic interaction is
described by the quantized Maxwell gauge field theory with the gauge
group $G=U(1)$.  Asymptotic freedom is a consequence of gluon
self-interactions.  Therefore, this is a very characteristic feature
of non-Abelian gauge theory.

\par
The purpose of this article is to demonstrate quark confinement
within QCD based on the Wilson criterion for quark confinement,
\cite{Wilson74} i.e., the area law of the Wilson loop.  The Wilson loop
is a gauge invariant quantity and hence the Wilson criterion is also
gauge invariant.   The formulation of lattice gauge theory proposed
by Wilson \cite{Wilson74} is manifestly gauge invariant and does not
need the gauge fixing. It is easy to show that the strong coupling
expansion in lattice gauge theory leads to the area law of the
Wilson loop.  However, this result has not yet been continued to the
weak coupling region, where the string tension is expected to obey
the scaling law suggested from the result of the renormalization
group based on loop calculations.  The first indication of the area
law of the Wilson loop for arbitrary coupling constant was found in a study 
based on numerical simulations within  the lattice gauge theory by
Creutz
\cite{Creutz} for $G=SU(2)$ and $SU(3)$.  Although the numerical
evidence of quark confinement was indeed a great progress toward a complete understanding of quark confinement, the analytical proof
is still lacking.  

\par
This work was initiated to
justify the dual superconductor picture of the QCD vacuum proposed in the
mid-1970s \cite{DSC} within the framework of continuum quantum
field theory.   For dual superconductivity to occur, magnetic
monopoles must be condensed, just as ordinary superconductivity
requires condensation of Cooper pairs.  In fact, the importance and the
validity of taking into account magnetic monopoles in quark
confinement has been demonstrated, at least for simplified
four-dimensional and lower-dimensional models, especially, by
Polyakov,
\cite{Polyakov} and recently for the four-dimensional Yang-Mills
theory and QCD with extended supersymmetries by Seiberg and Witten.
\cite{SW94}  In this scenario, quark confinement is realized due to
condensation of magnetic monopoles. Recent developments in numerical
simulations on the lattice 
\cite{review} confirm the existence of dual superconductivity
in QCD, at least, under a specific gauge fixing called the Abelian gauge.\cite{tHooft81} 
\par
This article gives a detailed exposition of the  results on quark
confinement that were announced in  a previous article
\cite{KT99a} for
$G=SU(3)$ together with  new results for
$G=SU(N)$ with arbitrary $N$.  They are extensions of the analyses of
the Yang-Mills theory in the maximal Abelian (MA) gauge given in a
series of articles,
\cite{KondoI,KondoII,KondoIII,KondoIV,KondoV,KondoVI} where the case
of $SU(2)$ was explicitly worked out. In this process, we have found that the extension
from $SU(2)$ to $SU(3)$ is non-trivial, but  the extension
from $SU(3)$ to
$SU(N)$, $N > 3$, is rather straightforward.  New features come out
when we begin to analyze the $SU(N)$ case with $N \ge 3$. It seems that
they have been overlooked to this time in the conventional approach based
on the maximal Abelian gauge.
\par
The MA gauge is a partial gauge fixing from the original
non-Abelian gauge group $G$ to its subgroup $H$ \cite{tHooft81} in which the
gauge degrees of  freedom of the coset $G/H$  are fixed.  Even
after the MA gauge, there is a residual gauge group $H$ which is
taken to be the maximal torus subgroup $H=U(1)^{N-1}$.  After the MA
gauge, the magnetic monopole is expected to appear, since the
Homotopy group 
$\pi_2(G/H)$ is non-trivial, i.e.,
\begin{equation}
 \pi_2(SU(N)/U(1)^{N-1}) = \pi_1(U(1)^{N-1}) = {\bf Z}^{N-1} .
\end{equation}
This implies that the breaking of gauge group $G \rightarrow H$ by partial gauge fixing 
leads to ($N-1$) species of magnetic monopoles.  However, we do not
necessarily need to consider the maximal breaking 
$SU(N) \rightarrow U(1)^{N-1}$, although the maximal torus group is desirable as a gauge group of the low-energy effective {\it Abelian} gauge theory.\cite{KondoI}   Actually, even if we restrict $H$ to a 
continuous subgroup%
\footnote{The possibility of a discrete subgroup has been extensively 
investigated recently from the viewpoint of the Abelian gauge, e.g., the
center $Z_{N}$ for $SU(N)$ (see e.g. Ref.\citen{Greensite}).  }
 of $G$, there are other possibilities for choosing $H$, e.g., we can 
choose a subgroup $\tilde H$ such that
\begin{equation}
  G \supset \tilde H \supset H := U(1)^{N-1} .
\end{equation}
The possible number of cases for choosing $\tilde H$ increases as
$N$ increases.  We have found \cite{KT99a} that the group $\tilde H$
may depend on the representation to which the quark
belongs when $N \ge 3$  and that it suffices to take $\tilde
H=U(N-1)$ for the fundamental quark to be confined in the sense of the 
area law of the Wilson loop under the partial gauge fixing.     Here
$\tilde H$ is equal to the maximal stability group specified by the
highest-weight state of the representation of the quark in the
Wilson loop. This is a new feature which does not show up in the
$SU(2)$ case.  Nevertheless, this does not mean that the choice of the
maximal torus does not lead  to quark confinement.  In fact, even if
we choose the maximal torus, the area law can be derived.  This is
because the coset $G/\tilde H$ is contained in $G/H$, i.e.,
$G/\tilde H \subset G/H$, so that the Wilson loop does not feel the
whole of $G/H$, but only feels the components of $G/\tilde H$ that
are contained in $G/H$.  In other words, the variables belonging to
$G/H - G/\tilde H$ are irrelevant for the expectation value of
the Wilson loop, as can be seen from the non-Abelian Stokes theorem
(NAST) that was presented in Ref.\citen{KT99a} and is derived in this
article.  Therefore, a single kind of magnetic monopole is sufficient
for confining a fundamental quark, since
\begin{equation}
 \pi_2(SU(N)/U(N-1)) = \pi_1(U(1)) = {\bf Z} .
\end{equation}
Our results show that two partial gauge fixings  $SU(3) \rightarrow
U(2)$ and 
$SU(3) \rightarrow U(1) \times U(1)$ lead  to the same 
result for confinement as far as the fundamental quarks are concerned.\footnote{See Ref.\citen{Wensley99}
for a result of the simulation on a lattice.}
\par
The NAST plays the crucial role in this article.  The NAST has a 
number of versions which have been derived by many authors.\cite{NAST}   The version of NAST derived in this article is based on
the idea of Dyakonov and Petrov,\cite{DP89} who derived an  $SU(2)$
version and suggested a method of generalization.  We derive a version
of NAST based on the coherent state representation
\cite{FGZ90} on the flag space, \cite{Perelomov87,KO95} not on
the method suggested by them.  The coherent state representation is
 used in a different fashion to derive an $SU(2)$ version of
the NAST in
Ref.\citen{KondoIV}, but the extension to $SU(N)$, $N \ge 3$, was a
non-trivial issue which prevented us from presenting immediate
publication of general
$SU(N)$ results.  The NAST is not only
mathematically (or technically) important but also physically
interesting as we now discuss. 
\par
First, the NAST enables us to write the
non-Abelian Wilson loop, 
\begin{equation}
  W^C [{\cal A}] :=  {1 \over {\cal N}}{\rm tr} \left[ {\cal P} 
 \exp \left( i g \oint_C {\cal A}
 \right) \right] ,
\end{equation}
in terms of the Abelian-field strength (curvature two-form)
$f=da$ with the Abelian gauge potential (connection one-form) $a$:
\begin{eqnarray}
  W^C[{\cal A}]  
 =  \int [d\mu(V)]_C 
 \exp \left( i g \oint_{C}  a \right) 
= \int [d\mu(V)]_C 
 \exp \left( i g \int_{S}  f \right)  .
\end{eqnarray}
Combining this fact with the Abelian-projected effective gauge 
theory  (APEGT) derived by one of the authors,\cite{KondoI} we can explain the
Abelian dominance \cite{EI82,SY90} in the Wilson loop.  The APEGT is
an Abelian gauge theory obtained by integrating out the massive
degrees of freedom, i.e., the off-diagonal  gluon gauge field
$A_\mu^a$ with mass $m_A$.  Hence, the APEGT can be written in terms of
the diagonal massless gauge field $a_\mu^i$ and the anti-symmetric
(Abelian) tensor field $B_{\mu\nu}^i$ together with the ghost and
anti-ghost fields $C^a$ and $\bar C^a$, where the index $i$ denotes the
diagonal components and $a$ the off-diagonal ones. Therefore the
APEGT is regarded as a low-energy effective theory (LEET) which is
valid in the long-distance (or low-energy) region $R>m_A^{-1}$.  
The Abelian gauge field $b_\mu^i$ obtained after the Hodge
decomposition of $B_{\mu\nu}^i$ can be identified with the Abelian
gauge field dual to $a_\mu^i$.  In fact, we can obtain the theory
with an action $S[b]$ written in terms of $b_\mu^i$ alone by
integrating out all the fields other than
$b_\mu^i$ in APEGT, and the theory can be rewritten into the
dual Ginzburg-Landau theory, i.e., the dual Abelian Higgs theory,
provided that  magnetic monopole condensation occurs, i.e.,
$\langle k_\mu k_\mu \rangle \not= 0$.
In the dual Ginzburg-Landau theory, the coupling constant
$g$ in the original theory is replaced by the inverse coupling
constant $1/g$ which is proportional to the magnetic charge. 
Therefore, the dual theory can be identified with the magnetic theory. 
On the other hand, the theory with an action
$S[a]$ written in terms of $a^i$ alone is an Abelian gauge theory,
but the scale dependence  of the coupling constant $g(\mu)$ is the
same as that in the original Yang-Mills theory.  Thus the low-energy
effective Abelian gauge theory exhibits asymptotic freedom
reproducing the original renormalization-group beta function
$\beta(g)$.   This is a manifestation of an approximate weak-strong
or electro-magnetic  duality between two low-energy effective
theories described by $S[a]$ and $S[b]$.
\par
 Next, the NAST is able to separate the piece $\omega$ which 
corresponds to the magnetic monopole in the Abelian field
$a=C+\omega$.  Indeed, we can write the $SU(N)$ version of the 't
Hooft-Polyakov tensor describing the magnetic monopole.\cite{tHP74}  Hence we can separate the contribution of the magnetic monopole in the area law of the Wilson loop and explain the magnetic monopole dominance in
quark confinement. In fact, our derivation of the area law estimates only the monopole 
contribution, $\Omega_K=d\omega$.
Moreover, the NAST tells us that the essential ingredient
in the area law lies in the geometric phase, which is concerned with the
holonomy group of $U(N-1)$.
Thus we see that quark confinement is intimately related to the geometry of 
Yang-Mills gauge theory, in sharp contrast with the conventional
wisdom.
\par
We present two methods to derive the area law of the Wilson loop by
making use of the NAST. 
The first method is to use the APEGT for estimating
the diagonal Wilson loop; for a sufficiently large Wilson loop ($|C|
\gg m_A^{-1}$),  the expectation value of the non-Abelian Wilson
loop in Yang-Mills theory is reduced to that of the Abelian Wilson loop
in APEGT:
\begin{eqnarray}
 \Big\langle W^C [{\cal A}] \Big\rangle_{YM} 
= \Big\langle \exp \left( i g \oint_{C}  a \right)
\Big\rangle_{YM} 
\rightarrow 
 \Big\langle \exp \left( i g \oint_{C}  a \right)
\Big\rangle_{APEGT}     .
\end{eqnarray}
Then we can apply the result of Ref.\citen{KondoIII}, confinement in
the {\it Abelian} gauge theory, to show quark confinement in
Yang-Mills theory. 
\par
The second method is to treat the
non-Abelian gauge theory directly, without going through the
effective Abelian gauge theory, based on the novel reformulation of 
Yang-Mills theory in the MA gauge which was proposed by one of
the authors.\cite{KondoII}  The novel reformulation regards the
Yang-Mills theory as the perturbative deformation of a topological
quantum field theory (TQFT).    An advantage of this reformulation
in the MA gauge is that the derivation of the area law of the
non-Abelian Wilson loop in the four-dimensional $SU(N)$ Yang-Mills
theory is reduced to that of the diagonal (Abelian) Wilson loop in
the two-dimensional coset ($G/H$) non-linear sigma (NLS) model, at
least when the Wilson loop is planar. Therefore the four-dimensional
problem is reduced to a two-dimensional problem.  This dimensional
reduction is a remarkable feature of the modified MA gauge
\footnote{We must modify the MA gauge slightly in order to keep the 
supersymmetry, where the supersymmetry is expressed by the orthosymplectic
group
$OSp(4|2)$.}
 caused by hidden supersymmetry.  
The Yang-Mills
coupling constant $g$ of the four-dimensional Yang-Mills theory is
mapped into the coupling constant in the two-dimensional NLS model. 
Hence the coupling constant is expected to run in the same  way as
in the original Yang-Mills theory, since  the perturbative deformation
part provides the necessary running, as is well known from the loop
calculation.   For the fundamental quark, we are allowed to restrict
the flag space
$F_{N-1}:=SU(N)/U(1)^{N-1}$ to the complex projective space
$CP^{N-1}:=SU(N)/U(N-1)$.  This greatly simplifies the actual
treatment.  
\par
Another advantage of this reduction is that the magnetic monopole 
contribution to the Wilson loop in the four-dimensional Yang-Mills
theory in the MA gauge is equal to the instanton
contribution in the corresponding two-dimensional NLS model.  Indeed,
the diagonal Wilson loop can be written as the area integral of the
instanton density
$\Omega_K$ over the area $S$ bounded by the loop $C$.  This
correspondence may shed more light on the strong correlation between
magnetic monopoles and instantons observed in the Monte Carlo
simulations, since the two-dimensional instanton is identified as a
subclass of the four-dimensional Yang-Mills instanton (see e.g.
Ref.\citen{KondoII}).
\par
In this article the expectation value of the Wilson loop is
estimated by combining the instanton  calculus and the large $N$
expansion. (See
Refs.\citen{Yaffe82,Coleman85,Das87,RCV98} for reviews of the large $N$
expansion.)  We focus on the $CP^{N-1}$ model corresponding to
the fundamental quark. First, the instanton calculus is performed
within the dilute gas approximation.  It is shown that the
calculation in the $SU(N)$ case reduces to that in the $SU(2)$ case.  It is well
known that the large
$N$ expansion is a non-perturbative technique which can be
systematically improved.  We derive the area law to leading
order in the large $N$ expansion,
namely, in the region of large $N$ and weak
coupling $g$.  We hope that our derivation of quark confinement
based on the dimensional reduction and the large
$N$ expansion may shed more light on the relationship between QCD and
string theory, as first suggested by 't Hooft.\cite{tHooft74a}
\par
This article is organized as follows.
In the first half, we give a derivation of the NAST and discuss its
implications. Sections 2 and 3 are preparations for $\S$ 4.
In $\S$ 2 we review the construct of the coherent state on 
the flag space for the general compact semi-simple group $G$.   In
$\S$ 3, we present the explicit form of the coherent state on the
flag space for $G=SU(N)$.  We define the maximal stability group
$\tilde H$, which is very important in the following discussion.   In
$\S$ 4, making use of the results of $\S\S$ 2 and 3, we derive a
new version of the non-Abelian Stokes theorem for $G=SU(N)$.  Although we
discuss only the case of
$SU(N)$ explicitly, it is straightforward to extend this theorem to
an arbitrary compact semi-simple group $G$.  This version of the 
non-Abelian Stokes theorem is very interesting not only from the
mathematical but also from the physical point of view, since the
non-Abelian Wilson loop is expressed as the surface integral of the
two-form (i.e., the generalized 't Hooft-Polyakov tensor), which
leads to the magnetic monopole.  This fact is intimately related
with the Abelian and magnetic monopole dominance in quark
confinement, as discussed in subsequent sections. 
\par
In the second half, we derive the area law of the Wilson loop.
In $\S$ 5  we discuss the magnetic monopole in $SU(N)$
Yang-Mills theory.  In order to specify the type of possible
magnetic monopoles, it turns out that the maximal stability group is
more important than the maximal torus group $H$. 
In $\S$ 6 
Abelian dominance in the Wilson loop is shown in the $SU(N)$
Yang-Mills theory in the maximal Abelian gauge based on the
Abelian-projected effective gauge theory and the non-Abelian Stokes
theorem. In $\S$ 7 we 
briefly review a novel reformulation of the Yang-Mills theory which
has been proposed by one of the authors \cite{KondoII} to derive
quark confinement. This reformulation is called the (perturbative)
deformation of the topological quantum  field  theory. 
We apply this reformulation to derive the area law of the Wilson loop in
$SU(N)$ Yang-Mills theory in $\S\S$ 8 and 9. In $\S$ 9 we show
within this reformulation that the area law of the Abelian Wilson
loop in the two-dimensional nonlinear sigma model for the flag space
$G/\tilde H$ is sufficient to derive the area law of the
four-dimensional Yang-Mills theory in the maximal Abelian gauge. At
the same time, this derivation leads to the magnetic monopole
dominance in the area law. In $\S$ 8 we demonstrate the area law
of the Wilson loop in the nonlinear sigma model in an approach based on naive
instanton calculus.  For the fundamental quark, we have only to deal
with the $CP^{N-1}$ model.  
In $\S$ 9  we derive the area law based on the large $N$
expansion. These results imply the area law of the non-Abelian
Wilson loop in the four-dimensional $SU(N)$ Yang-Mills theory.   The
final section contains the conclusion of this article.
\par
In Appendix A, we give derivations of the inner product of the 
coherent states and the invariant measure on the flag space, which
are presented in $\S$ 3. 
In Appendix B  we explain the method of obtaining $CP^1$ and $CP^2$ by gluing
the complex planes.
In Appendix C, we explain two ways to characterize
the element of the flag space and the manner of formulating the NLS model using
these parameterizations.  
In Appendix D  we summarize the large $N$ expansion of $CP^{N-1}$.
In Appendix E  supplementary material  on the
$1/N$ expansion is presented.

\section{Coherent state and maximal stability group }
\setcounter{equation}{0}

\par
First, we construct the {\it coherent state}
$
  |\xi, \Lambda \rangle 
$
corresponding to the
coset representatives $\xi \in G/\tilde H$. 
We follow the method of Feng, Gilmore and Zhang.\cite{FGZ90}  For
inputs, we prepare the following:
\begin{enumerate}
\item[(a)]
 the gauge group%
\footnote{
Note that any compact semi-simple Lie group is a direct product of
compact simple Lie group.  Therefore, it is sufficient to consider
the case of a compact simple Lie group.  In the following we assume
that
$G$ is a compact simple Lie group, i.e., a compact Lie group with no
closed connected invariant subgroup.
}
 $G$  and the Lie algebra
${\cal G}$ of $G$ with the generators $\{ T^A  \}$, which obey the
commutation relations 
\begin{equation}
 [T^A, T^B ] = i f^{AB}{}_C T^C ,
\end{equation} 
where the $f^{AB}{}_C$ are the structure constants of the Lie algebra. 
If the Lie algebra is semi-simple, it is more convenient to rewrite
the Lie algebra in terms of the Cartan basis 
$\{ H_i, E_\alpha, E_{-\alpha} \}$.  There are two types of basic 
operators in the Cartan basis, $H_i$ and $E_\alpha$.  The operators
$H_i$ may be taken as diagonal, while $E_\alpha$ are the
off-diagonal shift operators.  They obey the commutation relations 
\begin{eqnarray}
 [H_i, H_j] &=& 0,
\end{eqnarray}
\begin{eqnarray}
[ H_i, E_\alpha ] &=& \alpha_i E_\alpha,
\end{eqnarray}
\begin{eqnarray}
 [ E_\alpha, E_{-\alpha} ] &=& \alpha^i H_i,
\end{eqnarray}
\begin{eqnarray}
 [ E_\alpha, E_\beta ] &=& \cases{ N_{\alpha;\beta} E_{\alpha+\beta},
&($\alpha+\beta \in R$) \cr 0, &($\alpha+\beta \not\in R,
\alpha+\beta \not= 0$) }  
\end{eqnarray}
where $R$ is the root system, i.e., a set of root vectors
$\{ \alpha_1, \cdots, \alpha_r \}$, with $r$ the rank of $G$.
\item[(b)]
  The Hilbert space $V^\Lambda$ is a carrier (the
representation space) of the unitary irreducible representation
$\Gamma^\Lambda$ of $G$.   
\item[(c)]  
We use a reference state 
$|\Lambda \rangle$ within the Hilbert space $V^\Lambda$,
which can be normalized to unity:
$
 \langle \Lambda |\Lambda \rangle = 1 .
$
\end{enumerate}
\par
We define the {\it maximal stability} subgroup ({\it isotropy}
subgroup)
$\tilde H$ as a subgroup of
$G$ that consists of all the group elements $h$ that leave the
reference state $|\Lambda \rangle$ invariant up to a phase factor:
\begin{equation}
  h |\Lambda \rangle = |\Lambda \rangle e^{i\phi(h)}, h \in \tilde H .
\end{equation}
The phase factor is unimportant in the following discussion because
we consider the expectation value of operators in the coherent
state. 
Let $H$ be the Cartan subgroup of $G$, i.e., the maximal commutative
semi-simple subgroup in $G$,  and Let ${\cal H}$ be the Cartan subalgebra
in ${\cal G}$, i.e., the Lie algebra for the group $H$.
 The maximal stability subgroup $\tilde H$ includes the Cartan
subgroup $H=U(1)^r$, i.e., $H \subset \tilde H$.
\par
For every element
$g\in G$, there is a unique decomposition of
$g$ into a product of two group elements, 
\begin{equation}
 g = \xi h, \quad \xi \in G/\tilde H, \quad h \in \tilde H ,
\end{equation}
for $g \in G$.
We can obtain a unique coset space $G/\tilde H$ for a given $|\Lambda \rangle$.
The action of arbitrary group element $g\in G$  on 
$|\Lambda \rangle$ is given by
\begin{equation}
  g |\Lambda \rangle = \xi h  |\Lambda \rangle
  = \xi  |\Lambda \rangle  e^{i\phi(h)}.
\end{equation}
\par
The coherent state is constructed as
$
  |\xi, \Lambda \rangle = \xi |\Lambda \rangle .
$
This definition of the coherent state is in one-to-one
correspondence with the coset space $G/\tilde H$ and the coherent states
preserve all the algebraic and topological properties of the coset
space $G/\tilde H$.
\par
If $\Gamma^\Lambda({\cal G})$ is Hermitian, then $H_i^\dagger=H_i$ 
and $E_{\alpha}^\dagger = E_{-\alpha}$.  
Every group element $g \in G$ can be written as  the exponential of
a complex linear combination of diagonal operators $H_i$ and
off-diagonal shift operators $E_\alpha$.  
Let $|\Lambda \rangle$ be the highest-weight state, i.e., 
$H_j | \Lambda \rangle = \Lambda_j | \Lambda \rangle$, 
$E_\alpha | \Lambda \rangle = 0$ for $\alpha \in R_+$, 
where
$R_+ (R_-)$ is a subsystem of positive (negative) roots.
Then the coherent
state is given by \cite{FGZ90}
\begin{eqnarray}
  |\xi, \Lambda \rangle 
  = \xi |\Lambda \rangle
  = \exp \left[ 
  \sum_{\beta\in R_{-}} \left( \eta_\beta E_\beta - 
  \bar{\eta}_\beta E_{\beta}^{\dagger}\right) \right] |\Lambda \rangle,
\quad \eta_\beta \in \bf{C},
\label{cohrentdef}
\end{eqnarray}
such that the following hold:
\begin{enumerate}
\item[(i)]
 $|\Lambda \rangle$  is
annihilated by all the (off-diagonal) shift-up operators $E_{\alpha}$ with
$\alpha \in R_+$, 
$
 E_{\alpha} |\Lambda \rangle = 0 (\alpha \in R_+) ;
$
\item[(ii)] 
$|\Lambda \rangle$  is
mapped into itself by all diagonal operators $H_i$,
$
 H_{i} |\Lambda \rangle = \Lambda_i |\Lambda \rangle ;
$
\item[(iii)] 
$|\Lambda \rangle$  is
annihilated by some shift-down operators $E_{\alpha}$ with
$\alpha \in R_-$, not by other $E_{\beta}$ with $\beta \in R_-$:
$
 E_{\alpha} |\Lambda \rangle = 0 ({\rm some~} \alpha \in R_-) ;
$
$
 E_{\beta} |\Lambda \rangle = |\Lambda+\beta \rangle 
 ({\rm some~} \beta \in R_-) ;
$
\end{enumerate}
and the sum $\sum_{\beta}$ is restricted to those shift operators
$E_{\beta}$ which obey (iii).
\par
The coherent states are normalized to unity: 
\begin{equation}
 \langle \xi, \Lambda | \xi, \Lambda \rangle = 1 .
\end{equation}
The coherent state spans the entire space $V^\Lambda$.
However, the coherent states are non-orthogonal: 
\begin{equation}
 \langle \xi' , \Lambda | \xi, \Lambda \rangle \not= 0 .
\end{equation}
By making use of the the group-invariant measure $d\mu(\xi)$ of
$G$ which is appropriately normalized, we obtain
\begin{eqnarray}
  \int |\xi, \Lambda \rangle d\mu(\xi) 
  \langle \xi,  \Lambda |
  = I ,
\end{eqnarray}
which shows that the coherent states are complete, but in fact over-complete.
This resolution of identity is very important to obtain the path
integral formula of the Wilson loop given in $\S$ 4.
\par
The coherent states 
$|\xi, \Lambda \rangle$
are in one-to-one correspondence with the coset representatives $\xi \in G/\tilde H$:
\begin{equation}
 |\xi, \Lambda \rangle \leftrightarrow G/\tilde H .
\end{equation}
In other  words, $|\xi, \Lambda \rangle$
and $\xi \in G/\tilde H$ are topologically equivalent.

\section{Flag space and coherent state for $SU(N)$}
\setcounter{equation}{0}

\subsection{$SU(2)$ coherent state}

\par
\begin{figure}
\begin{center}
 \leavevmode
 \epsfxsize=80mm
 \epsfysize=40mm
 \epsfbox{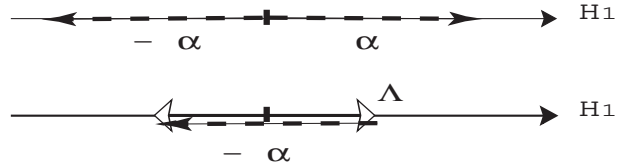}
\end{center} 
 \caption[]{Root diagram and weight diagram of the  fundamental representation of $SU(2)$ where  $\Lambda$ is the highest weight of the fundamental representation.}
 \label{root_SU2}
\end{figure}

In the case of $SU(2)$, the maximal stability group agrees with the maximal torus group $U(1)$ irrespective of the representation.
The $SU(2)$ case is well known (see, e.g., Ref.\citen{KondoIV}).
The weight and root diagrams are given in Fig.\ref{root_SU2}.

  The coherent
state for
$F_1:=SU(2)/U(1)$ is obtained as
\begin{eqnarray}
 |j, w \rangle = \xi(w) |j, -j \rangle 
 = e^{\zeta J_{+} - \bar \zeta J_{-}}  |j, -j \rangle 
 = {1 \over (1+|w|^2)^j}e^{ w J_{+}}  |j, -j \rangle ,
\end{eqnarray}
where $|j, -j \rangle $ is the lowest
state, 
$
 |j, m=-j \rangle $,
 of $|j, m \rangle ,$
and
\begin{equation}
  J_{+} = J_1+iJ_2 , \quad
  J_{-}=J_{+}^\dagger ,  \quad
 w = {\zeta \sin |\zeta| \over |\zeta| \cos |\zeta|} .
\end{equation}
Note that $(1+|w|^2)^{-j}$ is a normalization factor that ensures 
$
 \langle j, w |j, w \rangle = 1 ,
$
which is obtained from the Baker-Campbell-Hausdorff (BCH) formulas.
The invariant measure is given by
\begin{eqnarray}
 d\mu = {2j+1 \over 4\pi} {dwd\bar w \over (1+|w|^2)^2} .
\end{eqnarray}
For
$
J_A = {1 \over 2}\sigma^A (A=1,2,3)  
$
with Pauli matrices $\sigma^A$, we obtain
$
  J_{+} = \pmatrix{0 & 1 \cr 0 & 0} ,
$ 
and
\begin{equation}
 e^{ w J_{+}}  = \pmatrix{1 & w \cr 0 & 1} \in F_1 = CP^1=
SU(2)/U(1) \cong S^2.
\label{s2}
\end{equation}
\par
The complex variable $w$ is a $CP^1$ variable written as 
$
 w = e^{i\phi} \cot {\theta \over 2},
$
in terms of the polar coordinate on $S^2$ or Euler angles, see
Ref.\citen{KondoII}. We introduce the $O(3)$ vector ${\bf n}$ as
\begin{eqnarray}
n^1:=\sin{\theta}\cos{\phi},\quad n^2:=\sin{\theta}\sin{\phi},\quad
 n^3:=\cos{\phi}.
\end{eqnarray}
The relation 
\begin{equation}
 n^A(x) =   \bar \phi_a(x) \sigma_{ab}^A \phi_b(x) \quad (a,b=1,2)  
\end{equation}
is equivalent to
\begin{equation}
 n_1 = 2 \Re(\phi_1 \bar \phi_2), \quad
 n_2 = 2 \Im(\phi_1 \bar \phi_2), \quad
 n_3 = |\phi_1|^2 - |\phi_2|^2 .
 \label{defn}
\end{equation}
The complex coordinate $w$ obtained by the stereographic projection from the north pole is identical to the inhomogeneous local coordinates of $CP^1$ when $\phi_2 \not=0$,
\begin{equation}
 w = w^{(1)} + i w^{(2)}
 = {n_1 + i n_2 \over 1-n_3}
 = {2 \phi_1 \bar \phi_2 \over (|\phi_1|^2 + |\phi_2|^2)
 - (|\phi_1|^2 - |\phi_2|^2)}
 = {\phi_1 \over \phi_2} .
\end{equation}
The stereographic projection from the south pole leads to 
\begin{equation}
 w = {n_1 + i n_2 \over 1+n_3}
 = \left({\phi_2 \over \phi_1} \right)^*  
\end{equation}
if $\phi_1 \not=0$.
The variable $w$ is $U(1)$ gauge invariant.
Another representation of {\bf n} is obtained by using the parameterization 
(\ref{s2}) of the $F_1$ variable $\xi$:
\begin{equation}
 n^A =   \langle \Lambda | \xi(w)^\dagger \sigma^A \xi(w) | \Lambda \rangle 
 =  \pmatrix{\bar \phi_1 & 0}\pmatrix{1 & \bar w \cr 0 & 1}
 \sigma^A \pmatrix{1 & 0 \cr w & 1} \pmatrix{\phi_1 \cr 0} .
\end{equation}
This leads to 
\begin{equation}
 n_1 =  |\phi_1|^2(w+\bar w), \quad
 n_2 = -i |\phi_1|^2(w-\bar w), \quad
 n_3 =  |\phi_1|^2(1-w\bar w) .
\end{equation}
Indeed, this agrees with (\ref{defn}) if 
$w=({\phi_2 \over \phi_1})^*$.
The entire space of $F_1$ is covered by two charts,
\begin{equation}
 CP^{1} = U_1 \cup U_2, \quad
 U_a = \{ (\phi_1, \phi_2) \in CP^{1}; \phi_a \not= 0 \} .
\end{equation}

\subsection{$SU(3)$ coherent state}

\begin{figure}
\begin{center}
 \leavevmode
 \epsfxsize=90mm
 \epsfysize=50mm
 \epsfbox{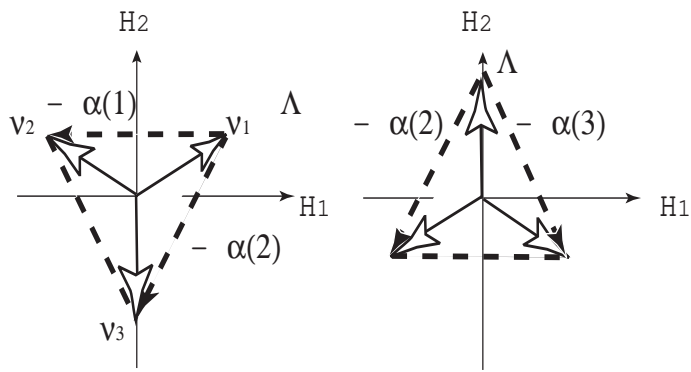}
\end{center} 
 \caption[]{The weight diagram and root vectors required to define
the coherent state in the fundamental representations $[1,0]={\bf
3}$, $[0,1]={\bf 3}^*$  of $SU(3)$ where  $\vec
\Lambda = \vec h_1=\nu^1:=({1
\over 2},{1 \over 2\sqrt{3}})$ is the highest weight of the
fundamental representation, and the other weights are 
 $\nu^2:=(-{1 \over 2},{1 \over 2\sqrt{3}})$ and
 $\nu^3:=(0,-{1 \over \sqrt{3}})$.}
 \label{fundamental-weight}
\end{figure}

\begin{figure}
\begin{center}
 \leavevmode
 \epsfxsize=60mm
 \epsfysize=60mm
 \epsfbox{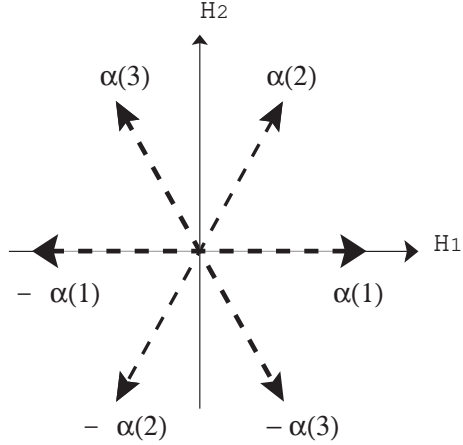}
\end{center} 
 \caption[]{The root diagram of $SU(3)$, where positive root vectors are given by 
$\alpha^{(1)}=(1,0)$,
$\alpha^{(2)}=({1 \over 2},{\sqrt{3} \over 2})$, and
$\alpha^{(3)}=({-1 \over 2},{\sqrt{3} \over 2})$.
 Here we have used the same weight ordering as in the $SU(N)$ case (see (\ref{order})) 
 in defining the simple roots. Then the two simple roots are given by $\alpha^1:=
 \alpha=\alpha^{(1)},
$
and
$ \alpha^2:=\beta=\alpha^{(3)}$.}
 \label{root}
\end{figure}

For concreteness, we first focus on the $SU(3)$ case.  
The general $SU(N)$ case will be discussed in the final part of this
section.
The highest weight $\Lambda$ of the representation specified
by the Dynkin index $[m,n]$ can be written as
\begin{equation}
  \vec \Lambda = m \vec h_1 + n \vec h_2 ,
 \label{Dynkin index}
\end{equation}
where $m$ and $n$ are non-negative integers for the highest weight and
$h_1$ and $h_2$ are the highest weights of the two fundamental representations of
$SU(3)$ corresponding to $[1,0]$ and $[0,1]$, respectively (see
Fig.\ref{fundamental-weight}) 
\begin{equation}
  \vec h_1 = \left({1 \over 2}, {1 \over 2\sqrt{3}} \right), \quad
  \vec h_2 = \left(0, {1 \over \sqrt{3}} \right) .
\end{equation}
Therefore, we obtain\footnote{
This choice of $h_2$ is different from that in Ref.\citen{KT99a} 
It is adopted so as to obtain the $SU(3)$ case when considering the $N=3$ case of 
$SU(N)$ case studied in the next subsecton.}
\begin{equation}
  \vec \Lambda =  \left({m \over 2}, {m+2n \over 2\sqrt{3}} \right)
.
\end{equation}
The generators  of SU(3) in the Cartan basis are written as
$\{ H_1, H_2, E_{\alpha}, E_{\beta}, E_{\alpha+\beta},$ $  E_{-\alpha}, E_{-\beta}, E_{-\alpha-\beta} \}$,
where $\alpha$ and $\beta$ are the two simple roots.
(See Fig.\ref{root} for the explicit choice.)

\begin{figure}
\begin{center}
 \leavevmode
 \epsfxsize=60mm
 \epsfysize=60mm
 \epsfbox{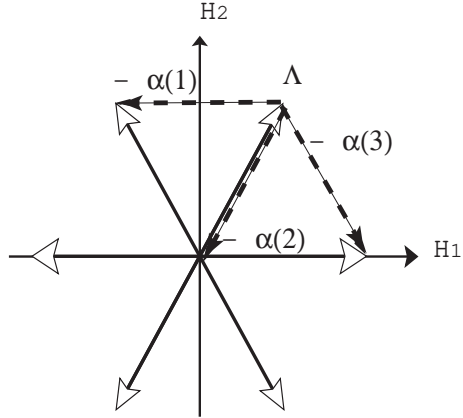}
\end{center} 
 \caption[]{The weight diagram and root vectors required to define
the coherent state in the adjoint representation $[1,1]={\bf 8}$ of
$SU(3)$, where  $\Lambda=({1 \over 2},{\sqrt{3} \over 2})$ is the highest weight of the adjoint
representation.}
 \label{adjoint-weight}
\end{figure}

If $mn=0$, ($m=0$ or $n=0$), the maximal stability group
$\tilde H$ is given by $\tilde H=U(2)$ with generators 
$\{ H_1, H_2, E_\beta, E_{-\beta} \}$ (case (I)). 
Such a degenerate case occurs when the highest-weight vector $\vec
\Lambda$ is orthogonal to some root vectors  
(see Fig.\ref{fundamental-weight}).
If $mn \not=0$ 
($m\not=0$ and $n\not=0$), $H$ is the maximal torus group 
$\tilde H=U(1) \times U(1)$  with generators 
$\{ H_1, H_2 \}$ (case (II)). This is a non-degenerate case (see Fig.\ref{adjoint-weight}).
Therefore, for the highest weight
$\Lambda$ in  case (I), the coset $G/\tilde H$ is given by
\begin{equation}
  SU(3)/U(2)=SU(3)/(SU(2)\times U(1))=CP^2,
\end{equation}
whereas in case (II), 
\begin{equation}
  SU(3)/(U(1)\times U(1))=F_2 .
\end{equation}
Here, $CP^{n}$ is the
complex projective space and $F_n$ is the flag space.\cite{Perelomov87}  Therefore, the two fundamental representations
belong to case (I), and hence the maximal stability group is $U(2)$,
rather than the maximal torus group $U(1) \times U(1)$.  
The implications of this fact for the mechanism of quark confinement
is discussed in subsequent sections.

\par
The coherent state for 
$F_2 = SU(3)/U(1)^2$ is given by
\begin{equation}
 | \xi, \Lambda \rangle = \xi(w) | \Lambda \rangle 
 := V^\dagger(w) | \Lambda \rangle ,
\end{equation}
with the highest- (lowest-) weight state $| \Lambda \rangle$, i.e.,
\begin{eqnarray}
 | \xi, \Lambda \rangle
 &=& \exp \left[\sum_{\alpha\in R_{+}}(\zeta_\alpha E_{-\alpha} -
\bar\zeta_\alpha E_{-\alpha}^\dagger ) \right] | \Lambda \rangle
\nonumber \\
 &=&  e^{-{1 \over 2}K(w,\bar w)}  
\exp \left[\sum_{\alpha\in R_{+}} \tau_\alpha E_{-\alpha} \right]  |
\Lambda \rangle ,
\label{cohe}
\end{eqnarray}
where
$e^{-{1 \over 2}K}$ is the normalization factor obtained from
the K\"ahler potential (explained below):
\begin{eqnarray}
  K(w,\bar w) 
  &:=& \ln [(\Delta_1(w,\bar w))^m(\Delta_2(w,\bar w))^n] ,
\\
 \Delta_1(w,\bar w) &:=& 1+|w_1|^2+|w_2|^2,
 \Delta_2(w,\bar w) := 1+|w_3|^2+|w_2-w_1 w_3|^2 . 
\label{Delta}
\end{eqnarray}
The coherent state 
$
 | \xi, \Lambda \rangle 
$
is normalized, so that
$
 \langle \xi, \Lambda | \xi, \Lambda \rangle = 1 .
$
We show in Appendix A that the inner product is given by
\begin{eqnarray}
 \langle \xi', \Lambda | \xi, \Lambda \rangle 
 =  e^{K(w, \bar w')} e^{-{1\over2}[K(w',\bar w')+K(w,\bar w)]} ,
 \label{norm}
\end{eqnarray}
where 
\begin{equation}
  K(w, \bar w') := 
 \ln [1+\bar w_1{}' w_1+\bar w_2{}'w_2]^m[1+\bar w_3{}' w_3
 +(\bar w_2{}'-\bar w_1{}'\bar w_3{}')(w_2-w_1w_3)]^n .
\end{equation}
Note that $K(w, \bar w')$ reduces to the K\"ahler potential $K(w,
\bar w)$ when
$w'=w$, in agreement with the normalization.
It follows from the general formula (see discussion of the $SU(N)$ case) that the
$SU(3)$ invariant measure is given (up to a constant factor) by
\begin{eqnarray}
 d\mu(\xi) = d\mu(w, \bar w)
 = D(m,n)[(\Delta_1)^m (\Delta_2)^n]^{-2} \prod_{\alpha=1}^{3}
dw_\alpha d\bar w_\alpha , 
\end{eqnarray}
where $D(m,n)={1 \over 2}(m+1)(n+1)(m+n+2)$ is the dimension of the
representation. For the choice of shift-up $(E_{+i})$ or shift-down 
$(E_{-i})$ operators
\begin{equation}
  E_{\pm 1} := \frac{\lambda_1\pm i\lambda_2}{2\sqrt{2}}, \quad
  E_{\pm 2} := \frac{\lambda_4\pm i\lambda_5}{2\sqrt{2}}, \quad
  E_{\pm 3} := \frac{\lambda_6\pm i\lambda_7}{2\sqrt{2}}, 
\end{equation}
with the Gell-Mann matrices $\lambda_A(A=1,\cdots,8)$,
we obtain
\begin{equation}
\exp \left[\sum_{i=1}^{3} \tau_{i} E_{-i} \right]  
 = \pmatrix{1 & w_1 & w_2 \cr 0 & 1 & w_3 \cr 0 & 0 & 1}^t 
 \in F_2 = SU(3)/U(1)^2 ,
\end{equation}
where we have used the abbreviation $E_{\pm i}\equiv E_{\pm\alpha ^{(i)}}
\ (i=1,2,3)$.
These two sets of three complex variables are related as (see Appendix A)
\begin{equation}
 w_1 = \frac{\tau_1}{\sqrt{2}}, \quad w_2 = \frac{\tau_2}{\sqrt{2}} + \frac
 {\tau_1\tau_3}{4}, \quad w_3 = \frac{\tau_3}{\sqrt{2}} ,
\end{equation}
or conversely
\begin{equation}
 \tau_1 = \sqrt{2}w_1,\quad \tau_2 = \sqrt{2}\left(w_2 - \frac{w_1 w_3}{2}\right), 
\quad \tau_3 = \sqrt{2}w_3 .
\end{equation}
The complex projective space $CP^2$ is covered by three complex planes
${\bf C}$ through holomorphic maps \cite{Ueno95} (see Appendix B).
The parameterization of $SU(3)$ in terms of eight angles is also
possible  in
$SU(3)$, just as $SU(2)$ is parameterized by three Euler angles (see
Ref.\citen{FiniteRotation}).

\par

\subsection{$SU(N)$ case}

\par
For $SU(N)=SU(n+1)$,  the flag space \cite{Perelomov87} is defined by
\begin{equation}
 F_n= SU(n+1)/U(1)^n \ni V  .
\label{Fndef1}
\end{equation}
We use $V$ to denote an element of
$F_n$ in this definition.
$F_n$ is a compact K\"ahler manifold,\cite{Nakahara90,GSW2} which 
is a homogeneous but nonsymmetric manifold of dimension
${\rm dim}_{\bf C} F_{n}=n(n+1)/2$.
\par
Since the flag manifold $F_n$ is a
K\"ahler manifold,\cite{Nakahara90,GSW2}  it
possesses complex {\it local} coordinates
$w_{\alpha}$, a Hermitian Riemannian metric,
\begin{equation}
ds^2=g_{\alpha\bar \beta}dw^{\alpha}d\bar w^{\beta},
\end{equation}
and a
corresponding two-form, called the 
K\"ahler form,%
\footnote{The imaginary unit $i$ is needed to make the K\"ahler two-form real, since
$
 {\bar{g_{\alpha \bar \beta}}}= g_{\bar \alpha \beta}
 = g_{\beta \bar \alpha } . 
$
}
\begin{equation}
\Omega_K=i g_{\alpha\bar \beta}dw^{\alpha} \wedge d\bar w^{\beta},
\label{K2f}
\end{equation}
which is closed, i.e., 
\begin{equation}
  d\Omega_K=0 .
  \label{closed}
\end{equation}
Any closed form
$\Omega_K$ is {\it locally} exact 
($\Omega_K= d\omega$), due to Poincar\'e's lemma. 
The condition ({\ref{closed}) is equivalent to 
\begin{equation}
 {\partial g_{\alpha \bar \beta} \over \partial w^\gamma}
 = {\partial g_{\gamma \bar \beta} \over \partial w^\alpha},
 \quad {\rm or} \quad
 {\partial g_{\alpha \bar \beta} \over \partial \bar w^\gamma}
 = {\partial g_{\alpha \bar \gamma} \over \partial \bar w^\beta} .
\end{equation}
This holds if and only if the metric $g_{\alpha\beta}$ can be
obtained from a real scalar function $K$ as
\begin{equation}
 g_{\alpha\bar\beta} = {\partial \over \partial w^\alpha}
 {\partial \over \partial \bar w^\beta} K ,
\end{equation}
where  $K=K(w,\bar w)$ is called the K\"ahler potential. 
Then the K\"ahler two-form is obtained from (\ref{K2f}) as
\begin{equation}
 \Omega_K  = i \partial \bar \partial K .
\label{OmegaK}
\end{equation}

\par
On the flag space, there transitively act two groups, $G=SU(n+1)$ and
 its complexification $G^c=SL(n+1,{\bf C})$. 
Any element of $F_{n}$ can be written as an upper
triangular $(n+1) \times (n+1)$ matrix, whose main diagonal elements are all $1$ and whose upper
$n(n+1)/2$ elements are complex numbers, $w_\alpha \in {\bf C}$: 
\begin{equation}
 \xi = \pmatrix{
 1 & w_1 & w_2     & \cdots & \cdots &w_n \cr
 0 &  1  & w_{n+1} & \cdots & \cdots & w_{2n-1} \cr
 0 & 0   & 1 & w_{2n}  & \cdots & w_{3n-3} \cr
   &   &  &  &   &   \cr
 \vdots & \vdots &\vdots &\vdots & \vdots & \vdots \cr
   &   &  &  &   &   \cr
 0 & 0   & \cdots & 0 &  1 & w_{n(n+1)/2} \cr 
 0 & 0   & \cdots & \cdots & 0 & 1 \cr 
 }  \in F_n .
\end{equation}
Therefore, we can write
\begin{equation}
 F_n = SL(n+1,{\bf C})/B_- \ni \xi ,
\label{Fndef2}
\end{equation}
where $B_- (B_+)$ is the Borel subgroup, i.e., the group of lower
(upper) triangular matrices with determinant equal to $1$ (Iwasawa
decomposition). 
This definition (\ref{Fndef2})
\footnote{Note that $\xi$ is not necessarily unitary as a matrix
under this definition.}
should be compared with the first definition
(\ref{Fndef1}). The mapping
$G/H
\rightarrow G^c/B_-$ is a generalization of the stereographic projection in the
$G=SU(2)$ case.\cite{KondoII}
The action of the group $SL(n+1,{\bf C})$ on $F_n$, 
$g: V \rightarrow V_g$  can be found through the Gauss decomposition,
\begin{equation}
 V \cdot g = TD V_g , \quad g \in SL(n+1, {\bf C}), \quad
T \in Z_-(n+1), \quad V_g \in Z_+(n+1) ,
\end{equation}
where $Z_+(n+1) (Z_-(n+1))$ is the set of upper (lower)
triangular matrices whose  main diagonal elements are all $1$ and $D$ is a
diagonal matrix with determinant equal to 1.  The elements of the
factors $T, D$ and $V_g$ are rational functions of the elements of $g$.%
\footnote{
For $n=1$, 
\begin{equation}
 V = \pmatrix{1 & w \cr 0 & 1}, \quad
 g = \pmatrix{a & b \cr c & d}, \quad
 V_g = \pmatrix{1 & {aw+b \over cw+d} \cr 0 & 1} .
\end{equation}
Hence $w$ is the complex one-dimensional representation of $SL(2,C)$.
}
\par
 The group $G=SU(N)$ has rank $N-1$, and the Cartan subalgebra is
constructed from $(N-1)$ diagonal generators $H_i$.  Hence, there
are $N(N-1)$ off-diagonal shift operators $E_\alpha$, since ${\rm dim}
SU(N):=N^2-1=(N-1)+N(N-1)$. Therefore, the total number of roots is
$N(N-1)$, of which there are $N-1$ simple roots. Other roots are
constructed as linear combination of the simple roots.   Also, there are
$N$ weight vectors.  An element of $SU(N)$ is represented by the
$N\times N$ unitary matrices with determinant $1$ that are
generated by traceless Hermitian matrices, $N^2-1$ linearly
independent generators
$T^A(A=1,\cdots,N^2-1)$.  The generators are normalized as
\begin{equation}
 {\rm tr}(T^A T^B) = {1 \over 2} \delta_{AB} .
\end{equation}
Each off-diagonal generator $E_\alpha$ has a single non-zero element
$1/\sqrt{2}$. The diagonal generator $H_m$ is
defined by
\begin{eqnarray}
 (H_m)_{ab} &=& {1 \over \sqrt{2m(m+1)}}
 (\sum_{k=1}^{m} \delta_{ak}\delta_{bk} - m \delta_{a,m+1}\delta_{b,m+1})
 \\
 &=& {1 \over \sqrt{2m(m+1)}}{\rm diag}(1,\cdots,1,-m,0,\cdots,0) .
\label{H}
\end{eqnarray}
For $m=1$ to $N-1$, the first $m$ diagonal elements (beginning from the upper left-hand corner) of $H_m$ are $1$, the next one is $-m$, and the rest of the diagonal elements are $0$.  Thus $H_m$ is traceless.
 The weight
vectors (eigenvectors of all
$H_i$: 
$H_j |\nu \rangle = \nu^j |\nu \rangle$) of the
fundamental representation
${\bf N}$ ($N$-dimensional irreducible representation of $SU(N)$) are
given  by
\cite{Georgi82}
\begin{eqnarray}
  \nu^1 &=& \left({1 \over 2}, {1 \over 2\sqrt{3}}, \cdots,
  {1 \over \sqrt{2m(m+1)}}, \cdots, {1 \over \sqrt{2(N-1)N}}\right) ,
  \nonumber\\
  \nu^2 &=& \left(-{1 \over 2}, {1 \over 2\sqrt{3}}, \cdots,
  {1 \over \sqrt{2m(m+1)}}, \cdots, {1 \over \sqrt{2(N-1)N}}\right) ,
  \nonumber\\
  \nu^3 &=& \left(0, -{1 \over \sqrt{3}}, {1
\over 2\sqrt{6}},\cdots,  {1 \over \sqrt{2(N-1)N}}\right) ,
  \nonumber\\
\vdots
  \nonumber\\
  \nu^{m+1} &=& \left(0,0, \cdots,0,
  -{m \over \sqrt{2m(m+1)}}, \cdots, {1 \over \sqrt{2(N-1)N}}
\right) ,
  \nonumber\\
\vdots
  \nonumber\\
  \nu^N &=& \left(0, 0, \cdots, 0, {-N+1 \over \sqrt{2(N-1)N}}
\right) .
\end{eqnarray}
All the weight vectors have the same length, and the angles between different weights are the same:
\begin{equation}
 \nu^i \cdot \nu^i = {N-1 \over 2N},
 \quad \nu^i \cdot \nu^j = -{1 \over 2N} \quad ({\rm for} i \not= j).
 \label{wr}
\end{equation}
The weights constitute a polygon in the $N-1$ dimensional
space.  This implies that any weight can be used as the highest
weight. 
A weight will be called positive if its {\it last} non-zero component
is positive. With this definition, the weights satisfy 
\begin{eqnarray}
 \nu^1 > \nu^2 > \cdots > \nu^N .
\label{order}
\end{eqnarray}
The simple roots are given by  
\begin{equation}
 \alpha^i = \nu^i - \nu^{i+1} \quad (i=1, \cdots, N-1) 
\end{equation}
Explicitly, we have
\begin{eqnarray}
  \alpha^1 &=& \left(1, 0, \cdots, 0 \right),
  \nonumber\\
  \alpha^2 &=& \left(-{1 \over 2}, {\sqrt{3} \over 2}, 0, \cdots,0 \right),
  \nonumber\\
  \alpha^3 &=& \left(0, -{1 \over \sqrt{3}}, \sqrt{{2 \over 3}},0, \cdots, 0 \right),
  \nonumber\\
\vdots
  \nonumber\\
  \alpha^{m} &=& \left(0,0, \cdots, 
  -\sqrt{{m-1 \over 2m}}, \sqrt{{m+1 \over 2m}}, 0, \cdots, 0  \right),
  \nonumber\\
\vdots
  \nonumber\\
  \alpha^{N-1} &=& \left(0, 0, \cdots,  -\sqrt{{N-2 \over 2(N-1)}},
\sqrt{{N \over 2(N-1)}} \right) .
\label{simple root}
\end{eqnarray}
As can be shown from (\ref{wr}),
all these roots have length 1, the angles between successive roots
are the same, $2\pi/3$, and other roots are orthogonal:
\begin{eqnarray}
   \alpha^j \cdot \alpha^j = 1,
  \quad \alpha^i \cdot \alpha^j = - {1 \over 2}, \quad (j=i\pm1) 
  \\
  \alpha^i \cdot \alpha^j = 0 . \quad (j \not= i, i\pm 1)
  \label{rr}
\end{eqnarray}
This fact is usually expressed by a Dynkin diagram 
(see Fig.\ref{dynkinSUN}).

\begin{figure}
\begin{center}
 \leavevmode
 \epsfxsize=80mm
 \epsfysize=20mm
 \epsfbox{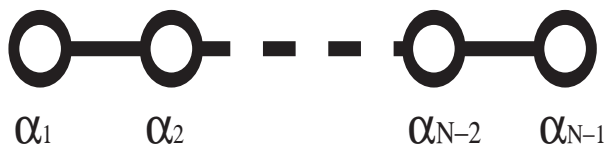}
\end{center} 
 \caption[]{The Dynkin diagram of $SU(N)$.}
 \label{dynkinSUN}
\end{figure}

If we choose $\nu^1$ as the highest-weight $\vec \Lambda$ of the
fundamental representation ${\bf N}$, some of the roots are
orthogonal to
$\nu^1$.    From the above construction, it is easy to see that only
one simple root
$\alpha_1$ is not-orthogonal to $\nu^1$, and that all the other
simple roots are orthogonal:
\begin{equation}
 \nu^1 \cdot \alpha^1 \not= 0,
 \quad \nu^1 \cdot \alpha^2 = \nu^1 \cdot \alpha^3
 = \cdots = \nu^1 \cdot \alpha^{N-1} = 0 .
\end{equation}
Therefore, all the linear combinations constructed from 
$\alpha^2, \cdots, \alpha^{N-1}$ are also orthogonal to $\nu^1$.
Non-orthogonal roots are obtained only when $\alpha^1$ is included
in the linear combinations.  It is not difficult to show that the
total number of non-orthogonal roots is $2(N-1)$, and hence there are 
$N(N-1)-2(N-1)=(N-2)(N-1)$ orthogonal roots.  The $(N-2)(N-1)$
shift operators $E_\alpha$ corresponding to these orthogonal roots
together with the $N-1$ Cartan subalgebra $H_i$ constitute  the maximal
stability subgroup
$\tilde H=U(N-1)$, since
$(N-2)(N-1)+(N-1)=(N-1)^2={\rm dim} U(N-1)$.  Thus, for the fundamental
representation, the stability subgroup $\tilde H$ of $SU(N)$ is
given  by $\tilde H=U(N-1)$. In order to describe the coset space
$G/\tilde H$, we need only $(N-1)$ complex numbers, since
\begin{equation}
 G/\tilde H = SU(N)/U(N-1) ,
\end{equation}
and ${\rm dim} G/\tilde H= 2(N-1)$.  We conclude that
$G/\tilde H=CP^{N-1}$, where $CP^{N-1}$ is the $(N-1)$-dimensional complex projective
space,\cite{Nakahara90,GSW2,Ueno95,Picken90}
which is a submanifold of the flag manifold $F_{N-1}$.
\par
The complex projective space $CP^n$ is the compact K\"ahler symmetric space with ${\rm dim}_{\bf C} CP^{n}=n$.  The $SU(n+1)$ group can act transitively on this manifold, and this manifold can be considered as a factor space:
\begin{equation}
 {\bf C}P^n =  SU(n+1)/(SU(n)\times U(1)) .
\end{equation}
An element of $CP^n$ can be expressed using the $n$ complex variables $w_1,
\cdots, w_n$ as
\begin{equation}
 \pmatrix{
 1 & w_1 & w_2     & \cdots & \cdots &w_n \cr
 0 &  1  & 0 & \cdots & \cdots & 0 \cr
 0 & 0   & 1 & 0  & \cdots & 0 \cr
   &   &  &  &   &   \cr
 \vdots & \vdots &\vdots &\vdots & \vdots & \vdots \cr
   &   &  &  &   &   \cr
 0 & 0   & \cdots & 0 &  1 & 0 \cr 
 0 & 0   & \cdots & \cdots & 0 & 1 \cr 
 }  \in CP^n .
\end{equation}
\par
The K\"ahler potential of $F_n$ ($n=N-1$) is obtained as follows. 
Let
$H$ be the Hermitian matrix defined by
$H=VV^\dagger, V \in F_n$.  Consider the Gauss decomposition\begin{equation}
  H = VV^\dagger = T D V,
\end{equation}
where $T \in Z_-(n+1)$ and $D$ is a diagonal matrix with determinant equal to 1.
$H$ has the upper principal minors $\Delta_\ell(VV^\dagger) 
(\ell=1,\cdots, n+1)$,  which are equal to the following products of
the elements
$\delta_\ell$ of the diagonal matrix $D={\rm diag}(\delta_1, \cdots,
\delta_n)$:
\begin{equation}
 \Delta_1 = \delta_1, \quad
 \Delta_2 = \delta_1 \delta_2, \quad \cdots, \quad
 \Delta_n = \delta_1 \delta_2 \cdots \delta_n, \quad
 \Delta_{n+1} = 1.
\end{equation}
Let $d_\ell$ $(\ell=1,\cdots,n)$ be the Dynkin index of $SU(n+1)$.  Then the K\"ahler potential of $F_n$ is given in the form,
\begin{equation}
 K(w,\bar w) = \sum_{\ell=1}^{n} d_\ell K_\ell(w,\bar w)
 = \sum_{\ell=1}^{n} d_\ell \ln \Delta_\ell(w,\bar w) 
 = \ln \left[\prod_{\ell=1}^{n} (\Delta_\ell(w,\bar w))^{d_\ell}\right] .
\end{equation}
The function $K(w,\bar w')$ can also be obtained from the Gauss decomposition of
$VV'^\dagger$:
\begin{equation}
 K(w,\bar w') = \ln \left[ \prod_{\ell=1}^{n} 
(\Delta_\ell(w,\bar w'))^{d_\ell} \right] .
\end{equation}
The $SU(n+1)$ invariant measure on $F_n$ is written, up to a
multiplicative factor, as \cite{Zelobenko73}
\begin{eqnarray}
  d\mu(V, \bar V) = \rho(V,\bar V) \prod_{\alpha=1}^{n(n+1)/2} dw_\alpha d\bar w_\alpha ,
  \\
  \rho(V,\bar V) = \left[ \prod_{\ell=1}^{n} \Delta_\ell^{d_\ell} \right]^{-2} = \left[ \prod_{\ell=1}^{n} \Delta_\ell \right]^{-2} ,
\end{eqnarray}
where $d_\ell=1$ for all $\ell$.
The density $\rho$ of the invariant measure is calculated \cite{Perelomov87} from
\begin{equation}
 \rho = \det(g_{\alpha \bar \beta}) 
 = \det \left( {\partial^2 K \over 
\partial w^\alpha \partial \bar w^\beta} \right)  .
\end{equation}
\par
The K\"ahler potential of the $CP^{N-1}$ manifold is given by
\begin{equation}
 K  = m \ln \left( 1 + \sum_{\alpha=1}^{N-1} |w_\alpha|^2 \right)
 = m \ln \left( 1 + |||w|||^2 \right) ,
 \label{KpCP}
\end{equation}
where
\begin{eqnarray}
 |||w|||^2 := \sum_{a=1}^{N-1} |w_a|^2  .
\end{eqnarray}
Hence, the metric reads
\begin{equation}
 g_{\alpha\bar \beta} = m {(1+|||w|||^2)\delta_{\alpha\beta} - \bar
w_\alpha w_\beta \over
 (1+|||w|||^2)^2} .
 \label{metricCP}
\end{equation}
\par
The above construction of the coherent state can be extended to an
arbitrary compact semi-simple Lie group (see Chapter 11 of Perelomov
\cite{Perelomov86}).

\section{Non-Abelian Stokes theorem}
\setcounter{equation}{0}

In this section we derive a new version of the non-Abelian Stokes
theorem (NAST) based on the coherent state obtained in the previous
section.  An advantage of this version of NAST is that it is
possible to separate the magnetic monopole contribution in the
Wilson loop and that it is very
helpful to understand the dual superconductor picture of quark
confinement in QCD. 

\subsection{Non-Abelian Stokes theorem for $SU(N)$}
\par
We consider the infinitesimal deviation $\xi'=\xi+d\xi$ (which is
sufficient to derive the path integral formula). 
From (\ref{cohe}), 
\begin{eqnarray}
 | \xi, \Lambda \rangle
 =   \xi| \Lambda \rangle
 =  e^{-{1 \over 2}K(w,\bar w)}  
\exp \left[\sum_{\alpha}^{} \tau_\alpha(w) E_{-\alpha} \right]  |
\Lambda \rangle ,
\end{eqnarray}
we find  
\begin{eqnarray}
 \langle \xi+d\xi, \Lambda | \xi, \Lambda \rangle
 &=& \exp [ i \omega + O((dw)^2) ] ,
 \\ \omega(x) 
 &:=& 
\langle \Lambda | i \xi^\dagger (x) d\xi(x) |\Lambda \rangle ,
\end{eqnarray}
where $d$ denotes the exterior derivative 
\begin{equation}
  d :=  dx^\mu {\partial \over \partial x^\mu}
:= dx^\mu \partial_\mu  .
\end{equation}
Then $\omega$ is the one-form 
\begin{equation}
  \omega  =  dx^\mu \omega_\mu ,
  \quad  \omega_\mu
  = \langle \Lambda | i \xi^\dagger (x) \partial_\mu \xi(x) |\Lambda
\rangle .
\end{equation}
Here the $x$ dependence of $\xi$ comes from that of $w(x)$ (the
local field variable $w(x)$), i.e., $\xi(x)=\xi(w(x))$.
\par
The exterior derivative is regarded as the operator 
\begin{equation}
 d = \partial + \bar \partial  
 = dw_\alpha {\partial \over \partial w_\alpha} 
 + d\bar w_\beta {\partial \over \partial \bar w_\beta},
\end{equation}
where the operators  $\partial$ and $\bar \partial$ are called
the Dolbeault operators.\cite{Nakahara90}  From the inner product
(\ref{norm}),
\begin{eqnarray}
 \langle \xi', \Lambda | \xi, \Lambda \rangle 
 =  e^{K(w, \bar w')} e^{-{1\over2}[K(w',\bar w')+K(w,\bar w)]} ,
\end{eqnarray}
we obtain another expression for $\omega$ using the K\"ahler
potential $K$:
\begin{equation}
 \omega = {i \over 2}(\partial - \bar \partial) K
=  {i \over 2} \left( dw_\alpha {\partial \over \partial w_\alpha} 
 - d\bar w_\beta {\partial \over \partial \bar w_\beta} \right) K . 
 \label{omega}
\end{equation}

\par
The Wilson loop operator $W^C[{\cal A}]$ is defined as the trace of
the path-ordered exponent along the closed loop $C$ as
\begin{equation}
  W^C [{\cal A}] :=  {1 \over {\cal N}}{\rm tr} \left[ {\cal P} 
 \exp \left( i g \oint_C {\cal A}
 \right) \right] ,
\end{equation}
where ${\cal N}$ 
is the dimension of the representation  
(${\cal N}={\rm dim}({\bf 1}_R)={\rm tr}({\bf 1}_R)$),
and  ${\cal A}$ is the (Lie-algebra
valued) connection one-form 
\begin{equation}
 {\cal A}(x) = {\cal A}_\mu^A(x)T^A dx^\mu = {\cal A}^A(x) T^A .
\end{equation}
Consider a curve starting from $x_0$ and ending at $x$.   We
parameterize this curve by the parameter $t$ and define
\begin{equation}
 W_{ab}(t,t_0) := \left[ {\cal P} \exp \left( i g
 \int_{x_0(t_0)}^{x(t)} dx^\mu {\cal A}_\mu(x) \right) \right]_{ab} 
 =  \left[ {\cal P} \exp \left( ig \int_{t_0}^{t} dt {\cal A}(t)
\right) \right]_{ab} ,
\label{W}
\end{equation}
where 
\begin{equation}
   {\cal A}(t) := {\cal A}_\mu(x) dx^\mu/dt .
\end{equation}
Then the wavefunction defined by
\begin{equation}
 \psi_a(t) = W_{ab}(t,t_0) \psi_b(t_0) 
\end{equation}
is a solution of the Schr\"odinger equation 
\begin{equation}
 \left[ i {d \over dt} + g{\cal A}(t) \right]_{ab} \psi_b(t) = 0 .
\end{equation}
Note that the Wilson loop operator is obtained by taking the trace
of (\ref{W}) for the closed loop, say $C$:
\begin{equation}
W^C[{\cal A}]
= {1 \over {\cal N}}{\rm tr}(W(t,0)) 
= {1 \over {\cal N}}\sum_{a=1}^{{\rm dim of rep}} W_{aa}(t,0) .
\end{equation}
This implies \cite{DP89} that it is possible to write the path
integral representation of the Wilson loop operator if we identify
${\cal A}(t)$ with the Hamiltonian as
\begin{equation}
  H(t) = -g {\cal A}(t) = -g {\cal A}_\mu(x) dx^\mu/dt .
\end{equation}
First, we define the path-ordered exponent by
discretizing the time interval $t$ into $N$ infinitesimal steps and
subsequently taking the limit 
$N \rightarrow \infty, \epsilon \rightarrow 0$ with 
$N\epsilon=t$ fixed as
\begin{eqnarray}
 {\rm tr} \left\{ {\cal P} \exp \left[  i g \int_0^t dt {\cal
A}(t)\right] \right\}
 = \lim_{N \rightarrow \infty, \epsilon \rightarrow 0} {\rm tr}
 \left\{ {\cal P} \prod_{n=0}^{N-1}
 [1+i \epsilon g{\cal A}(t_n)] \right\} ,
 \label{defW}
\end{eqnarray}
where $t_n:=n\epsilon, \epsilon:=t/N$.  For simplicity, we set
$t_0=0$ and $t_N=t$. 
Next, we use the coherent state $| \xi, \Lambda \rangle$, following Ref.\citen{KondoIV}.
On the right-hand-side of (\ref{defW}), we replace the trace with 
\begin{equation}
{1 \over {\cal N}} {\rm tr}(\cdots) = \int d\mu(\xi_N) \langle
\xi_N,\Lambda|(\cdots) | \xi_N, \Lambda \rangle ,
\end{equation}
and insert  the complete set  (resolution of unity) 
\begin{equation}
 I = \int | \xi_n, \Lambda \rangle d\mu(\xi_n) 
\langle \xi_n,\Lambda| .  \quad (n=1,2, \cdots, N-1)  
\end{equation}
Then we obtain
\begin{eqnarray}
&& \ \ {1 \over {\cal N}}{\rm tr} \left\{ {\cal P} \exp \left[  i g
\int_0^t dt {\cal A}(t)\right] \right\}
\nonumber\\
 &=& \lim_{N \rightarrow \infty, \epsilon \rightarrow 0} 
\int \cdots \int d\mu(\xi_N)  
\langle \xi_N,\Lambda|[1+i \epsilon g{\cal A}(t_{N-1})] | \xi_{N-1},
\Lambda
\rangle d\mu(\xi_{N-1})
\nonumber\\
&& \times
\langle \xi_{N-1},\Lambda|[1+i \epsilon g{\cal A}(t_{N-2})] |
\xi_{N-2},
\Lambda \rangle d\mu(\xi_{N-2}) 
\nonumber\\
&&
\cdots d\mu(\xi_{1})
\langle \xi_1,\Lambda|[1+i \epsilon g{\cal A}(t_0)] | \xi_N, \Lambda
\rangle
\nonumber\\
&=& \lim_{N \rightarrow \infty, \epsilon \rightarrow 0}
\prod_{n=1}^{N} \int d\mu(\xi_n) \prod_{n=0}^{N-1}
\langle \xi_{n+1},\Lambda|[1+i \epsilon g{\cal A}(t_{n})] | \xi_{n},
\Lambda \rangle  
\nonumber\\
&=& \lim_{N \rightarrow \infty, \epsilon \rightarrow 0}
\prod_{n=1}^{N} \int d\mu(\xi_n) 
\prod_{n=0}^{N-1} \left[ 1+i \epsilon  g\bar A(t_{n}) \right] 
\prod_{n=0}^{N-1}
\langle \xi_{n+1},\Lambda | \xi_{n}, \Lambda \rangle  
\nonumber\\
&=& \lim_{N \rightarrow \infty, \epsilon \rightarrow 0}
\prod_{n=1}^{N} \int d\mu(\xi_n) 
\exp \left[ i \epsilon \sum_{n=0}^{N-1} g\bar A(t_{n}) \right] 
\prod_{n=0}^{N-1}
\langle \xi_{n+1},\Lambda | \xi_{n}, \Lambda \rangle ,
\end{eqnarray}
where we have used $\xi_0=\xi_N$ 
and have defined  
\begin{equation}
 \bar A(t_n) := {\langle \xi_{n+1},\Lambda|{\cal A}(t_n) 
| \xi_{n}, \Lambda \rangle \over 
\langle \xi_{n+1},\Lambda| \xi_{n}, \Lambda \rangle} .
\end{equation}
Up to $O(\epsilon^2)$, we find
\begin{equation}
 \bar A(t_n) :=  \langle \xi_{n},\Lambda|{\cal A}(t_n) 
| \xi_{n}, \Lambda \rangle + O(\epsilon^2)
=   \langle \Lambda| \xi(t_{n})^\dagger {\cal A}(t_n) \xi(t_{n})  
| \Lambda \rangle + O(\epsilon^2),
\end{equation}
and
\begin{eqnarray}
 \langle \xi_{n+1},\Lambda | \xi_{n}, \Lambda \rangle 
&=& \langle \xi(t_{n}),\Lambda | \xi(t_{n}), \Lambda \rangle 
+ \epsilon \langle \dot \xi(t_{n}),\Lambda | \xi(t_{n}), \Lambda
\rangle  +   O(\epsilon^2) 
\nonumber\\
&=& \exp [ \epsilon \langle \dot \xi(t_{n}),\Lambda 
| \xi(t_{n}), \Lambda \rangle  +   O(\epsilon^2) ] 
\nonumber\\
&=& \exp [ - i \epsilon \langle \Lambda |
 i \dot \xi(t_{n})^\dagger \xi(t_{n}) | \Lambda \rangle  +  
O(\epsilon^2) ] 
\nonumber\\
&=& \exp [ i \epsilon \langle \Lambda |
 i \xi(t_{n})^\dagger \dot \xi(t_{n}) | \Lambda \rangle  +  
O(\epsilon^2) ] ,
\end{eqnarray}
where we have used 
$\langle \xi(t_{n}),\Lambda | \xi(t_{n}), \Lambda \rangle=1$.
Therefore we arrive at the expression 
\begin{eqnarray}
 W^C[{\cal A}]
&=& \lim_{N \rightarrow \infty, \epsilon \rightarrow 0}
\prod_{n=1}^{N} \int d\mu(\xi(t_n)) 
\nonumber\\
&\times& \exp \left\{ i g \epsilon \sum_{n=0}^{N-1} \langle \Lambda|
[ \xi(t_{n})^\dagger {\cal A}(t_n) \xi(t_{n})  
+   ig^{-1} \xi(t_{n})^\dagger \dot \xi(t_{n}) ]| \Lambda
\rangle  
\right\} .
\end{eqnarray}
Thus we obtain the path integral representation of the
Wilson loop,
\begin{eqnarray}
 W^C[{\cal A}] &=& \int [d\mu(\xi)]_C \exp \left( 
i g \oint_C  \langle \Lambda | \left[ V{\cal A} V^\dagger 
+ {i \over g} V d V^\dagger \right]
|\Lambda \rangle \right) ,
\end{eqnarray}
where $[d\mu(\xi)]_C$ is the
product measure of
$d\mu(w(x),\bar w(x))$ along the loop $C$:
\begin{eqnarray}
  [d\mu(\xi)]_C := 
\lim_{N \rightarrow \infty, \epsilon \rightarrow 0}
\prod_{n=1}^{N} \int d\mu(\xi_n) .
\end{eqnarray}
Using the (usual) Stokes theorem,
$
 \oint_{C=\partial S} \omega = \int_S d \omega  ,
$
we obtain the non-Abelian Stokes theorem (NAST):
\begin{eqnarray}
 W^C[{\cal A}] 
&=& \int [d\mu(\xi)]_C \exp \left( 
i g \oint_C  \left[ n^A {\cal A}^A  
+ {1 \over g} \omega \right] \right)  
\nonumber\\
&=& \int [d\mu(\xi)]_C \exp \left( 
ig \int_{S:\partial S=C} 
\left[ d(n^A {\cal A}^A) + {1 \over g} \Omega_K \right] \right) .
\label{NAST}
\end{eqnarray}
Here we have defined
\begin{eqnarray}
 n^A(x) &:=&
 \langle \Lambda | \xi^\dagger(x) T^A \xi(x) |\Lambda \rangle ,
\label{ndef}
\\
  \omega(x)  &:=& 
  \langle \Lambda | i \xi^\dagger (x) d\xi(x) |\Lambda
\rangle ,
\label{omegadef}
\end{eqnarray}
and
\begin{equation}
 \Omega_K := d\omega.
\end{equation}
Taking into account (\ref{omega}), we find that this $\Omega_K$ is
identical to the K\"ahler two-form, in agreement  with the general
statement (\ref{OmegaK}), i.e.,
\begin{equation}
 \Omega_K = d\omega = (\partial + \bar \partial)
 {i \over 2}(\partial - \bar \partial) K
 = i \partial \bar \partial K ,
\end{equation}
since the identity $d^2=0$ leads to
$
  \partial^2 = 0 = \bar \partial^2, 
  \partial \bar \partial + \bar \partial \partial = 0 .
$
Therefore, the second term, $\omega$ or $\Omega_K$, in the exponent of
the NAST is entirely determined from the K\"ahler potential of the
flag manifold.
\par
For $SU(N)$, the topological part, 
\begin{equation}
  \gamma := \oint_C \omega = \int_S \Omega_K ,
\end{equation}
corresponding to the residual $U(N-1)$ invariance is interpreted as
the geometric phase of the  Wilczek-Zee holonomy, \cite{WZ84}  just
as in the $SU(2)$ case it is interpreted as the Berry-Aharonov-Anandan
phase for the residual $U(1)$ invariance. The details of this point will be given in
a subsequent article.\cite{KS00a}

\subsection{Fundamental representation of $SU(N)$ and $CP^{N-1}$
variable}

Defining
\begin{equation}
  \omega_a(x) := (V^\dagger(x) |\Lambda \rangle)_a , \quad 
  (a=1, \cdots, N) 
\end{equation}
we can write
\begin{equation}
 n^A(x) := \langle \Lambda | V(x) T^A V^\dagger(x) |\Lambda \rangle
 =  \bar \omega_a(x) (T^A)_{ab} \omega_b(x) .
\end{equation}
In the $CP^{N-1}$ case, in particular,
the highest-weight state is given by a column vector,
\begin{equation}
 |\Lambda \rangle = \pmatrix{1 \cr 0 \cr \vdots \cr 0} ,
\end{equation}
and then we can write
\begin{eqnarray}
 n^A(x) := \langle \Lambda | U(x) T^A U^\dagger(x) |\Lambda \rangle
 =  \bar \phi_a(x) (T^A)_{ab} \phi_b(x),
 \label{CPrep}
\end{eqnarray}
where
$U\in SU(N)$ and
\begin{eqnarray}
 \phi_a(x) &:=& (U^\dagger(x) |\Lambda \rangle)_a
 = {\bar U}_{1a}(x) .
 \label{CPrep2}
 \end{eqnarray}
 Then $n^A(x)$ can be rewritten as
\begin{eqnarray}
 n^A(x) &=&  (U(x) T^A U^\dagger(x))_{11} .
\end{eqnarray}
Note that the $CP^{N-1}$ variables $\phi_a (a=1, \cdots, N)$ are
subject to the constraint 
\begin{equation}
  \sum_{a=1}^{N} |\phi_a(x)|^2  = 1 .
\end{equation}
This is clearly satisfied by the unitarity of $U$,
$
 \sum_{a=1}^{N} U_{1a}(x) {\bar U}_{1a}(x) 
= (U(x)U^\dagger(x))_{11} = 1.
$
\par
 Now, we
examine another expression (the adjoint orbit representation),
\begin{equation}
  n^A(x)=2{\rm tr}(U^\dagger(x) {\cal H} U(x) T^A) ,
 \label{adjrep0}
\end{equation}
or equivalently,
\begin{equation}
 {\bf n}(x) := n^A(x) T^A = U^\dagger(x) {\cal H} U(x) .
 \label{adjrep}
\end{equation}
Here  ${\cal H}$ is defined by
\begin{equation}
{\cal H} = \vec \Lambda \cdot (H^1,\cdots, H^{N-1})
= \sum_{i=1}^{N-1} \Lambda^i H^i= {1 \over 2}{\rm diag}\left(\frac{N-1}{N},
-\frac{1}{N},\cdots,-\frac{1}{N}\right) ,
\label{gH}
\end{equation}
where we have used (\ref{H}) and 
\begin{equation}
 \Lambda^i = {1 \over \sqrt{2i(i+1)}} .
\end{equation}

For $SU(3)$, 
when the Dynkin index $[m,n]=[1,0]$ or $[0,1]$, (\ref{gH}) reduces to
\begin{equation}
 {\cal H} = \vec \Lambda \cdot \left( \frac{\lambda^3}{2},\frac{\lambda^8}{2} \right) =
 {1 \over 2}{\rm diag} \left( {2 \over 3},-{1 \over 3},-{1 \over 3} \right)  ,
\quad {\rm or} \quad
 {1 \over 2}{\rm diag} \left( {1 \over 3},{1 \over 3},-{2 \over 3} \right)  .
\end{equation}
Note that two elements agree with each other.  Hence the adjoint
orbit cannot cover all the flag space $F_2$.  This is the $CP^2$
case. It is easy to see that the two definitions (\ref{CPrep}) and
(\ref{adjrep0}) are equivalent,
\begin{equation}
n^A(x) 
= (U(x) T^A U^\dagger(x))_{11} 
=  2{\rm tr}({\cal H} U(x) T^A U^\dagger(x)) ,
\end{equation}
since $U T^A U^\dagger$ is traceless.%
\footnote{When  $[m,n]=[1,1]$, on the other hand,
$
 {\cal H} = {\rm diag}(1,-1,0)  ,
$
and all the diagonal elements are different.
Therefore ${\bf n}(x)$ moves on the whole flag space, $F_2$.}
\par
The correspondence between the $F_{N-1}$ variables $w_a$ and the
$CP^{N-1}$ variables $\phi_a$ is given, e.g., by
\begin{equation}
\phi_1=w_1, \quad \phi_2=w_2, \quad \cdots, \quad \phi_{N-1}=w_{N-1}, \quad \phi_N=1 .
\end{equation}
Thus $\omega$ is an inhomogeneous coordinate, 
$\omega_a=\phi_a/\phi_N=w_{a}$ ($a=1, \cdots, N-1$), by definition.
In the $CP^{N-1}$ case (for the fundamental representation), we can
perform the replacement 
\begin{equation}
 \langle \Lambda |f(V) |\Lambda \rangle
 = 2{\rm tr}[{\cal H} f(U)]  
\end{equation}
if $f(V)$ belongs to the Lie algebra of $G$ (and hence  $f(V)$ is traceless, i.e., ${\rm tr}f(V)=0$).
Thus, we obtain another expression for $\omega$,
\begin{equation}
 \omega(x) := 2{\rm tr}[{\cal H} i U(x) d U^\dagger(x)]
 = -  i2{\rm tr}[{\cal H}dU(x)U^\dagger(x)] ,
\end{equation}
which is a diagonal piece of the Maurer-Cartan one-form, 
\begin{equation}
 \vartheta := dU U^{-1} .
 \label{MCf}
\end{equation}
It turns out that the two-form $\Omega_K$ is the symplectic two-form, 
\begin{equation}
 \Omega_K =  d\omega = 2{\rm tr}({\cal H}[U^{-1}dU,U^{-1}dU])
 = 2{\rm tr}({\bf n} [d{\bf n},d{\bf n}]).
 \label{symp}
\end{equation}
Our choice of ${\cal H}$ is the most economical one (see Ref.\citen{FN98}
for different choices and more discussion of the related issues).
\par
From the K\"ahler potential of $CP^{N-1}$ (\ref{KpCP}) and the relation (\ref{omega}),
the connection one-form $\omega$ reads
\begin{eqnarray}
\omega := {i \over 2}{
 \bar w_\alpha d w_\alpha - d \bar w_\alpha w_\alpha
 \over 1 + \bar w_\alpha  w_\alpha} ,
\end{eqnarray}
which is equal to 
\begin{eqnarray}
\omega :=  i  {
 \bar w_\alpha d w_\alpha 
 \over 1 + \bar w_\alpha  w_\alpha} ,
\end{eqnarray}
up to the total derivative.
By taking the exterior derivative, we obtain
\begin{eqnarray}
\Omega_K = d\omega 
= i {(1+|||w|||^2) \delta_{\alpha\beta}- \bar w_\alpha w_\beta
\over (1+|||w|||^2)^2} dw^\alpha \wedge d\bar w^\beta ,
\label{K2fN}
\end{eqnarray}
which agrees with the metric (\ref{metricCP}).

\subsection{An implication of the NAST}

\par
The NAST (\ref{NAST}) implies that  the
Wilson loop operator is given by
\begin{eqnarray}
 W^C[{\cal A}]  
 =    \int [d\mu(\xi)]_C  \exp \left( i g \oint_{C} a \right)
 =   \int [d\mu(\xi)]_C  \exp \left( 
i g \int_{S:C=\partial S} f \right) .
\label{NAST2}
\end{eqnarray}
First, $a$ is the connection one-form 
\begin{eqnarray}
 a := n^A {\cal A}^A  + {1 \over g} \omega 
 = \langle \Lambda | {\cal A}^V |\Lambda \rangle ,
 \label{ap}
\end{eqnarray}
where ${\cal A}^V$ is obtained as the gauge transformation of ${\cal A}$ by
$V \in F_{N-1}$:
\begin{equation}
{\cal A}^V:=V{\cal A} V^\dagger + {i \over g}V d V^\dagger 
= \xi^\dagger{\cal A} \xi + {i \over g}\xi^\dagger d \xi .
\end{equation}
For a quark in the fundamental representation, we can write
\begin{equation}
 a  = 2 {\rm tr}({\cal H}{\cal A}^V) .
\end{equation}
Therefore, the one-form $a$ is equal to the diagonal
piece of the gauge-transformed potential ${\cal A}^V$.
This fact is very useful to derive the Abelian dominance in the
low-energy physics of QCD (see  $\S$ 6).
\par
Next, $f$ is the curvature two-form,
\begin{eqnarray}
f := da 
= dC +  {1 \over g} d\omega 
= dC + {1 \over g} \Omega_K ,
 \label{apf}
\end{eqnarray}
where we have defined the one-form 
\begin{eqnarray}
  C := n^A {\cal A}^A .
\end{eqnarray}
The anti-symmetric tensor $f_{\mu\nu}$ can be called the generalized
't Hooft-Polyakov tensor for the following reasons:
(1) it gives a non-vanishing magnetic monopole (current), where only
the second term $\Omega_K$ gives a non-trivial contribution;
(2)  it is invariant under the full gauge transformation, although
it is an Abelian field strength.
These facts are demonstrated as follows.
\par
First, we characterize the flag space in complex
coordinates.  More precisely, the target space at each space-time point 
$x \in R^D$ is parameterized by the complex variables,
$w^\alpha = w^\alpha(x)$. The K\"ahler two-form is rewritten as
\begin{equation}
  \Omega_K
 = i g_{\alpha\bar \beta}
 \partial_\mu w^{\alpha} \partial_\nu \bar w^{\beta}
 dx^\mu \wedge dx^\nu .
\end{equation}
On the other hand,
\begin{equation}
 \Omega_K := d\omega = 
 {1 \over 2}(\partial_\mu \omega_\nu - \partial_\nu \omega_\mu ) 
dx^\mu \wedge dx^\nu 
:= {g \over 2} f_{\mu\nu}^\Omega  dx^\mu \wedge dx^\nu .
\end{equation}
Then the second piece $g^{-1} \Omega_K$ of $f$ can be written as
\begin{equation}
   f_{\mu\nu}^\Omega 
= {1 \over g} (\partial_\mu \omega_\nu - \partial_\nu \omega_\mu )
=   {i \over g} g_{\alpha\bar \beta}
 \partial_\mu w^{\alpha} \partial_\nu \bar w^{\beta} ,
\end{equation}
and hence 
\begin{equation}
    f_{\mu\nu}  =  \partial_\mu C_\nu - \partial_\nu C_\mu  
    + f_{\mu\nu}^\Omega  .
\end{equation}
The magnetic monopole current $k_\mu$ is obtained as the divergence
of the dual tensor ${}^*f_{\mu\nu}^\Omega$:
\begin{equation}
  k_\mu := \partial_\nu {}^*f_{\mu\nu} ,
\end{equation}
where the Hodge dual of $f_{\mu\nu}$ in four dimensions is defined by
\begin{equation}
   {}^*f_{\mu\nu}  := {1 \over 2} \epsilon_{\mu\nu\rho\sigma}
  f_{\rho\sigma} .
\end{equation}
The first piece $dC$ in $f$ does not contribute to the magnetic
current, due to the Bianchi identity.
On the other hand, the second term $\Omega_K$ in $f$
can lead to a non-vanishing magnetic current, as shown shortly.  Here it should be remarked that the expression for $\Omega_K$ given in terms of the {\it local} coordinate $w_\alpha$ leads to a vanishing magnetic current.  In fact, if the metric is given by the K\"ahler potential we have
\begin{eqnarray}
  k_\mu &=& {i \over 2g} \epsilon_{\mu\nu\rho\sigma}
   \partial_\nu (g_{\alpha\bar \beta}
 \partial_\rho w^{\alpha} \partial_\sigma \bar w^{\beta})
 \nonumber\\
&=& {i \over 2g} \epsilon_{\mu\nu\rho\sigma}
    \partial_\nu g_{\alpha\bar \beta}
 \partial_\rho w^{\alpha} \partial_\sigma \bar w^{\beta} 
 \nonumber\\
&=& {i \over 2g} \epsilon_{\mu\nu\rho\sigma}
    \left( {\partial g_{\alpha\bar \beta} \over \partial w^{\gamma}}
    \partial_\nu w^{\gamma}
+ {\partial g_{\alpha\bar \beta} \over \partial \bar w^{\gamma}}
    \partial_\nu \bar w^{\gamma} \right)
 \partial_\rho w^{\alpha} \partial_\sigma \bar w^{\beta}   
 \nonumber\\
&=& {i \over 2g} \epsilon_{\mu\nu\rho\sigma}
    \left( {\partial K \over \partial w^{\gamma}\partial w^\alpha
\partial \bar w^\beta}
+  {\partial K \over \partial \bar w^{\gamma}\partial w^\alpha
\partial \bar w^\beta} \right)
    \partial_\nu \bar w^{\gamma}
 \partial_\rho w^{\alpha} \partial_\sigma \bar w^{\beta}  = 0 ,
\end{eqnarray}
where we have used the antisymmetric property of 
$\epsilon_{\mu\nu\rho\sigma}$ under the exchange of $\nu$ and
$\rho$, and $\nu$ and $\sigma$. However, this does not imply the
vanishing total magnetic flux or magnetic charge.
\par
We  recall that this situation is similar to that of the Wu-Yang monopole
\cite{WY75} compared with the original Dirac monopole.\cite{Dirac31} There are two ways to describe the Dirac magnetic
monopole. One is to use a single vector potential with (line)
singularities, called the Dirac string, where the singularities are
distributed on a semi-infinite line going through the origin of the
space coordinates. In the absence of singularities, the vector
potential gives a vanishing magnetic charge, due to the Bianchi identity,
\begin{equation}
  \Phi = \oint {\bf B} \cdot d{\bf S}
  = \oint {\rm curl} {\bf A} \cdot d{\bf S}
  = \int_{D^3} {\rm div}{\rm curl}  {\bf A} dV = 0 .
  \label{Gauss}
\end{equation}
Therefore, the singularities must produce a magnetic charge which has the same magnitude as the magnetic charge at the origin, but     opposite sign.
Another way is to introduce more than one vector potential to avoid singularities.  Each vector potential $A^\alpha$ is defined in a sub-region $U_\alpha$ of the sphere $S^2$ such that $A^\alpha$ is regular in each region $U_\alpha$ and the union of the sub-regions covers the whole sphere. Thus, the Bianchi identity leads to zero magnetic flux in each sub-region.
Note that we can not apply the Gauss theorem 
${\rm div}{\rm curl}  {\bf A}=0$, since $U_\alpha$ is not a closed surface.  The total magnetic flux  is recovered by summing up all the contributions of the differences of vector potentials on the boundary  $B_{\alpha,\beta}$ between two regions $U_\alpha$ and $U_\beta$ are
\begin{equation}
  \Phi = \sum_{\alpha} \oint_{U_\alpha} 
  {\rm curl} {\bf A}^\alpha \cdot d{\bf S}
  = \sum_{\alpha,\beta} \oint_{B_{\alpha,\beta}} ({\bf A}^\alpha - {\bf A}^\beta) d{\bf \ell} ,
\end{equation}
where the minus sign follows from the fact that the orientation of the boundary is opposite for neighboring regions.
The difference is given by the gauge transformation, 
$
 {\bf A}^\alpha - {\bf A}^\beta = \nabla \Lambda_{\alpha,\beta} .
$
This recovers the same magnetic flux as in the former case.
\par
The variable $w^\alpha$ corresponds to ${\bf A}^\alpha$ in the case of the Wu-Yang monopole.  Therefore, to show the existence of a non-zero magnetic flux, we must specify the method of gluing different coordinate  patches on the boundary.
These subtleties are avoided by using a different parameterization.
This generalizes the argument given by 't Hooft and
Polyakov for the $SU(2)$ magnetic monopole. 
The antisymmetric tensor $f_{\mu\nu}$ given by
(\ref{apf}) is the $SU(N)$ generalization of the 't Hooft-Polyakov
tensor for $SU(2)$.
In the $SU(2)$ case, $a=2{\rm tr}(T^3{\cal A}^V)$ for any
representation, and the two-form $f:=da$ is the Abelian field strength, 
which is invariant under the $SU(2)$ transformation. 
Hence the two-form $f$
is identically the 't Hooft-Polyakov tensor, 
\begin{equation}
 f_{\mu\nu}(x) :=  \partial_\mu(n^A(x){\cal A}_\nu^A(x)) 
   - \partial_\nu(n^A(x){\cal A}_\mu^A(x))
   - {1 \over g}{\bf n}(x) \cdot (\partial_\mu {\bf n}(x) \times
\partial_\nu {\bf n}(x)) ,
\end{equation}
describing the magnetic flux emanating from the magnetic monopole,
if we identify $n^A$ with the direction of the Higgs field:
\begin{equation}
 \hat \phi^A :=\phi^A/|\phi|, \quad |\phi|:= \sqrt{\phi^A \phi^A} . 
\end{equation}
The complex coordinate representation reads 
\begin{equation}
  f_{\mu\nu}^\Omega(x) = {1 \over g} {1 \over (1+|w(x)|^2)^2} 
\partial_\mu w(x) \partial_\nu \bar w(x)  .
\end{equation}
\par
In general, the (curvature)
two-form 
$f=d(n^A {\cal A}^A) + \Omega_K$ in the NAST is the Abelian field
strength, which is invariant  under the full non-Abelian  gauge
transformation of $G=SU(N)$:%
\footnote{The normalization 
\begin{equation}
 {\rm tr}(T^A T^B) = {1 \over 2} \delta_{AB}   
\label{trT2}
\end{equation}
holds for any group.  
For $SU(2)$, 
$
 {\rm tr}(T^A T^B T^C) = {1 \over 4} i \epsilon_{ABC} .
$
For $SU(3)$, 
$
 {\rm tr}(T^A [T^B, T^C]) = {1 \over 4}  i f_{ABC} ,
$
while
\begin{equation}
 {\rm tr}(T^A T^B T^C) = {1 \over 4} (i f_{ABC}+d_{ABC}) .
\label{trT3}
\end{equation}
Here we have used
$
T^B T^C = {1 \over 2}[T^B,T^C] + {1 \over 2}\{T^B,T^C\},
$
$
 [T^B,T^C] = i f_{BCD} T^D ,
$
and
$
 \{T^B,T^C\} = {1 \over 3} \delta_{AB} I + d_{BCD} T^D,
$
where $d_{ABC}$ is completely symmetric in  the indices.
Furthermore, we find
\begin{equation}
 {\rm tr}(T^A T^B T^C T^D) = {1 \over 12} \delta_{AB} \delta_{CD}
- {1 \over 8} f_{ABE}f_{CDE} + {1 \over 8} d_{ABE}d_{CDE}
+ {i \over 8}(f_{ABE}d_{CDE} + f_{CDE}d_{ABE}) .
\label{trT4}
\end{equation} 

}
\begin{equation}
 f_{\mu\nu}(x) :=  \partial_\mu(n^A(x){\cal A}_\nu^A(x)) 
   - \partial_\nu(n^A(x){\cal A}_\mu^A(x))
   + {i \over g}{\bf n}(x) \cdot [\partial_\mu {\bf n}(x),
\partial_\nu {\bf n}(x)] .
\label{tHPtensor}
\end{equation} 
The invariance of $f$ is obvious from the NAST (\ref{NAST2}), since
$W^C[{\cal A}]$ is gauge invariant
and the measure 
$[d\mu(\xi)]_C$ is also invariant under the $G$ gauge
transformation. 
In the case of the fundamental representation,  the invariance is easily
seen, because  it is possible to rewrite  (\ref{apf}) or
(\ref{tHPtensor})  into the manifestly gauge-invariant form%
\footnote{An explicit derivation of this form is given by Hirayama and Ueno in Ref.\citen{HU99}, where the corresponding expression in the adjoint representation is also given.  }
\begin{equation}
 f_{\mu\nu}(x) :=  2{\rm tr}\left({\bf n}(x) {\cal F}_{\mu\nu}(x)
   + {i \over g}{\bf n}(x)
 [D_\mu {\bf n}(x), D_\nu {\bf n}(x)] \right) ,
\end{equation}
where
\begin{equation}
{\cal F}_{\mu\nu}(x)
   := \partial_\mu {\cal A}_\nu(x) - \partial_\nu {\cal A}_\mu(x)
   - ig [{\cal A}_\mu(x), {\cal A}_\nu(x)]  
\end{equation}
and
\begin{equation}
  D_\mu {\bf n}(x) := \partial_\mu  {\bf n}(x)
- i g [{\cal A}_\mu(x), {\bf n}(x)] .
\end{equation}

In fact, we obtain the magnetic charge 
\begin{eqnarray}
  g_m &=& \int_{D^3} d^3x k_0
\\  &=& \int_{D^3} d^3x {1 \over 2} \epsilon_{ijk}
\partial_i f_{jk}
  \quad (i,j,k=1,2,3)
\nonumber\\   &=& \int_{D^3} d^3x {1 \over 2} \epsilon_{ijk}
\partial_i 
  ({i \over g} {\bf n} \cdot [\partial_j {\bf n}, \partial_k {\bf
n}])
\nonumber\\   &=&  \int d^2 \sigma_i  {1 \over 2} \epsilon_{ijk} 
  {i \over g} {\bf n} \cdot [\partial_j {\bf n}, \partial_k {\bf n}]
\nonumber\\   &=&  \int_{S^2} d^2 x  {1 \over 2} \epsilon_{ab} 
  {i \over g} {\bf n} \cdot [\partial_a {\bf n}, \partial_b {\bf n}]
  \quad (a,b = 1,2)
\nonumber\\   &=& {2 \over g} \int \Omega_K =: {2 \over g} \pi Q ,
\label{charge}
\end{eqnarray}
where we have used (\ref{symp}) in the last step.
Here $Q$ is the integer-valued instanton charge in the NLS model in
two-dimensional space, $S^2={\bf R}^2 \cup \{ \infty \}$ (see
$\S$~8).  The contribution from the magnetic monopole is replaced
by the instanton in the two-dimensional NLS model. The magnetic charge
satisfies the Dirac quantization  condition,
\begin{eqnarray}
  g_m g = 2\pi Q = 2\pi n . \quad (n=0,\mp 1,\mp 2, \cdots)  
\end{eqnarray}

\subsection{Explicit forms of $\omega$ and $\Omega_K$ for $SU(3)$ and
$SU(2)$} 
\par
For $SU(3)$, we find that $\omega$ is given by
\begin{eqnarray}
  \omega &=& im {w_1d\bar w_1+w_2d\bar w_2 \over \Delta_1(w, \bar w)} 
\nonumber \\
 && + in {w_3d\bar w_3+(w_2-w_1 w_3)(d\bar w_2-\bar w_1d\bar w_3-\bar
w_3d\bar w_1)
\over \Delta_2(w, \bar w)} ,
\end{eqnarray}
up to the total derivative.
Hence, we obtain
\begin{eqnarray}
  \Omega_K &=&   d\omega 
  = im (\Delta_1)^{-2}[(1+|w_1|^2)dw_2 \wedge d\bar w_2
  - \bar w_2 w_1 dw_2 \wedge d\bar w_1
  \nonumber\\&&
  - w_2 \bar w_1 dw_1 \wedge d\bar w_2
  + (1+|w_2|^2) dw_1 \wedge d\bar w_1]
  \nonumber\\&&
  + in (\Delta_2)^{-2}[\Delta_1 dw_3 \wedge d\bar w_3
  - (w_1+\bar w_3 w_2) dw_3 \wedge (d\bar w_2 - \bar w_3 d\bar w_1)
    \nonumber\\&&
  - (\bar w_1+w_3 \bar w_2)(dw_2-w_3 dw_1) \wedge d\bar w_3
  \nonumber\\&&
  + (1+|w_3|^2)(dw_2-w_3 dw_1)(d\bar w_2 - \bar w_3 d\bar w_1)] .
  \label{K2fb}
\end{eqnarray}
The K\"ahler potential for $F_2$ is given by
\begin{equation}
  K(w,\bar w) = \ln [(\Delta_1)^m(\Delta_2)^n] .
\end{equation}
For $CP^2$, it reads
\begin{equation}
  K(w,\bar w) = \ln [(\Delta_1)^m] ,
\end{equation}
which is obtained as a special case of $F_2$ by setting $w_3=0$ and
$n=0$. Hence, we obtain
\begin{eqnarray}
  \omega = im {w_1d\bar w_1+w_2d\bar w_2 \over \Delta_1(w, \bar w)} ,
\end{eqnarray}
up to the total derivative, and
\begin{eqnarray}
  \Omega_K &=&   d\omega 
  = im (\Delta_1)^{-2}[(1+|w_1|^2)dw_2 \wedge d\bar w_2
  - \bar w_2 w_1 dw_2 \wedge d\bar w_1
  \nonumber\\&&
  - w_2 \bar w_1 dw_1 \wedge d\bar w_2
  + (1+|w_2|^2) dw_1 \wedge d\bar w_1] .
  \label{CP2K2f}
\end{eqnarray}
\par
This should be compared with the case $F_1=CP^1$, 
\begin{equation}
  K(w,\bar w) = m \ln [(1+|w|^2)], \quad m=2j .
\end{equation}
For $SU(2)$, we reproduce the well-known results,
\begin{eqnarray}
 \omega = i m {wd\bar w \over 1+|w|^2}   ,
\end{eqnarray}
and
\begin{eqnarray}
 \Omega_K = i g_{w\bar w} dw \wedge d\bar w
 = i m  {dw \wedge d\bar w  \over (1+|w|^2)^{2}}.
\end{eqnarray}
The explicit form of $\Omega_K$ is necessary to carry out the
instanton calculus in the following.

\section{Magnetic monopole in SU(N) Yang-Mills theory}
\setcounter{equation}{0}

\par
In the dual superconductor picture of quark confinement, the magnetic
monopoles give the dominant contribution to the area law of the
Wilson loop or the string tension.      
Following the 't Hooft argument, \cite{tHooft81} the partial gauge
fixing $G \rightarrow H$ realizes the magnetic monopole in
Yang-Mills gauge theory, even in the absence of an elementary scalar
field. In the conventional approach based on the MA gauge, the
residual gauge group was chosen to be the maximal torus subgroup
$H=U(1)^{N-1}$ for $G=SU(N)$.  This choice immediately determines the
type of magnetic monopoles.  
We now re-examine this issue.  

We have learned that the magnetic monopole
which is responsible for the area law of the Wilson loop is determined by
the maximal stability group $\tilde H$ rather than the residual gauge
group $H$. This is a new feature that appears in 
$SU(N)$ for  $N \ge 3$.  It seems that this possibility has been overlooked in the lattice community, as far as we know. Indeed, this
situation occurs only for
$SU(N)$ with $N \ge 3$.  Therefore, we must distinguish the maximal
stability group
$\tilde H$  from the residual gauge group $H$.
In general, the maximal stability group $\tilde H$ is larger than
the maximal torus subgroup: 
$H=U(1)^{N-1} \subset \tilde H$.  Hence, the coset space is smaller than in
the maximal torus case, i.e.,
$G/\tilde H \subset G/H$.
\par
 The existence of magnetic
monopoles is suggested by the non-trivial Homotopy groups
$\pi_2(G/H)$. In case (II), we have
\begin{equation}
 \pi_2(F_2) = \pi_2(SU(3)/(U(1) \times U(1)))  
=\pi_1(U(1) \times U(1)) ={\bf Z}+{\bf Z} .
\end{equation}
On the other hand, in case (I), i.e., [m,0] or [0,n], we have
\begin{equation}
 \pi_2(CP^2) = \pi_2(SU(3)/U(2)) = \pi_1(U(2)) 
= \pi_1(SU(2)\times U(1)) = \pi_1(U(1))={\bf Z} .
\end{equation}
Note that the $CP^{N-1}$ model possesses only local $U(1)$ invariance
for any
$N\ge 2$.  It is this $U(1)$ invariance that corresponds to a single
kind of Abelian magnetic monopole appearing in case (I).
This magnetic monopole may be related to the non-Abelian magnetic
monopole proposed by  Weinberg et al. \cite{Weinberg96} 
The explicit solution for the magnetic monopole in $SU(3)$ gauge
theories are discussed in
Ref.\citen{Sinha76}.
\par
This situation should be compared with the $SU(2)$ case, where the
maximal stability group is always given by the maximal torus $H=U(1)$, 
irrespective of the representation.  Therefore, the coset is given by
\begin{equation}
G/H=SU(2)/U(1)=F_1=CP^1 \cong  S^2 
\end{equation}
and 
\begin{equation}
 \pi_2(SU(2)/U(1)) = \pi_2(F_1) = \pi_2(CP^1) = {\bf Z} ,
\end{equation}
 for {\it arbitrary} representation.
\par
For $SU(N)$, our results suggest that  the fundamental quarks are  to
be confined when the maximal stability group $\tilde H$ is given by
$\tilde H=U(N-1)$ and 
\begin{equation}
\pi_2(G/\tilde H) =  \pi_2(SU(N)/U(N-1)) = \pi_2(CP^{N-1}) = {\bf Z} ,
\end{equation}
while the adjoint quark is
related to the maximal torus
$\tilde H=U(1)^{N-1}$ and
\begin{equation}
\pi_2(G/H) =  \pi_2(SU(N)/U(1)^{N-1}) = \pi_2(F_{N-1}) = {\bf Z}^{N-1} .
\end{equation}
This observation is in sharp contrast with the
conventional treatment, in which the $(N-1)$ species of magnetic
monopoles corresponding to the residual maximal torus group
$U(1)^{N-1}$ of $G=SU(N)$ are taken into account on equal footing.
In fact, the NAST derived in this article
shows that the fundamental quark feels only the $U(1)$ that
is embedded in the maximal stability group
$U(N-1)$ as a magnetic monopole.   This is a component along the
highest weight.

\section{Abelian dominance in $SU(N)$ gauge theory}
\setcounter{equation}{0}

\subsection{APEGT as a low-energy effective theory}
\par
The Abelian dominance in $SU(N)$ Yang-Mills theory can be explained
as follows.   
First, we adopt the maximal Abelian (MA) gauge.
The MA gauge for $SU(N)$ is defined as follows.
Consider the Cartan decomposition of ${\cal A}$ into diagonal
($H$) and off-diagonal ($G/H$) pieces,
\begin{equation}
 {\cal A}(x) = {\cal A}^A(x) T^A
 = a^i(x) H^i + A^a(x) T^a , \quad 
 (A=1, \cdots, N^2-1) .
\end{equation}
In particular, for $G=SU(3)$,
\begin{equation}
 H^1={\lambda^3 \over 2}, \quad H^2={\lambda^8 \over 2}, \quad
 T^a = {\lambda^a \over 2} . \ (a=1,2,4,5,6,7) 
\end{equation}
The MA gauge is obtained by minimizing the functional of
off-diagonal fields,
\begin{equation}
 {\cal R} := \int d^4x {1 \over 2} A_\mu^a(x) A^\mu{}^a(x) 
 := \int d^4x  {\rm tr}_{G/H}
({\cal A}_\mu(x) {\cal A}^\mu(x)),
\end{equation}
under the gauge transformation.  Under the infinitesimal gauge
transformation $\Lambda$, ${\cal R}$ transforms as
\begin{eqnarray}
 \delta_\Lambda {\cal R} 
&=& \int d^4x  A^\mu{}^a \delta_\Lambda A_\mu^a 
\nonumber\\
&=& \int d^4x  A^\mu{}^a
 (\partial_\mu \Lambda^a + g f^{aij} a_\mu^i \Lambda^j
 + g f^{aib} a_\mu^i \Lambda^b
 + g f^{abC} A_\mu^b \Lambda^C) 
\nonumber\\
&=& - \int d^4x 
 (\partial_\mu A^\mu{}^a + g f^{aib} a_\mu^i A^\mu{}^b) \Lambda^a  ,
\end{eqnarray}
since $f^{ABC}$ is completely antisymmetric in the indices and  
$f^{aij}=0$ ($T^i$ and $T^j$ commute). Therefore, the
condition
$\delta_\Lambda {\cal R}=0$ for arbitrary $\Lambda$ leads to the
differential MA gauge given by
\begin{equation}
 \partial_\mu A^\mu{}^a(x) - g f^{abi}a_\mu^i(x) A^\mu{}^b(x)  
:= (D_\mu[a] A^\mu)^a = 0 .
\end{equation}
The $SU(N)$ Yang-Mills theory in the MA gauge%
\footnote{In order to fix the residual Abelian gauge group $H$, we add an additional $GF+FP$ term, e.g.,
$
- \int d^4x i \delta_B \left[ \bar C^i \left( \partial_\mu a^\mu + {\beta \over 2} B \right)^i   \right] .
$
}
is given by
\begin{eqnarray}
  S_{YM}^{total} &=& S_{YM} +S_{GF+FP},
\\
  S_{YM} &=& \int d^4 x {-1 \over 4} 
{\cal F}_{\mu\nu}^A {\cal F}^{\mu\nu}{}^A ,
\\
  S_{GF+FP} &=& - \int d^4x i \delta_B 
\left[ \bar C^a \left( D_\mu[a]A^\mu + {\alpha \over 2} B \right)^a
 \right] ,
\label{MA}
\end{eqnarray}
where $\delta_B$ is the Becchi-Rouet-Stora-Tyupin (BRST)
transformation,
\begin{eqnarray}
   \delta_B {\cal A}_\mu(x)  
   &=&  {\cal D}_\mu[{\cal A}] {C}(x)
   := \partial_\mu {C}(x) 
   - ig [{\cal A}_\mu(x),  {C}(x)],
    \nonumber\\
   \delta_B {C}(x)  
   &=& i{1 \over 2}g[{C}(x), {C}(x)],
    \nonumber\\
   \delta_B \bar {C}(x)  &=&   i B(x)  ,
    \nonumber\\
   \delta_B B(x)  &=&  0 ,
    \label{BRST0}
\end{eqnarray}
which is nilpotent, i.e., $\delta_B^2 \equiv 0$.
The generating functional is given by
\begin{eqnarray}
  Z_{YM}[J] = \int [d{\cal A}] [dC] [d\bar C] [dB] \exp
(iS_{YM}^{total}+iS_J) .
\end{eqnarray}

\par
Now we proceed to derive the effective Abelian gauge theory in the
MA gauge by integrating out the off-diagonal gauge fields (together
with the ghost and anti-ghost fields), $A^a, C^a, \bar C^a$ and $B^a$, as done in Refs.\citen{QR97,KondoI}.  
Then the $SU(N)$ Yang-Mills theory can be reduced to the
$U(1)^{N-1}$ Abelian gauge theory, which is written in terms of the
diagonal fields, $a^i, C^i, \bar C^i$ and $B^i$, alone: 
\begin{eqnarray}
  Z_{YM}[J] = \int [da^i] [dC^i] [d\bar C^i] [dB^i] \exp
(iS_{APEGT}^{total}+i\tilde S_J) ,
\end{eqnarray}
where
\begin{eqnarray}
  \exp (iS_{APEGT}^{total}+i\tilde S_J)
  = \int [dA^a] [dC^a] [d\bar C^a] [dB^a] \exp
(iS_{YM}^{total}+iS_J) .
\end{eqnarray}
In particular, the partition function reads
\begin{eqnarray}
  Z_{YM}[0] = \int [da^i] [dC^i] [d\bar C^i] [dB^i] \exp
(iS_{APEGT}^{total})  := Z_{APEGT} .
\end{eqnarray}
 The Abelian gauge theory obtained in this way is called the
``Abelian-projected effective gauge theory'' (APEGT). 
It has been shown that the APEGT is an Abelian gauge theory whose 
 gauge coupling constant $g$ runs according to the same
renormalization-group beta function as in the original $SU(N)$
Yang-Mills theory,
\begin{eqnarray}
  S_{APEGT}^{total}[a^i,C^i,\bar C^i,B^i] 
  &=&  \int d^4 x {-1 \over 4g^2(\mu)} (da^i, da^i) 
  + S_{GF}' ,
  \label{APEGT}
  \\
  {1 \over g^2(\mu)} &=&  {1 \over g^2(\mu_0)}
  + {b_0 \over 8\pi^2} \ln {\mu \over \mu_0}, \quad
  b_0 = {11N \over 3} > 0 ,
  \label{runc}
\end{eqnarray}
up to the
one-loop level.%
\footnote{
See Ref.\citen{KondoI} for details.  The result of Ref.\citen{QR97,KondoI} for $SU(2)$ can be generalized to
$SU(N)$ in straightforward way, at least at the one-loop level
\cite{KS00b}. At the two-loop level, this is not trivial.  The two-loop
result will be given in Ref.\citen{KS00b}.
}
Hence
\begin{eqnarray}
  \Big\langle f[a^j] \Big\rangle_{YM} 
&=& Z_{YM}^{-1} \int [d{\cal A}] [dC] [d\bar C] [dB] \exp
(iS_{YM}^{total}) f[a^j]
\\
  &\cong& Z_{APEGT}^{-1} \int  [da^i] [dC^i] [d\bar C^i] [dB^i]
 \exp ( i S_{APEGT}^{total} ) f[a^j] 
\\&=& \Big\langle f[a^j] \Big\rangle_{APEGT}  .
\label{formula}
\end{eqnarray}

\subsection{Modified MA gauge}
\par
In the MA gauge, it is expected \cite{EI82} that the off-diagonal gauge fields $A_\mu^a$ have
 non-zero mass $m_A (m_A\not=0)$, whereas the diagonal gauge
fields
$a_\mu^i$ remain massless ($m_a=0$).
Therefore, the APEGT obtained in this way can be regarded as the low-energy
effective gauge theory of Yang-Mills theory.
In the framework of lattice gauge theory, this prediction was confirmed in numerical calculations by Amemiya
and Suganuma \cite{AS99} for $G=SU(2)$ and $SU(3)$. 
In the framework of the continuum gauge field theory, on the other hand, it is more efficient to modify the MA gauge into the $OSp(D|2)$
invariant form by making use of the BRST $\delta_B$ and anti-BRST $\bar \delta_B$
transformations, as proposed by one of the authors:\cite{KondoII} 
\begin{eqnarray}
  S_{GF+FP}[{\cal A}, C, \bar C, B]
  := \int d^4x \ i \delta_B \bar \delta_B 
  {\rm tr}_{G/H} \left[ {1 \over 2} {\cal A}_\mu {\cal A}^\mu
  -{\alpha \over 2}i C \bar C \right] ,
  \label{mMA}
\end{eqnarray}
where $\alpha$ corresponds to the gauge fixing parameter. 
Note that (\ref{mMA}) is obtained  from (\ref{MA}) by adding the
ghost self-interaction terms and by adjusting the parameter for the ghost self-interaction term,
since
\begin{eqnarray}
 &&- \bar \delta_B \left[ {1 \over 2} A_\mu^a A^\mu{}^a
  -{\alpha \over 2} i C^a \bar C^a \right]
\nonumber\\
&=& \bar C^a \left( D_\mu[a] A^\mu + {\alpha \over 2}B \right)^a 
- i {\alpha \over 2} f^{abi} \bar C^a \bar C^b C^i
- i {\alpha \over 4} f^{abc}  C^a \bar C^b \bar C^c  .
\end{eqnarray}
The special case $\alpha=-2$ is discussed in previous articles.\cite{KondoII,KondoIV}
 For $G=SU(2)$, the last two terms reduce to 
$
 2i \bar C^1 \bar C^2 C^3 = - 2 C^3 \bar C^+ \bar C^- 
$ 
in agreement with the previous result\cite{KondoII} 
(See Ref.\citen{KondoII} for details).
\par
In the modified MA gauge (\ref{mMA}), the non-zero mass generation of  off-diagonal components was demonstrated analytically
(at least in the topological sector), using dimensional
reduction to the two-dimensional coset non-linear (NLS) sigma model (see section IV.C of Ref.\citen{KondoII}). 
Quite recently, dynamical mass generation of the off-diagonal gluons has been shown to take place due to ghost--anti-ghost condensation caused by the attractive quartic ghost interaction (contained in the gauge fixing term of the modified MA gauge), which is necessary to maintain renormalizability in four dimensions.  
The off-diagonal gluon mass obtained in this way can be written in terms of the intrinsic scale of the gluodynamics, $\Lambda_{QCD}$, and hence it is a renormalization-group invariant quanity.
(For more details, see Ref.\citen{KS00a}.) 
Therefore, integration of the massive off-diagonal
gauge fields is interpreted as a step of the Wilsonian
renormalization group,%
\footnote{Rigorously speaking, all high-energy modes should be
integrated out in the Wilsonian renormalization group.  Hence, we
must integrate out the high-energy mode of the diagonal fields,
$a_\mu^i$, too.  The result is the same as the above, at least at
the one-loop level.  At the two-loop level, we must be more careful
in dealing with the high-energy mode (see Ref.\citen{KS00b}).
 }
and the APEGT can describe  low-energy physics on the length
scale
$R>m_A^{-1}$.  In this sense, the APEGT can be regarded as the
low-energy effective gauge theory of Yang-Mills theory.

\subsection{Abelian dominance} 
\par
With the NAST for $SU(N)$ just derived, Abelian dominance in
$SU(N)$ Yang-Mills theory is explained as follows, in a manner based on
the same argument as that in the $SU(2)$ case.\footnote{
 Abelian dominance in the low-energy region of $SU(2)$ QCD is demonstrated in Ref.\citen{KondoIV} using the result in Ref.\citen{KondoI}
combined with the $SU(2)$ NAST.\cite{DP89,KondoIV} 
The monopole dominance is derived for $SU(2)$ in
Ref.\citen{KondoIV} by showing that the dominant contribution to the
area law comes from the monopole piece alone,
$
 \Omega_K  = d\omega = {\rm tr}({\bf n} [d{\bf n},d{\bf n}]).
$
}

The NAST (\ref{NAST}) implies that the expectation value of the
Wilson loop in the $SU(N)$ Yang-Mills theory is given by
\begin{eqnarray}
 \langle W^C[{\cal A}] \rangle_{YM}
 =  \int [d\mu(V)]_C 
\Big\langle \exp \left( i g \oint_{C}  a \right)
\Big\rangle_{YM}   ,
\label{W1}
\end{eqnarray}
where the one-form $a$ is written as
\begin{eqnarray}
 a  = dx^\mu a_\mu = 2{\rm tr}({\cal H}{\cal A}^V)  
\end{eqnarray}
for a quark in the fundamental representation.
Thus, the one-form $a$ is equal to the diagonal
piece of the gauge-transformed potential ${\cal A}^V$, i.e., a
component along the highest-weight vector of the fundamental
representation of
$SU(N)$.%
\footnote{
Therefore, the component $a$ along the highest-weight is obtained as
an appropriate linear combination of
$a^i$. Other components orthogonal to the highest-weight do not
contribute to the expectation value of the Wilson loop,
since the APEGT (\ref{APEGT}) is an Abelian gauge theory without
self-interaction among the gauge fields.
}
 By applying the above result 
(\ref{formula}) to (\ref{W1}), we obtain
\begin{eqnarray}
  \Big\langle W^C[{\cal A}]  \Big\rangle_{YM} 
 &=&  \Big\langle \exp \left( i g \oint_{C}  a \right)
\Big\rangle_{YM}   
 \\
  &\cong& \Big\langle \exp \left( i g \oint_{C}  a \right)
\Big\rangle_{APEGT}
\\
  = Z_{APEGT}^{-1} \int &[da^i]&[dC^i] [d\bar C^i] [dB^i] 
  \exp ( i S_{APEGT}^{total}[a^i]) 
\exp \left( i g \oint_{C}  a \right)   ,
\end{eqnarray}
where we have used in the first equality the fact that the the
integration over
$V$ is redundant after having taken the expectation value
$\langle  \cdot \rangle_{YM}$.
 This implies the Abelian dominance for the large Wilson loop in
$SU(N)$ gauge theory in the sense that the expectation value of the
non-Abelian Wilson loop in Yang-Mills theory is nearly equal to (or
dominated by) that of the Abelian Wilson loop in the APEGT, where the
Wilson loop
$C$ is so large that the APEGT is valid in that region, i.e.,  
$R,T > m_A^{-1}$ for a rectangular Wilson loop with sides $R$ and
$T$.  The APEGT is valid in the range $\Lambda_{QCD} < \mu \sim
R^{-1} < m_A$.

\subsection{Monopole dominance and the area law}
\par
In our framework, the Abelian dominance and the monopole dominance
are understood as implying the first and the second relations below, respectively:
\begin{eqnarray}
  \Big\langle W^C[{\cal A}]  \Big\rangle_{YM} 
  &\cong& \Big\langle \exp \left( i g \oint_{C}  a \right)
\Big\rangle_{APEGT} 
\\
&\cong& \Big\langle \exp \left( i \oint_{C}  \omega \right)
\Big\rangle_{APEGT} .
\end{eqnarray}
Numerical simulations show that the monopole part obeys the area law and $\sigma_{Abel}$ exhausts the full string tension
obtained from  the non-Abelian Wilson loop (i.e., monopole dominance
in the string tension or area law),
\begin{eqnarray}
  \Big\langle \exp \left( i \oint_{C}  \omega \right)
\Big\rangle_{APEGT} 
 \sim& \exp (- \sigma_{Abel} |S|) ,
\end{eqnarray}
while 
$
\langle \exp \left( i g \oint_{C}  a - i \oint_C \omega \right) \rangle_{APEGT}
$
does not obeys the area law.
This result implies the area law of the original non-Abelian
Wilson loop,
\begin{eqnarray}
  \Big\langle W^C[{\cal A}]  \Big\rangle_{YM}  
 \sim& \exp (- \sigma |S|)  \quad \sigma \cong \sigma_{Abel} .
\end{eqnarray}

\par
In Ref.\citen{KondoV}, the monopole dominance and the area law of the
Wilson loop have been demonstrated using the APEGT for $G=SU(2)$.
Now this scenario can be extended  to $G=SU(N)$.
The important remarks are in order.
\begin{enumerate}
\item
 The APEGT has a running coupling constant which increases as the
relevant energy decreases (asymptotic freedom), and thus the APEGT is in
the strong coupling region in the low-energy regime (see Fig.\ref{running}).

\item
  The Abelian gauge group in APEGT is a compact group embedded in
the compact non-Abelian gauge group $SU(N)$.  It is the compactness
that causes the phase transition in the APEGT at 
$
 \alpha_c = {\pi \over 4},
$
which separates the Coulomb (conformal) phase ($\alpha<\alpha_c$)
from the strong coupling phase ($\alpha>\alpha_c$).  This follows
from the Berezinski-Kosterlitz-Thouless (BKT) phase transition in the
two-dimensional $O(2)$ NLS model obtained through dimensional reduction.

\item
 In the low-energy region such that $\alpha(\mu) >\alpha_c$, the
APEGT is in the strong coupling phase which confines the quark due
to vortex condensation.
The above described strategy is schematically depicted below.
\vskip 0.5cm
\begin{center}
\unitlength=1cm
\thicklines
 \begin{picture}(12,6)
 \put(2,5){\framebox(8,1){D-dim. U(1) Gauge Theory}}
 \put(6.2,5){\vector(0,-1){0.8}}
 \put(7,4.5){Covariant Lorentz gauge fixing}
 \put(-0.2,2.8){\framebox(12.4,1.4){}}
 \put(0,3){\framebox(5,1){D-dim. Perturbative U(1)}}
 \put(6,3.5){$\bigotimes$}
 \put(5.5,3){deform}
 \put(7,3){\framebox(5,1){D-dim. TQFT}}
 \put(8.5,3){\vector(0,-1){1}}
 \put(9,2.4){Dimensional reduction}
 \put(-0.2,0.8){\framebox(12.4,1.4){}}
 \put(0,1){\framebox(5,1){D-dim. Perturbative U(1)}}
 \put(6,1.5){$\bigotimes$}
 \put(5.5,1){deform}
 \put(7,1){\framebox(5,1){(D-2)-dim. O(2) NLSM}}
 \end{picture}
\end{center}

\end{enumerate}

\begin{figure}
\begin{center}
 \leavevmode
 \epsfxsize=70mm
 \epsfysize=50mm
 \epsfbox{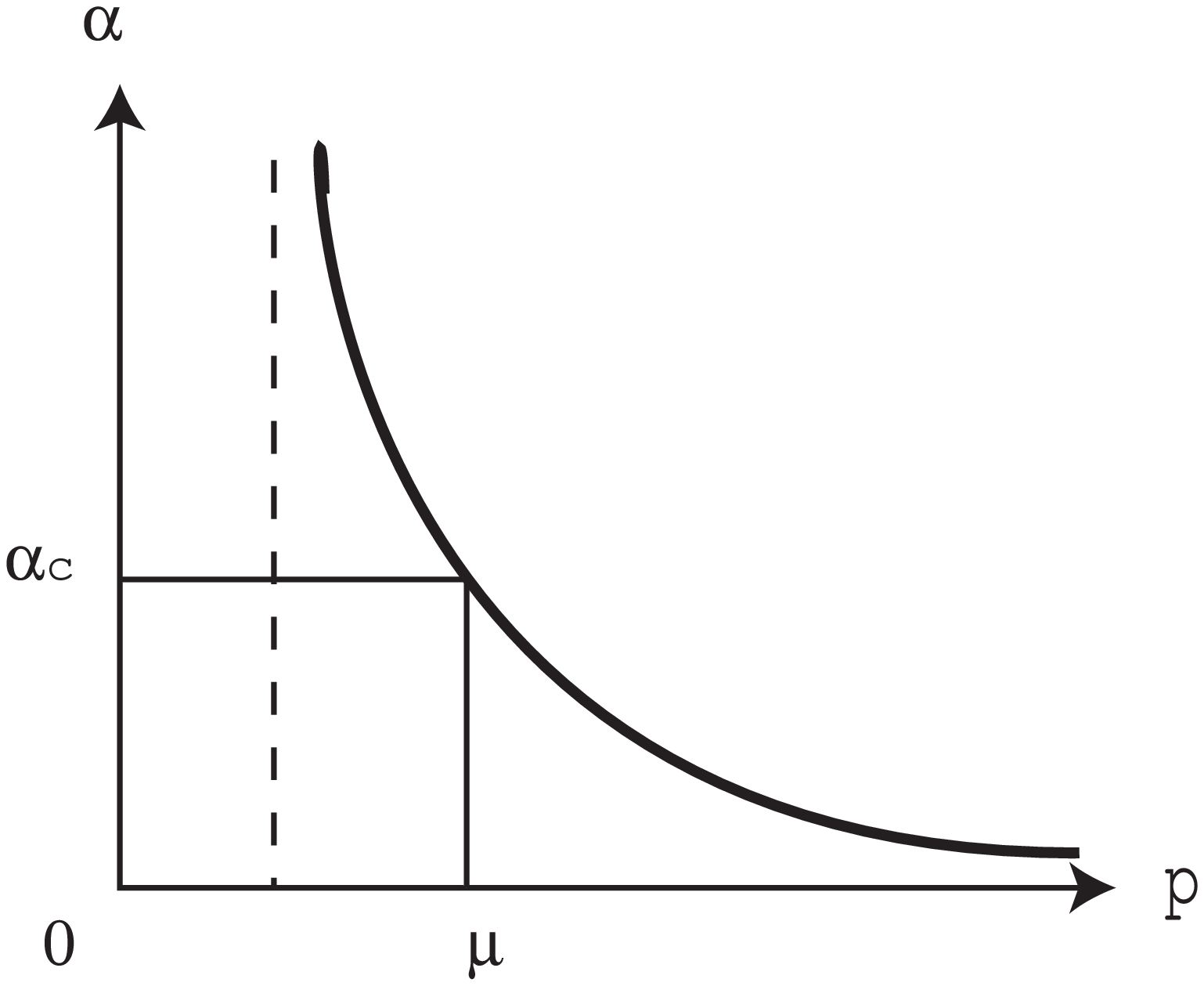}
\end{center} 
 \caption[]{The running coupling constant in the APEGT.}
 \label{running}
\end{figure}

\section{Reformulation of Yang-Mills theory}
\setcounter{equation}{0}
 
In this section we summarize the novel reformulation of the Yang-Mills
theory proposed in Ref.\citen{KondoII} and elaborated in
Ref.\citen{KondoVI}.  This material is necessary for subsequent
sections. 

\subsection{Deformation of topological field theory}
\par
We consider the quantization of Yang-Mills theory on the topological
background field.
We decompose the connection ${\cal A}$ as
\begin{eqnarray}
 {\cal A}_\mu(x)  = \Omega_\mu(x)   + {\cal Q}_\mu(x)   
 \label{deco}
\end{eqnarray}
and identify ${\cal Q}$ with the quantum fluctuation field on the
background field
$\Omega$.
For arbitrary but fixed background field $\Omega$, the generating
functional is given by
\begin{eqnarray}
 \tilde Z[J, \Omega] 
 &:=& \int [d{\cal Q}] [d\tilde C][d\bar {\tilde C}][d\tilde B]
 \exp \{ iS_{YM}[\Omega+{\cal Q}] 
\nonumber \\
 &&+ i\tilde S_{GF}[Q, \tilde C, \bar {\tilde C}, \tilde B]
 + i(J_\mu \cdot {\cal Q}_\mu) \} ,
\label{ZBGFM}
\end{eqnarray} 
where $S_{YM}[{\cal A}]$ is the usual Yang-Mills action,
\begin{eqnarray}
  S_{YM}[{\cal A}] = - \int d^D x {1 \over 4} 
  {\cal F}_{\mu\nu}^A[{\cal A}] {\cal F}^{\mu\nu}{}^A[{\cal A}] ,
\end{eqnarray}
and $\tilde S_{GF}$ is the gauge-fixing and FP ghost term for the
quantum fluctuation field
${\cal Q}$:
\begin{eqnarray}
  \tilde S_{GF}[{\cal Q}, \tilde C, \bar {\tilde C}, \tilde B]
  &:=& - \int d^Dx \ i  \tilde \delta_B \ {\rm tr}_G
  \left[ \bar {\tilde C}\left( \tilde F[{\cal Q}]+
  {\tilde \alpha \over 2}\tilde B \right) 
  \right] .
  \label{GF0}
\end{eqnarray}
We wish to retain gauge invariance for the background field
$\Omega$ even after the gauge fixing for ${\cal Q}$. 
This is realized by choosing
the background field (BGF) gauge fixing condition 
\begin{eqnarray}
  \tilde F^A[{\cal Q}] := {\cal D}_\mu^{AB}[\Omega] {\cal Q}^\mu{}^B 
  = 0  .
  \label{BGFgauge}
\end{eqnarray}
In fact, in the BGF gauge, $\tilde Z[J, \Omega]$ is invariant under
the gauge transformation of the background field; the infinitesimal
version is given by
$
 \delta \Omega_\mu = {\cal D}_\mu[\Omega] \omega .
$
Hence, the theory with the action 
\begin{eqnarray}
 \tilde S_{eff}[J, \Omega] := - i \ln \tilde Z[J, \Omega] ,
\end{eqnarray}
is defined only on the space of the gauge orbit.  
Suppose that the background field satisfies the equation  
$
  F^A[\Omega]  = 0 .
$
In order to consider the quantized Yang-Mills theory on all possible
  background fields satisfying the equations 
$
  F^A[\Omega]  = 0 ,
$
we define the total generating functional
\begin{eqnarray}
 && Z[J]  
 = \int [d\Omega_\mu][dC][d\bar C][dB] 
 \tilde Z[J, \Omega] 
  \exp (i S_{TQFT}[\Omega, C, \bar C, B]) 
 \exp [i(J_\mu \cdot \Omega^\mu)]
\nonumber \\ 
 &=& \int [d\Omega_\mu] [dC][d\bar C][dB]
 \exp \{ i \tilde S_{eff}[J, \Omega]
 +iS_{TQFT}[\Omega, C, \bar C, B]
+i(J_\mu \cdot \Omega^\mu)\} ,
\end{eqnarray}
where $S_{TQFT}[\Omega, C, \bar C, B]$ corresponds to the
gauge-fixing term for  the background field $\Omega_\mu$.
In order to describe the magnetic monopole as a topological
background field in Yang-Mills theory, we choose the MA gauge
for $\Omega_\mu$, 
\begin{eqnarray}
  S_{TQFT}[\Omega, C, \bar C, B]
  &:=& - \int d^Dx \ i  \delta_B \ {\rm tr}_{G/H}
  \left[ \bar C \left( F[\Omega]+{\alpha \over 2} B \right) 
  \right] ,
  \label{GF1}
\end{eqnarray}
where
\begin{eqnarray}
  F^{a}[\Omega] := D_\mu{}^{ab}{}[\Omega^i] \Omega^\mu{}^b
  := (\partial_\mu \delta^{ab} 
   -  g f^{abi} \Omega_\mu{}^i) \Omega^\mu{}^b 
    \quad  (i = 1, \cdots, N-1) .
   \label{dMAG2}
\end{eqnarray}
Note that the trace is taken on the coset $G/H$, not on the entire
$G$, in the MA gauge.
\par
Under the identification 
\begin{eqnarray}
  \Omega_\mu(x) := {i \over g} U(x) \partial_\mu U^\dagger(x), \quad
  {\cal Q}_\mu(x)  := U(x) {\cal V}_\mu(x) U^\dagger(x) ,
  \label{cov}
\end{eqnarray}
we assume that all the topologically non-trivial configurations come 
from $\Omega$, whereas ${\cal V}$ denotes  topologically trivial
configurations.  Therefore, ${\cal V}$ changes under the small gauge
transformation, while $\Omega$ includes the effect of large gauge
transformations.  Therefore, we must take into account the finite gauge
rotation $U$, without restriction to the infinitesimal gauge
transformation. 
Under the identification (\ref{cov}), the Yang-Mills action is
invariant,
\begin{eqnarray}
  S_{YM}[{\cal A}] = S_{YM}[\Omega+{\cal Q}]
  = S_{YM}[{\cal V}] 
  = \int d^D x {-1 \over 4} 
  {\cal F}_{\mu\nu}^A[{\cal V}] {\cal F}^{\mu\nu}{}^A[{\cal V}],
\end{eqnarray}
while the gauge-fixing term (\ref{GF0}) is changed \cite{KondoVI}
into
\begin{eqnarray}
  \tilde S_{GF}[{\cal V}, \gamma, \bar \gamma, \beta]
  &:=& - \int d^Dx \ i  \tilde \delta_B \ {\rm tr}_G
  \left[ \bar \gamma \left( \partial^\mu {\cal V}_\mu +{\tilde \alpha
\over 2}\beta \right)   \right] .
\label{GF2}
\end{eqnarray}
This implies that the background gauge for ${\cal Q}$ is changed into
the Lorentz gauge for ${\cal V}$, 
$\partial_\mu {\cal V}_\mu=0$.
Then the generating functional
in the background field $\Omega$ is cast into
\begin{eqnarray}
 \tilde Z[J, \Omega] 
 = \int [d{\cal V}] [d\gamma][d\bar \gamma][d\beta]
 \exp \left\{ iS_{YM}[{\cal V}] 
 + i\tilde S_{GF}[{\cal V}, \gamma, \bar \gamma, \beta]
 + i(J_\mu \cdot U{\cal V}^\mu U^{\dagger})
\right\} ,
 \nonumber\\ 
\end{eqnarray}
where ${\cal V}_\mu, \gamma, \bar \gamma$ and $\beta$ are defined by the
adjoint rotations of 
${\cal Q}_\mu, \tilde C, \bar {\tilde C}$ and $\bar B$ respectively:
\begin{eqnarray}
{\cal V}_\mu :=  U^\dagger {\cal Q}_\mu U, \quad
 \gamma := U^\dagger \tilde C U, \quad
 \bar \gamma := U^\dagger \bar {\tilde C} U, \quad
 \beta := U^\dagger \tilde B U  .
 \label{adjr}
\end{eqnarray}
Thus the total generating functional reads 
\begin{eqnarray}
 && Z[J] = \int [dU][dC][d\bar C][dB] 
 \tilde Z[J, \Omega] 
  \exp (i S_{GF}[\Omega, C, \bar C, B]) 
 \exp [i(J_\mu \cdot \Omega^\mu)]
 \nonumber\\
 &=& \int [dU] [dC][d\bar C][dB]
 \exp \{ i \tilde S_{eff}[J, \Omega]
 + i S_{TQFT}[\Omega, C, \bar C, B]
 +i(J_\mu \cdot \Omega^\mu) \} ,\nonumber\\
\end{eqnarray}
where we have made a change of variable from $\Omega$ to $U$.
The measure $[dU]$ is invariant under the multiplication 
\begin{eqnarray}
   U(x) \rightarrow \tilde U(x) U(x) ,
\end{eqnarray}
which leads to the finite gauge transformation of the background
field,
\begin{eqnarray}
  \Omega(x) \rightarrow \tilde U(x) \Omega(x) \tilde U^\dagger(x)
  + {i \over g} \tilde U(x) d \tilde U^\dagger(x).
\end{eqnarray}
The modified MA gauge proposed in Ref.\cite{KondoII} leads to the TQFT with the action 
\begin{eqnarray}
  S_{TQFT}[\Omega, C, \bar C, B]
  := \int_{{\bf R}^D} d^Dx \ i \delta_B \bar \delta_B 
  {\rm tr}_{G/H} \left[ {1 \over 2} \Omega_\mu \Omega^\mu
  -{\alpha \over 2}i C \bar C \right] .
  \label{TQFT}
\end{eqnarray}
\par
The expectation value of the functional $f({\cal A})$ of
${\cal A}$ is calculated as follows. If 
the functional is of the form 
$f({\cal A}) = g({\cal V}_\mu, U) h(U)$, then the expectation value is calculated according to%
\footnote{See Remark 10.1.} 
\begin{eqnarray}
  \langle f({\cal A}) \rangle_{YM} 
  = \langle \langle g({\cal V}_\mu, U)
\rangle_{pYM}^{{\cal V}} h(U) \rangle_{TQFT}^U ,
\label{expec}
\end{eqnarray}
where the sector of the perturbative deformation is defined by
\begin{eqnarray}
  \langle  (\cdots) \rangle_{pYM}^{{\cal V}} 
  &=&  Z_{pYM}^{-1} \int [d{\cal V}] [d\gamma][d\bar \gamma][d\beta]
 \exp \left\{ i  S_{YM}[{\cal V}] 
 + i \tilde S_{GF}[{\cal V}, \gamma, \bar \gamma, \beta]  
\right\}  (\cdots) ,
\nonumber\\
 Z_{pYM} &:=& \int [d{\cal V}] [d\gamma][d\bar \gamma][d\beta]
 \exp \left\{ i  S_{YM}[{\cal V}] 
 + i \tilde S_{GF}[{\cal V}, \gamma, \bar \gamma, \beta]  
\right\} ,
\end{eqnarray}
and the sector topological field theory is defined by
\begin{eqnarray}
 \langle  (\cdots) \rangle_{TQFT}^U 
&:=& Z_{TQFT}^{-1} \int [dU] [dC][d\bar C][dB]
 \exp \{ i  S_{TQFT}[\Omega, C, \bar C, B]  \} 
 (\cdots) ,
\nonumber\\
 Z_{TQFT}  &:=&  \int [dU] [dC][d\bar C][dB]
 \exp \{  i S_{TQFT}[\Omega, C, \bar C, B] \} .
\end{eqnarray}
This reformulation of the Yang-Mills theory is called the
``perturbative deformation of a topological quantum field theory''.
The expectation value 
$
 \langle  (\cdots) \rangle_{pYM}^{{\cal V}}
$
for the field ${\cal V}$ is calculated using a perturbation theory in terms of the coupling constant $g$.  
On the other hand, the expectation value
$
 \langle  (\cdots) \rangle_{TQFT}^U 
$
should be calculated in a non-perturbative way to incorporate the
topological contribution.  Here $U$ is a compact gauge variable
corresponding to a finite gauge transformation.   In the instanton
calculus, the integration measure
$[dU]$ is replaced with the (finite-dimensional) integration over the
collective coordinates of the instanton.

\subsection{Dimensional reduction to the NLS model in the MA gauge}
\par
It has been shown \cite{KondoII} that, due to the hidden supersymmetry
 $OSp(D|2)$,  the TQFT part (\ref{TQFT}) is reduced to the
$(D-2)$-dimensional coset ($G/H$) nonlinear sigma (NLS) model,
\begin{eqnarray}
  S_{NLSM}[U, C, \bar C]
  := \alpha \pi \int_{{\bf R}^{D-2}} d^{D-2}x    \
  {\rm tr}_{G/H} \left[ {1 \over 2} \Omega_\mu \Omega_\mu
  -{\alpha \over 2}i C \bar C \right] ,
  \label{GF'}
\end{eqnarray}
where, for the matrix element $\Omega_{ab}$ (see Appendix C),
\begin{eqnarray}
  {\rm tr}_{G/H} \left[ {1 \over 2} \Omega_\mu(x) \Omega_\mu(x)
\right]
  = \sum_{a,b: a<b} (\Omega_\mu(x))_{ab}(\Omega_\mu(x))_{ab} . 
\end{eqnarray}
By making use of the complex coordinates in the flag space $G/H$,
the action can be rewritten as (see Appendix C)
\begin{eqnarray}
  S_{NLSM} &=& {\alpha \pi \over 2g^2} \int_{{\bf R}^{D-2}} d^{D-2}x
g_{\alpha\bar \beta}
  {\partial w^{\alpha} \over
\partial x_a}{\partial \bar w^{\beta} \over \partial x_a} ,
\quad (a=1, \cdots, D-2) 
\end{eqnarray}
where we have omitted to write the decoupled ghost term, $C(x) \bar
C(x)$. In particular, for $D=4$,
\begin{eqnarray}
  S_{NLSM} = {\alpha \pi \over 2g^2} \int_{{\bf C}} dzd\bar z \
g_{\alpha\bar
\beta} 
 \left(
 {\partial w^\alpha \over \partial z}{\partial \bar w^\beta \over
\partial \bar z}
  + {\partial w^\alpha \over \partial \bar z}{\partial \bar w^\beta
\over \partial z}
  \right) ,
\end{eqnarray}
where $z=x+iy=x_1+ix_2 \in  {\bf C} \cong {\bf R}^2$
and
$dxdy=dx_1 dx_2={i \over 2}dz d\bar z$.
The  $G=SU(2)$ case is analyzed in Ref.\citen{KondoII}, and in that case we have
\begin{eqnarray}
  S_{NLSM} = {\alpha \pi \over 2g^2} \int_{{\bf C}} dzd\bar z \
{1 \over (1+|w|^2)^2}
 \left(
 {\partial w \over \partial z}{\partial \bar w \over
\partial \bar z}
  + {\partial w \over \partial \bar z}{\partial \bar w
\over \partial z}
  \right) .
\end{eqnarray}

The above described strategy is schematically depicted below.%
\footnote{When we encounter the NLS model obtained through dimensional reduction in the following sections, the coupling constant $g$ should be replaced as
\begin{equation}
 g^2 \rightarrow {2 \over \alpha} g^2 ,
\end{equation}
since we do not express the $\alpha$ dependence explicitly.
}
\vskip 0.5cm
\begin{center}
\unitlength=1.0cm
\thicklines
 \begin{picture}(12,6)
 \put(2,5){\framebox(8,1){D-dim. QCD with a gauge group $G$}}
 \put(6.2,5){\vector(0,-1){0.8}}
 \put(7,4.5){MA gauge}
 \put(-0.2,2.8){\framebox(12.4,1.4){}}
 \put(0,3){\framebox(5,1){D-dim. Perturbative QCD}}
 \put(6,3.5){$\bigotimes$}
 \put(5.5,3){deform}
 \put(7,3){\framebox(5,1){D-dim. TQFT}}
 \put(8.5,3){\vector(0,-1){1}}
 \put(9,2.4){Dimensional reduction}
 \put(-0.2,0.8){\framebox(12.4,1.4){}}
 \put(0,1){\framebox(5,1){D-dim. Perturbative QCD}}
 \put(6,1.5){$\bigotimes$}
 \put(5.5,1){deform}
 \put(7,1){\framebox(5,1){(D-2)-dim. G/H NLSM}}
 \end{picture}
\end{center}

\section{Area law of the Wilson loop (I)}
\setcounter{equation}{0}

In this section we derive the area law of the Wilson loop
based on the instanton calculus. 
\citen{Luscher78,Gross78,BL79} 
A more systematic estimation is given in the next section based
on the large $N$ expansion.
\par
The static potential $V(R)$  is evaluated from the
rectangular Wilson loop $C$ with sides $T$
 and $R$ according to
\begin{eqnarray}
 V(R) :=   - \lim_{T \rightarrow \infty} {1 \over T} \ln  
 \langle W^C[{\cal A}] \rangle_{YM_4} .
\label{st}
\end{eqnarray}
The (full) string tension $\sigma$ is defined by
\begin{eqnarray}
  \sigma := - \lim_{A(C) \rightarrow \infty}
  {1 \over A(C)} \ln \langle W^C[{\cal A}] \rangle_{YM_4} ,
\end{eqnarray}
where $A(C)$ is the minimal area of the surface spanned by the Wilson
loop
$C$.  Of course, the rectangular loop  has minimal area:
$A(C)=TR$.
\par
Using the NAST for the Wilson loop operator,
\begin{eqnarray}
 W^C[{\cal A}] 
= \int [d\mu(\xi)]_C \exp \left( 
i g \oint_C   n^A {\cal A}^A  
+ i \oint_C   \omega  \right)  ,
\label{NAST3}
\end{eqnarray}
we can write its expectation value in the Yang-Mills theory as
\cite{KondoIV,KondoVI}
\begin{eqnarray}
 && \langle W^C[{\cal A}] \rangle_{YM_4} 
\nonumber\\  
&=& \Biggr\langle \Biggr\langle 
  \exp \left[ i g \oint_C dx^\mu n^A(x) {\cal V}_\mu^A(x) \right]
\Biggr\rangle_{pYM_4}
  \exp \left[ i \oint_C \omega \right]
\Biggr\rangle_{TQFT_4} ,
\label{NASTf}
\end{eqnarray}
where the measure $\int [d\mu(\xi)]_C$ gives a redundant factor which does not affect the string tension in the area law, as shown in Section IV.C of Ref.~\citen{KondoIV} and hence it is omitted in the following.
\subsection{Perturbative expansion and dimensional reduction}
\par
\par
In our reformulation, the field ${\cal V}_\mu^A(x)$ is identified
with the perturbative deformation. Thus we expand the first
exponential of (\ref{NASTf}) in powers of the coupling constant $g$:
\begin{eqnarray}
&& \Biggr\langle 
  \exp \left[ i g \oint_C dx^\mu n^A(x) {\cal V}_\mu^A(x) \right]
\Biggr\rangle_{pYM_4}
\nonumber\\ &=& 1 + \sum_{n=1}^{\infty} {(ig)^n \over n!}
\oint_C dx_1^{\mu_1} \oint_C dx_2^{\mu_2} \cdots \oint_C
dx_n^{\mu_n}   n^{A_1}(x_1) n^{A_2}(x_2) \cdots n^{A_n}(x_n)
\nonumber\\&& \times
\langle {\cal V}_{\mu_1}^{A_1}(x_1) {\cal V}_{\mu_2}^{A_2}(x_2) 
\cdots {\cal V}_{\mu_n}^{A_n}(x_n) \rangle_{pYM_4} .
\end{eqnarray}
Hence, we obtain
\begin{eqnarray}
 && \langle W^C[{\cal A}] \rangle_{YM_4} 
\nonumber\\&&  
= \Biggr\langle  \exp \left[ i \oint_C \omega \right]
\Biggr\rangle_{TQFT_4}
 + \sum_{n=1}^{\infty} {(ig)^n \over n!}
\oint_C dx_1^{\mu_1} \oint_C dx_2^{\mu_2} \cdots \oint_C
dx_n^{\mu_n} 
\nonumber\\&& \times 
\Biggr\langle n^{A_1}(x_1) n^{A_2}(x_2) \cdots n^{A_n}(x_n) 
\exp \left[ i \oint_C \omega \right] \Biggr\rangle_{TQFT_4}
\nonumber\\&& \times
\langle {\cal V}_{\mu_1}^{A_1}(x_1) {\cal V}_{\mu_2}^{A_2}(x_2) 
\cdots {\cal V}_{\mu_n}^{A_n}(x_n) 
\rangle_{pYM_4} .
\nonumber\\
\\
&=& \Biggr\langle  \exp \left[ i \oint_C \omega \right]
\Biggr\rangle_{TQFT_4} 
\nonumber\\
&&
\times \Biggr[ 1 +
\sum_{n=1}^{\infty} {(ig)^n \over n!}
\oint_C dx_1^{\mu_1} \oint_C dx_2^{\mu_2} \cdots \oint dx_n^{\mu_n}
\langle {\cal V}_{\mu_1}^{A_1}(x_1) {\cal V}_{\mu_2}^{A_2}(x_2) 
\cdots {\cal V}_{\mu_n}^{A_n}(x_n) 
\rangle_{pYM_4}  
\nonumber\\&& \times
{\langle  n^{A_1}(x_1) n^{A_2}(x_2) \cdots n^{A_n}(x_n)
 \exp \left[ i \oint_C \omega \right]
\rangle_{TQFT_4} \over
\langle  \exp \left[ i \oint_C \omega \right] \rangle_{TQFT_4}}  
\Biggr] .
\label{wlexpe}
\end{eqnarray}
We restrict the Wilson loop to a planar loop.  This choice has
the following advantage.
 For the planar Wilson loop $C$, the Parisi-Sourlas dimensional
reduction leads to the following identities:\cite{KondoII} 
\begin{equation}
 \Big\langle \exp \left[ i \oint_C \omega \right]
\Big\rangle_{TQFT_4} 
=  \Big\langle \exp \left[ i \oint_C \omega \right]  
\Big\rangle_{NLSM_2} ,
\end{equation}
and
\begin{equation}
 \Biggr\langle  n^{A_1}(x_1) \cdots n^{A_n}(x_n)
 \exp \left[ i \oint_C \omega \right]
\Biggr\rangle_{TQFT_4} 
=  \Biggr\langle  n^{A_1}(x_1) \cdots n^{A_n}(x_n)
 \exp \left[ i \oint_C \omega \right]
\Biggr\rangle_{NLSM_2} ,
\end{equation}
where $x_1, \cdots, x_n \in C \subset {\bf R}^2$. 
This is because, in the case of the fundamental representation of
$SU(N)$, $n^A$ and $\omega$ can be written in terms of $U$ as (see
(\ref{ndef}), (\ref{omegadef}))
\begin{eqnarray}
 n^A(x) &=& U_{1a}(x)(T^A)_{ab}\bar U_{1b}(x)
 = \bar \phi_a(x) (T^A)_{ab} \phi_b(x) ,
 \\
 \omega(x) &=&  {i \over 2}  
  [U_{1a}(x) d \bar U_{1a}(x) - d U_{1a}(x) \bar U_{1a}(x)]
\nonumber\\
  &=&  {i \over 2} 
  [\bar \phi_a(x) d \phi_a(x) - d \bar \phi_a(x) \phi_a(x)]  .
\end{eqnarray}

\par
Taking the logarithm of the Wilson loop (\ref{wlexpe}) and expanding
it in powers of the coupling constant, we obtain
\begin{eqnarray}
 && \ln \langle W^C[{\cal A}] \rangle_{YM_4} 
\\
&=& \ln \Biggr\langle  \exp \left[ i \oint_C \omega \right]
\Biggr\rangle_{NLSM_2} 
\nonumber\\
&&
+ \ln \left[ 1
-  {g^2 \over 2} \oint_C dx^\mu \oint_C dy^\nu G_{\mu\nu}^{AB}(x,y)
{\langle  n^A(x) n^B(y) \exp \left[ i \int_{S} \Omega_K
\right]
\rangle_{NLSM_2} \over
\langle  \exp \left[ i \oint_C \omega \right]
\rangle_{NLSM_2} }   + O(g^4)  \right]  
\nonumber\\
&=& \ln \Biggr\langle  \exp \left[ i \oint_C \omega \right]
\Biggr\rangle_{NLSM_2} 
\nonumber\\&&  
-  {g^2 \over 2} \oint_C dx^\mu \oint_C dy^\nu G_{\mu\nu}^{AB}(x,y)
{\langle  n^A(x) n^B(y) \exp \left[ i \oint_C \omega \right]
\rangle_{NLSM_2} \over
\langle  \exp \left[ i \oint_C \omega \right]
\rangle_{NLSM_2} }   + O(g^4)  ,
\label{expansion}
\end{eqnarray}
where we have defined the two-point function 
 \begin{eqnarray}
 G_{\mu\nu}^{AB}(x,y) :=
  \langle {\cal V}_\mu^A(x) {\cal V}_\nu^B(y) \rangle_{pYM_4} .
\end{eqnarray}

In the rest of this section we focus on the first term in
(\ref{expansion}). The remaining terms will be estimated in the next
section.

\subsection{The instanton in $F_{N-1}$ and $CP^{N-1}$ models}
\par
We wish to demonstrate the area law for the expectation value,
\begin{eqnarray}
&& \Big\langle \exp \left( i \oint_C \omega \right)
\Big\rangle_{NLSM_2}
\equiv \Big\langle \exp \left( i \int_{S} \Omega_K \right)
\Big\rangle_{NLSM_2}
\\
 &=& Z_{NLSM_2}^{-1} \int [d\mu(w,\bar w)] \exp (-S_{NLSM_2}[w,\bar
w]) 
 \exp \left( i \int_{S} \Omega_K \right) ,
 \label{2dexpec}
\end{eqnarray}
where the two-dimensional NLS model is defined by the action 
\begin{eqnarray}
  S_{NLSM_2} &=& {\pi \over g^2(\mu)} \int_{{\bf R}^2} d^2x
g_{\alpha\bar \beta}
  {\partial w^{\alpha} \over
\partial x_a}{\partial \bar w^{\beta} \over \partial x_a} 
\\
&=& {\pi \over g^2(\mu)} \int_{{\bf C}} dzd\bar z \ g_{\alpha\bar
\beta} 
 \left(
 {\partial w^\alpha \over \partial z}{\partial \bar w^\beta \over
\partial \bar z}
  + {\partial w^\alpha \over \partial \bar z}{\partial \bar w^\beta
\over \partial z}
  \right) ,
\end{eqnarray}
where 
\begin{equation}
z=x+iy=x_1+ix_2 \in  {\bf C} \cong {\bf R}^2, 
\quad dxdy=dx_1 dx_2={i \over 2}dz d\bar z ,
\end{equation}
and $g(\mu)$ is the
Yang-Mills coupling constant.\footnote{
The running of the coupling constant is given by the
perturbative deformation in four-dimensional Yang-Mills theory, as in
(\ref{runc}). 
}
\par
Note that the action satisfies the inequality,
\begin{eqnarray}
  S_{NLSM} &=& {\pi \over 2g^2(\mu)} \int_{{\bf R}^2} d^2x
g_{\alpha\bar \beta}
(\partial_a w^{\alpha} \pm i \epsilon_{ab} \partial_b w^{\alpha})(\partial_a w^{\beta} \pm i \epsilon_{ac} \partial_c w^{\beta})^*
  \nonumber\\
  && \pm   i
 {\pi \over g^2(\mu)} \int_{{\bf R}^2} d^2x
\epsilon_{ab} g_{\alpha\bar \beta}
  {\partial w^{\alpha} \over
\partial x_a}{\partial \bar w^{\beta} \over \partial x_b} 
\\
&& \ge \pm   i
 {\pi \over g^2(\mu)} \int_{{\bf R}^2} d^2x
\epsilon_{ab} g_{\alpha\bar \beta}
  {\partial w^{\alpha} \over
\partial x_a}{\partial \bar w^{\beta} \over \partial x_b} .
\end{eqnarray}
This inequality is saturated when $w^\alpha$ satisfies
the equation
$
\partial_a w^{\alpha} \pm i \epsilon_{ab} \partial_b w^{\alpha}=0,
$
which is equivalent to the Cauchy-Riemann equation,
\begin{equation}
 \bar \partial_z w^{\alpha} 
:= (\partial_1 + i \partial_2) w^{\alpha} = 0 .
\label{CR}
\end{equation}
The solution $w^\alpha=f^\alpha(z)$ is an arbitrary rational
function of $z$. 
\par
The finite action configuration of the coset NLS model is
provided by the instanton solution, which is a solution of the
Cauchy-Riemann equation (\ref{CR}).
 It is known
\cite{Perelomov87} that the integer-valued topological charge $Q$ of
the instanton in the $F_{N-1}$ NLS model is given by the
integral of the K\"ahler 2-form over ${\bf R}^2$:
\begin{eqnarray}
 Q = {1 \over \pi} \int \Omega_K 
&=& \int_{{\bf R}^2} {d^2x \over \pi} 
\epsilon_{ab} g_{\alpha\bar \beta}
  {\partial w^{\alpha} \over
\partial x_a}{\partial \bar w^{\beta} \over \partial x_b} 
\nonumber\\
&=&  \int_{{\bf C}} {dzd\bar z \over \pi} \ g_{\alpha\bar \beta} 
 \left(
 {\partial w^\alpha \over \partial z}{\partial \bar w^\beta \over
\partial \bar z}
  - {\partial w^\alpha \over \partial \bar z}{\partial \bar w^\beta
\over \partial z}
  \right) .
\end{eqnarray}
This is a generalization of the $F_1$ case ($N=2$), where
\begin{eqnarray}
  Q = {i \over 2\pi}\int_{{\bf C}} {dwd\bar w \over (1+|w|^2)^2}
  =  {i \over 2\pi}\int_{S^2} {dzd\bar z \over (1+|w|^2)^2}
 \left(
 {\partial w \over \partial z}{\partial \bar w \over \partial \bar z}
  - {\partial w \over \partial \bar z}{\partial \bar w \over \partial
z}
  \right) .
\end{eqnarray}
\par
Thus the instanton solution is characterized by the
integral topological charge $Q \in {\bf Z}$.  
For instanton ($Q>0$) and anti-instanton ($Q<0$) configurations 
with a topological charge $Q$, the action has 
\begin{equation}
 S_{NLSM} = {\pi^2 \over g^2}|Q| .
\end{equation}
\par
The metric in the K\"ahler manifold $F_{N-1}$ is obtained 
in accordance with the relation
$
  g_{\alpha\bar\beta} = {\partial \over \partial w^\alpha}
 {\partial \over \partial \bar w^\beta} K 
$
from  
the K\"ahler potential,
\begin{eqnarray}
 K(w,\bar w) = \sum_{\ell=1}^{N-1} d_\ell K_\ell(w,\bar w)
 = \sum_{\ell=1}^{N-1} d_\ell \ln \Delta_\ell (w,\bar w) ,
\end{eqnarray}
with the Dynkin indices $d_\ell (\ell=1,\cdots,N-1)$.
Then the integral
$\int \Omega_K$ over the whole two-dimensional space reads
\begin{equation}
 \int_{{\bf R}^2} \Omega_K = \pi Q 
 = \pi \sum_{\ell=1}^{N-1} d_\ell Q_\ell ,
\end{equation}
where $Q_\ell$ are integer-valued topological charges.
Hence, $\Omega_K(x)/\pi$ is identified with the density of the
topological charge (up to the weight due to the index $d_\ell$).

\par

Now we consider the $CP^{N-1}$ model.
If we identify $w_\alpha$ with the inhomogeneous coordinates, e.g., 
$
 w_\alpha := {\phi_\alpha \over \phi_N} ,
$
the metric (\ref{metricCP}) can be rewritten as
\begin{equation}
 g_{\alpha\bar \beta}(\phi) =  {(||\phi||^2)\delta_{\alpha\beta} -
\bar
\phi_\alpha \phi_\beta \over
 (||\phi||^2)^2} ,
 \label{metric3CP}
\end{equation}
where 
\begin{equation}
 ||\phi||^2   :=  \sum_{\alpha=1}^{N} |\phi_\alpha|^2 
 = |\phi_N|^2 \left( 1 + |||w|||^2 \right) .
\end{equation}
The action of the $CP^{N-1}$ model is given by
\begin{eqnarray}
 S_{CP^{N-1}} = {\pi \over g^2}
 \int d^d x   g_{\alpha\bar \beta}(w) \partial_\mu w^\alpha
\partial_\mu \bar w^\beta ,
\end{eqnarray}
or equivalently,
\begin{eqnarray}
 S_{CP^{N-1}} = {\pi \over g^2}
 \int d^d x   g_{\alpha\bar \beta}(\phi) \partial_\mu \phi^\alpha
\partial_\mu \bar \phi^\beta .
\end{eqnarray}
Under the constraint 
$
 ||\phi||^2 = 1,
$
the action can be  written as
\begin{eqnarray}
 S_{CP^{N-1}}
&=& {\beta_g \over 2} \int d^d x   (\delta_{\alpha\beta}-\bar
\phi_\alpha
\phi_\beta)\partial_\mu \phi^\alpha \partial_\mu \bar \phi^\beta 
\nonumber\\
 &=& {\beta_g \over 2} \int d^d x   [(\partial_\mu \bar \phi^\alpha
\partial_\mu \phi^\alpha + (\bar \phi^\alpha \partial_\mu
\phi^\alpha)(\bar \phi^\beta \partial_\mu \phi^\beta) ]
 .
\end{eqnarray}
This agrees with the action of the $CP^{N-1}$ model presented in
Ref.\citen{KondoII}. (See Appendix C for more details).

\subsection{Area law in the dilute instanton-gas approximation}

\begin{figure}
\begin{center}
 \leavevmode
 \epsfxsize=50mm
 \epsfysize=50mm
 \epsfbox{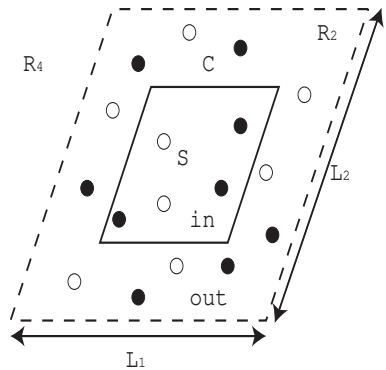}
\end{center} 
 \caption[]{Instanton and anti-instanton configuration and the 
Wilson loop $C$ in a finite-volume region $L_1 \times L_2$ in the
two-dimensional plane ${\bf R}^2$ embedded in ${\bf R}^4$.}
 \label{2d-instanton}
\end{figure}

\par
If the Wilson loop is large compared with the typical size of
the instanton,
$\int_S \Omega_K(x)/\pi$ in (\ref{2dexpec}) counts the number of
instantons
$n_{+}^{in}$ minus anti-instantons $n_{-}^{in}$ that are contained
inside the area $S \subset {\bf R}^2$ bounded by the loop $C$:
\begin{eqnarray}
  \int_S \Omega_K = \pi (n_{+}^{in} - n_{-}^{in}) .
\end{eqnarray}
(See Fig.\ref{2d-instanton}.)
Thus, the expectation value 
$
\langle \exp \left( i \int_{S} \Omega_K \right) \rangle_{NLSM} 
$
is calculated by summing over all the possible cases of instanton and
anti-instanton configurations (i.e., by the integration over the
instanton moduli).
In this calculation, we use  
\begin{equation}
 S_{NLSM} = {\pi^2 \over g^2}|Q|
 = {\pi^2 \over g^2} (n_{+} + n_{-}),
 \quad n_{\pm} = n_{\pm}^{in} + n_{\pm}^{out} .
\end{equation}
where $n_+^{out}$ ($n_-^{out}$) is the number of instantons
(anti-instantons) outside $S$, and $n_+$ ($n_-$) is the
total number of instantons (anti-instantons).
For the quark in the fundamental representation ${\bf N}$
($d_1=1,d_2=d_3=\cdots =d_{N-1}=0$), this is easily carried out as
follows. 
\par
For the $SU(3)$ case with $[1,0]$, an element
$\xi \in F_2$ is independent of $w_3$, so that $w_3$ is redundant in
this case.  Hence, it suffices to consider the $CP^2$ model
for the fundamental quark (up to Weyl symmetry). 
For $CP^2$, the K\"ahler two-form is given by (\ref{CP2K2f}) 
\begin{eqnarray}
  \Omega_K &=&  im (\Delta_1)^{-2}[(1+|w_1|^2)dw_2 \wedge d\bar w_2
  - \bar w_2 w_1 dw_2 \wedge d\bar w_1
  \nonumber\\&&
  - w_2 \bar w_1 dw_1 \wedge d\bar w_2
  + (1+|w_2|^2) dw_1 \wedge d\bar w_1] .
  \label{CP2K2f2}
\end{eqnarray}
When $w_2=0$,  $\Omega_K$ reduces to
\begin{equation}
  \Omega_K  
  = i  (1+|w_1|^2)^{-2} dw_1 \wedge d\bar w_1 .
\end{equation}
Similarly, when $w_1=0$,  we have
\begin{equation}
 \Omega_K  
  = i  (1+|w_2|^2)^{-2} dw_2 \wedge d\bar w_2  .
\end{equation}
For a polynomial $w_\alpha=w_\alpha(z)$  in $z=x+iy$ of order
$n$, we find the instanton charge 
\begin{equation}
\int \Omega_K = \pi Q, \quad Q \in {\bf Z} .
\end{equation}
This is the same situation as that encountered in
$SU(2)$, in which case
\begin{equation}
\int \Omega_K =2j \pi Q ,
\end{equation}
where $j=1/2$ corresponds to the fundamental representation.\cite{KondoII,KondoIV} 
Thus the Wilson loop can be estimated by the naive instanton calculus.  
In fact, the dilute instanton gas approximation leads to
the area law for the Wilson loop (see Ref.\citen{KondoII}).  
Here the factor $\pi$ is very important.   The integral of the K\"ahler
two-form $\Omega_K$ is a multiple of $\pi$.
If we had a factor $2\pi$, the area law would not hold, just as in the $j=1$
case of $SU(2)$.
\par

For the $SU(N)$ case with Dynkin index, 
$[1,0,\cdots,0]$, it suffices to consider the $CP^{N-1}$ model.
When $w_a\not=0$ and $w_b=0$ for all $b\not=a$, the K\"ahler
two-form (\ref{K2fN}) for $CP^{N-1}$ reduces to
\begin{equation}
 \Omega_K  
  = i  (1+|w_a|^2)^{-2} dw_a \wedge d\bar w_a  ,
\end{equation}
with no summation over $a$.  Therefore the above argument can be
applied to $SU(N)$ for any $N$.  This implies the confinement of
fundamental quarks in $SU(N)$ Yang-Mills theory within the
approximation of a dilute instanton gas. This naive instanton
calculation can be improved by including fluctuations from the
instanton solutions following
Ref.\citen{BL79}, and this issue will be discussed in detail in a subsequent article.\cite{Kondo00}

\section{Area law of the Wilson loop (II)}
\setcounter{equation}{0}

The derivation of the area law of the Wilson loop in the
four-dimensional Yang-Mills theory in the MA gauge is reduced
to demonstrating the area law of the diagonal Wilson loop in the
two-dimensional coset NLS model.   
In this section we complete a derivation of the area law of the Wilson 
loop in the fundamental representation.  
Here we use the large $N$ expansion
\cite{DVL78,DVL79,Witten79,Munster82,CR92} for the coset NLS model. 
(See, e.g.,
Refa.\citen{Yaffe82,Coleman85,Das87,RCV98} for reviews of the large $N$
expansion.)


To perform the large N expansion, it is convenient to introduce the
new variables 
\begin{eqnarray}
  P_{ab}(x)  &:=&  \bar \phi_a(x) \phi_b(x) 
= U_{1a}(x) \bar U_{1b}(x) ,
\\
 {\cal V}_\mu^{ab}(x) &:=&  {\cal V}_\mu^A(x) (T^A)_{ab} ,
\end{eqnarray}
which are used to rewrite  
\begin{equation}
  C_\mu(x) := n^A(x) {\cal V}_\mu^A(x) 
= P_{ab}(x) {\cal V}_\mu^{ab}(x) ,
\end{equation}
where $A=1, \cdots, N^2-1$ and $a,b = 1, \cdots N$ for $SU(N)$.

\par

In terms of the above variables, the expansion (\ref{wlexpe}) of the
expectation value of the Wilson loop operator in powers of the
coupling constant $g$ is rewritten as
\begin{eqnarray}
 \langle W^C[{\cal A}] \rangle_{YM_4} 
&=& \Biggr\langle  \exp \left[ i \oint_C \omega \right]
\Biggr\rangle_{TQFT_4}
 + \sum_{n=1}^{\infty} {(ig)^n \over n!} 
\oint_C dx_1^{\mu_1} \oint_C dx_2^{\mu_2} \cdots \oint_C dx_n^{\mu_n} 
\nonumber\\&& 
 \times  \langle 
{\cal V}_{\mu_1}^{a_1 b_1}(x_1) {\cal V}_{\mu_2}^{a_2 b_2}(x_2)
\cdots {\cal V}_{\mu_n}^{a_n b_n}(x_n)
\rangle_{pYM_4} 
\nonumber\\&& 
\times  \Biggr\langle P_{a_1 b_1}(x_1) P_{a_2 b_2}(x_2) \cdots P_{a_n
b_n}(x_n) \exp \left[ i \oint_C \omega
\right]
\Biggr\rangle_{TQFT_4} 
\\
&=& \Biggr\langle  \exp \left[ i \oint_C \omega \right]
\Biggr\rangle_{TQFT_4} 
\nonumber\\
&&
\times \Biggr[ 1 +
\sum_{n=1}^{\infty} {(ig)^n \over n!} 
\oint_C dx_1^{\mu_1} \oint_C dx_2^{\mu_2} \cdots \oint_C dx_n^{\mu_n}
\nonumber\\&& \times 
 \langle 
{\cal V}_{\mu_1}^{a_1 b_1}(x_1) {\cal V}_{\mu_2}^{a_2 b_2}(x_2)
\cdots {\cal V}_{\mu_n}^{a_n b_n}(x_n)
\rangle_{pYM_4}
\nonumber\\&&
\times 
 {\langle P_{a_1 b_1}(x_1) P_{a_2 b_2}(x_2) \cdots P_{a_n b_n}(x_n)
\exp \left[ i \oint_C \omega \right]
 \rangle_{TQFT_4} \over 
 \langle \exp \left[ i \oint_C \omega \right] \rangle_{TQFT_4}}
  \Biggr] .
\label{NASTW}
\end{eqnarray}
\par
The diagrams needed to calculate this expectation value are drawn in
Fig.\ref{WilsonLoopDiagram} based on the Feynmann rule given in
Fig.\ref{largeNrule}.
Here it should be remarked that the definition of the Wilson loop
operator  
\begin{equation}
 \langle W^C [{\cal A}] \rangle :=  {1 \over {\cal N}}
\Biggr\langle {\rm tr} \left[ {\cal P} 
 \exp \left( i g \oint_C dx^\mu {\cal A}_\mu(x)
 \right) \right] \Biggr\rangle_{YM_4} ,
\label{Wldef2}
\end{equation}
includes the normalization factor ${\cal N}^{-1}$ and that the
expectation value (\ref{Wldef2}) of the Wilson loop may have a
well-defined large
$N$ limit.  In particular, in the zero coupling limit,
the expectation value reduces to one.

\par
\begin{figure}
\begin{center}
 \leavevmode
 \epsfxsize=100mm
 \epsfbox{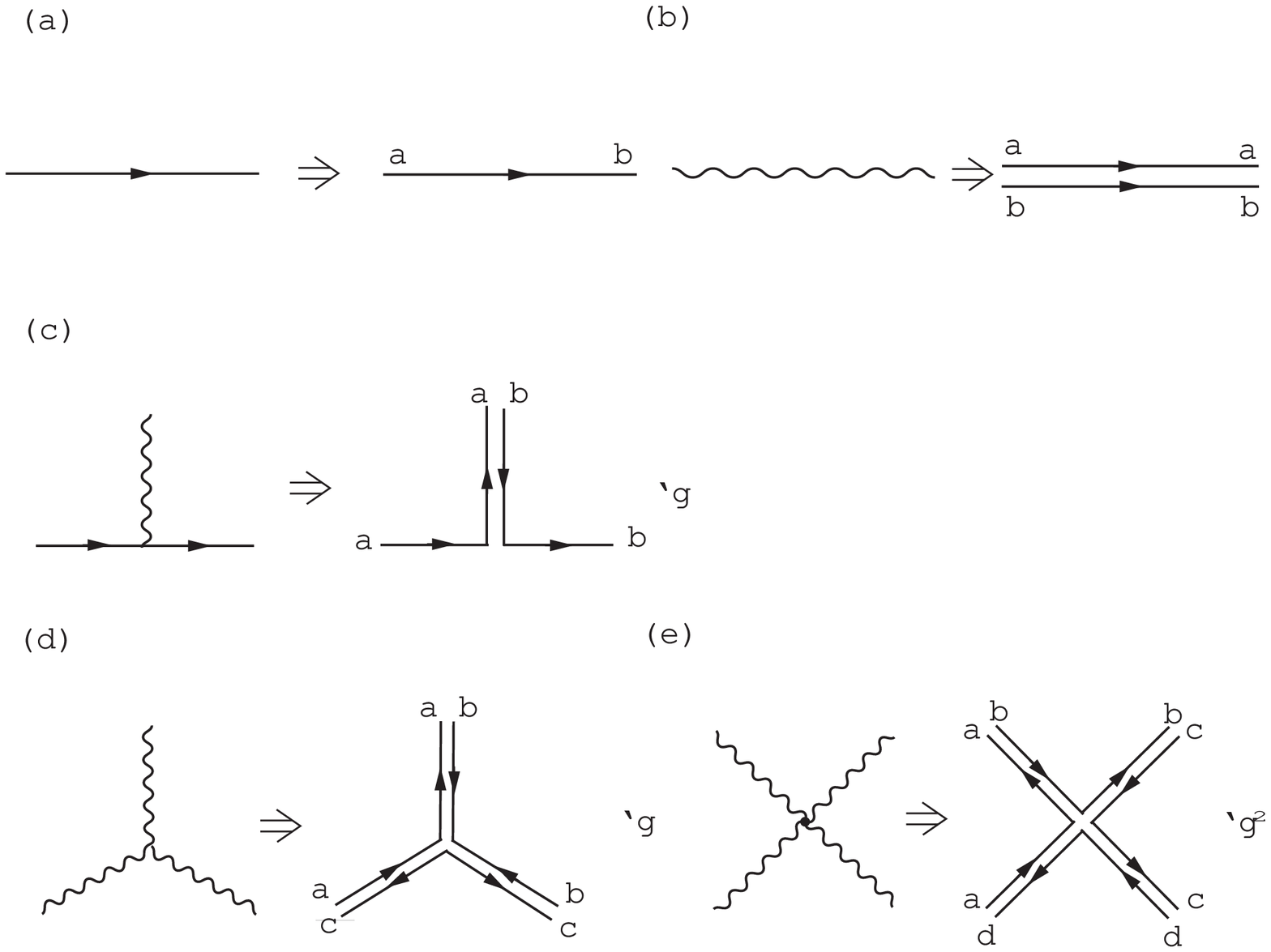}
\end{center} 
 \caption[]{Feynmann rule and the corresponding large $N$ rule
(double line notation due to 't Hooft) in QCD. Propagators: 
(a) quark propagator, (b) gluon propagator.
Vertices:  
(c) quark-gluon vertex ($g_{YM}$), (d) three-gluon vertex
($g_{YM}$), (e) four-gluon vertex ($g_{YM}^2$). }
 \label{largeNrule}
\end{figure}

\par
\begin{figure}[h]
\begin{center}
 \leavevmode
 \epsfxsize=120mm
 \epsfbox{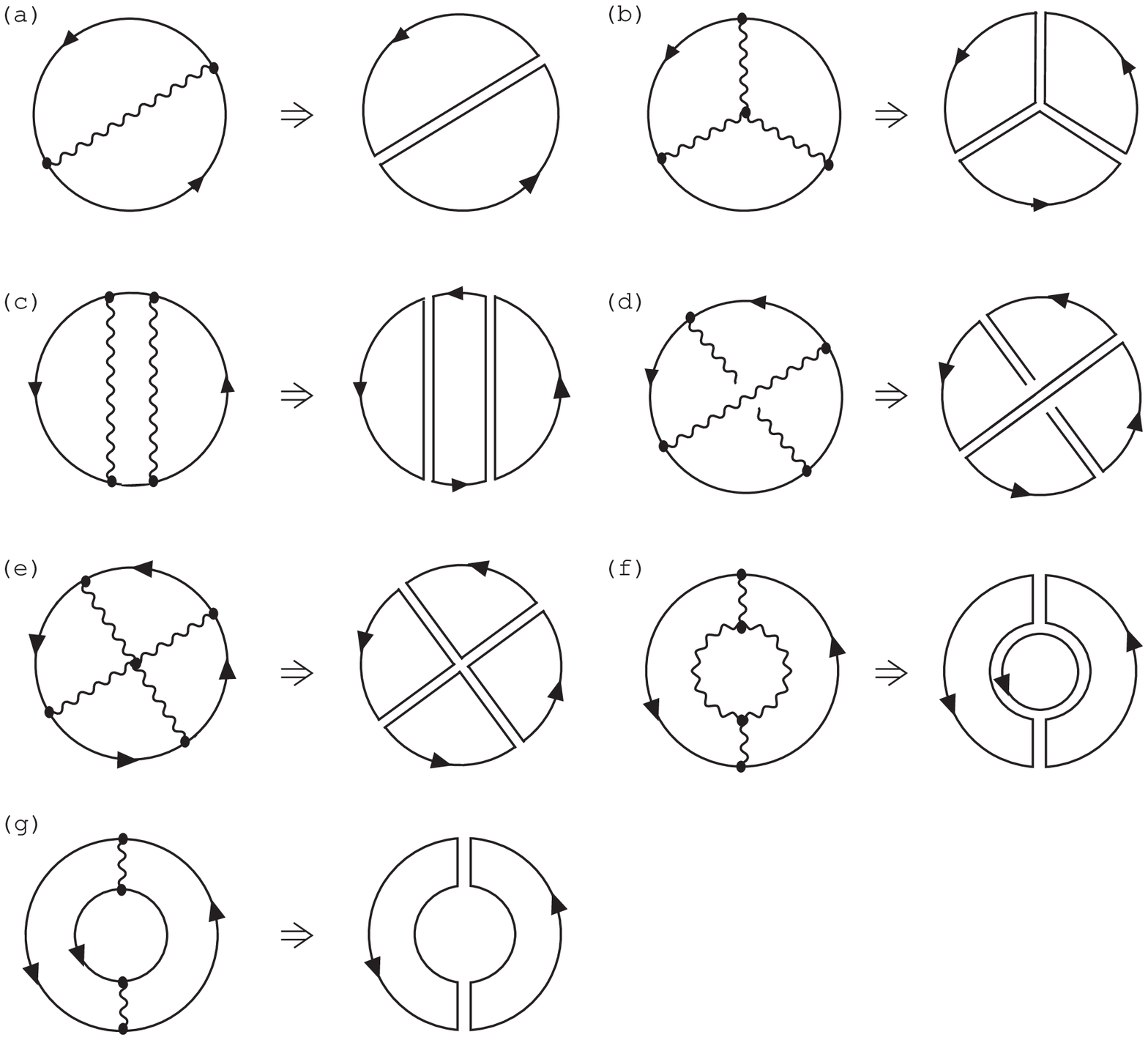}
\end{center} 
 \caption[]{Examples of Feynmann diagrams and the corresponding
double line notations that appear in calculating the
expectation value of the Wilson loop operator.  The order of each diagram
is estimated using the rule given in Fig.~\ref{largeNrule} as 
(a) $g_{YM}^2N^2 = \lambda N$, (b) $g_{YM}^4 N^3=\lambda^2 N$, 
(c) $g_{YM}^4 N^3 = \lambda^2 N$, (d) $g_{YM}^4 N = \lambda^2/N$,
(e) $g_{YM}^6 N^4 = \lambda^3 N$, (f) $g_{YM}^4 N^3 = \lambda^2 N$,
and (g) $g_{YM}^4 N^2 = \lambda^2 N^0$.
[Here note that each contribution should be divided by the
normalization factor $N$  for the fundamental quark, in agreement
with the definition (\ref{Wldef2}).]
In the leading order of large
$N$ expansion, the leading contributions come from the planar
diagrams, e.g., (a), (b), (c), (e) and (f), which are furthermore
classified by the order of
$\lambda$. Note that the contribution from a nonplanar diagram (d)
is suppressed for large $N$.  
The diagram (g) is the vacuum polarization diagram due to
quark--anti-quark pair creation and annihilation, which is neglected
in the pure Yang-Mills theory without dynamical quarks. 
}
 \label{WilsonLoopDiagram}
\end{figure}

\subsection{Large N expansion and dimensional reduction}
\par
\par
It is known \cite{tHooft74a} that only the planar diagrams contribute
to the expectation value
\begin{equation}
 \langle 
{\cal A}_{\mu_1}^{a_1 b_1}(x_1) {\cal A}_{\mu_2}^{a_2 b_2}(x_2)
\cdots {\cal A}_{\mu_n}^{a_n b_n}(x_n)
\rangle_{YM_4}
\end{equation}
in the leading order of the large $N$ expansion.
See Fig.~\ref{WilsonLoopDiagram}.
However, it is extremely difficult to sum up the infinite number
of terms belonging to the leading order of the large $N$ expansion
and to obtain a closed expression in the four-dimensional 
case. Of course,
this does not exclude the possibility that the closed expression
obtained by summing up all the leading diagrams may exhibit  the area
law. In fact, this strategy has been applied in the
two-dimensional case and has successfully lead to the area law (see,
e.g., Ref.\citen{Gross92}).  
\par
 For the planar Wilson loop $C$, we have already shown that the
Parisi-Sourlas dimensional reduction occurs and that the TQFT sector
reduces to the two-dimensional coset NLS model, i.e. the NLS model
on the flag space $F_{N-1}$. Hence, we obtain
\begin{equation}
 \Big\langle e^{ i \oint_C \omega }
\Big\rangle_{TQFT_4} 
=  \Big\langle e^{ i \oint_C \omega }
\Big\rangle_{NLSM_2} ,
\end{equation}
and
\begin{equation}
 \Biggr\langle P_{a_1 b_1}(x_1) \cdots P_{a_n b_n}(x_n) 
e^{ i \oint_C \omega }
 \Biggr\rangle_{TQFT_4}
 =
 \Biggr\langle P_{a_1 b_1}(x_1) \cdots P_{a_n b_n}(x_n) 
e^{ i \oint_C \omega }
 \Biggr\rangle_{NLSM_2} ,
\end{equation}
where $x_1, \cdots , x_n \in C \subset {\bf R}^2$. 
For the quark in the fundamental representation of $SU(N)$, the
relevant NLS model can be restricted to the $CP^{N-1}$ model.  
\par
We now return to the expression (\ref{NASTW}) obtained by way of the
NAST and apply the large $N$ expansion to the (perturbative)
deformation sector and the TQFT sector simultaneously.  
Taking the logarithm of the Wilson loop, therefore, we obtain
\begin{eqnarray}
 && \ln \langle W^C[{\cal A}] \rangle_{YM_4} 
\nonumber\\
&=& \ln \Biggr\langle  \exp \left[ i \oint_C \omega \right]
\Biggr\rangle_{CP^{N-1}_2} 
+ \ln \left[ 1
-   f[C]   + O(\lambda^2/N^2)  \right]  
\nonumber\\
&=& \ln \Biggr\langle  \exp \left[ i \oint_C \omega \right]
\Biggr\rangle_{CP^{N-1}_2}
-   f[C] + O(\lambda^2/N^2)  ,
\end{eqnarray}
where we have defined the two-point correlation function
 \begin{eqnarray}
 G_{\mu\nu}^{ab,cd}(x,y) &:=&
  \langle {\cal V}_\mu^{ab}(x) {\cal V}_\nu^{cd}(y) \rangle_{pYM_4} ,
\end{eqnarray}
and  
\begin{eqnarray}
  f[C] := {g^2 \over 2} \oint_C dx^\mu \oint_C dy^\nu
G_{\mu\nu}^{ab,cd}(x,y) {\langle  P_{ab}(x) P_{cd}(y) \exp
\left[ i \oint_C \omega \right]
\rangle_{CP^{N-1}_2} \over
\langle  \exp \left[ i \oint_C \omega \right]
\rangle_{CP^{N-1}_2} } .
\label{2ndterm}
\end{eqnarray}
\par
It should be remarked that the $f[C]$ is $O(\lambda/N)$.
This is different from the result of the usual large $N$ expansion
of the Yang-Mills theory, i.e., diagram (a) of
Fig.~\ref{WilsonLoopDiagram}, which is $O(\lambda)$.
This fact is shown as follows.
In a previous article,\cite{KondoVI}  it
is shown that the perturbative sector obeys the Lorentz-type gauge
fixing, 
$
 \partial_\mu {\cal V}_\mu= 0,
$
by virtue of the background gauge.  Here
we adopt the Feynman gauge to simplify the calculation.  Then the
propagator for ${\cal V}_\mu^A$ reads
\begin{eqnarray}
\langle {\cal V}_\mu^{A}(x) {\cal V}_\nu^{B}(y) \rangle_{pYM_4}
= \delta^{AB} \delta_{\mu\nu} D(x,y)  ,
\quad
 D(x,y) := {1 \over 4\pi^2} {1 \over |x-y|^2} .
\end{eqnarray}
Then we find
 \begin{eqnarray}
 G_{\mu\nu}^{ab,cd}(x,y) &:=&
  \langle {\cal V}_\mu^{ab}(x) {\cal V}_\nu^{cd}(y) \rangle_{pYM_4}
\\
&=& \langle {\cal V}_\mu^{A}(x) {\cal V}_\nu^{B}(y) \rangle_{pYM_4} 
(T^A)_{ab} (T^B)_{cd} 
\\
&=& \delta_{\mu\nu} D(x,y) \sum_{A=1}^{N^2-1} (T^A)_{ab} (T^A)_{cd} ,
\end{eqnarray}
where for $G=SU(N)$,
\begin{eqnarray}
 \sum_{A=1}^{N^2-1} (T^A)_{ab} (T^A)_{cd} 
 = {1 \over 2} \left( \delta_{ad} \delta_{bc} - {1 \over N}
 \delta_{ab} \delta_{cd} \right) .
\label{formula2}
\end{eqnarray}
This implies that 
\begin{equation}
 \sum_{A=1}^{N^2-1} (T^A T^A)_{ab} 
 =  {N^2-1 \over 2N}  \delta_{ab}  = C_2(R)  \delta_{ab} ,
\end{equation}
where $C_2(R)$ is the quadratic (second order) Casimir invariant of the fundamental representation.
If $G=U(N)$, the relation is simplified as
$
 \sum_{A=1}^{N^2} (T^A)_{ab} (T^A)_{cd} 
 = {1 \over 2} \delta_{ad} \delta_{bc} . 
$
The difference between $SU(N)$ and $U(N)$ disappears in the large
$N$ limit.
To leading order, we can set (see Appendix E)
\begin{eqnarray}
  {\langle  P_{ab}(x) P_{cd}(y) \exp \left[ i \oint_C \omega
\right]\rangle_{CP^{N-1}_2} \over
\langle  \exp \left[ i \oint_C \omega \right]
\rangle_{CP^{N-1}_2} } 
&\cong&
\langle  P_{ab}(x) P_{cd}(y) \rangle_{CP^{N-1}_2} .
\end{eqnarray}
Thanks to the $SU(N)$ invariance, it is easy to see that
\begin{eqnarray}
\langle  P_{ab}(x) P_{cd}(y) \rangle_{CP^{N-1}_2} 
= \left( \delta_{ad} \delta_{bc} - {1 \over N}
 \delta_{ab} \delta_{cd} \right)   Q(x,y)    ,
\label{GC0}
\end{eqnarray}
where
\begin{eqnarray}
 Q(x,y) = 
{\langle \bar \phi_a(x) \phi_b(x) \bar \phi_b(y) \phi_a(y)
\rangle_{CP^{N-1}_2}  - {1 \over N}
\langle \bar \phi_a(x) \phi_a(x) \bar \phi_b(y) \phi_b(y)
\rangle_{CP^{N-1}_2} 
\over N^2-1} .
\nonumber\\
\end{eqnarray}
This leads to
\begin{eqnarray}
&& g^2 G_{\mu\nu}^{ab,cd}(x,y) {\langle  P_{ab}(x) P_{cd}(y) \exp
\left[ i \oint_C \omega \right]
\rangle_{CP^{N-1}_2} \over
\langle  \exp \left[ i \oint_C \omega \right]
\rangle_{CP^{N-1}_2} }  
\nonumber\\
&=& \delta_{\mu\nu} D(x,y)
g^2  \sum_{A=1}^{N^2-1} (T^A)_{ab} (T^A)_{cd}
\left( \delta_{ad} \delta_{bc} - {1 \over N}
 \delta_{ab} \delta_{cd} \right)  Q(x,y)    
\nonumber\\
&=& \delta_{\mu\nu} D(x,y) g^2
 \sum_{A=1}^{N^2-1} \left[ {\rm tr}(T^A T^A)-{1 \over N}
{\rm tr}(T^A){\rm tr}(T^A) \right]
  Q(x,y)  
\nonumber\\
&=& \delta_{\mu\nu} D(x,y)  g^2{N^2-1 \over 2}
  Q(x,y)   .
\label{2ndterm4}
\end{eqnarray}
Note that $Q(x,x)$ is  $O(N^{-2})$, and hence
  $f[C]$ is  $O(\lambda/N)$.
This is because we consider the large $N$ expansion of the $CP^{N-1}$
model (see Appendix D) with the Lagrangian (see(\ref{CPNLag})) 
\begin{eqnarray}
  {\cal L}_{CP^{N-1}}    =  {N \over g_0^2}
|(\partial_\mu + i V_\mu(x)) \phi^{\alpha}(x)|^2 ,
  \quad g_0^2 := {2g_{YM}^2N \over \pi} ,
\label{CPNL}
\end{eqnarray}
where $V_\mu(x)$ is the composite gauge field,
\begin{eqnarray}
 V_\mu(x) = {i \over 2}
(\bar \phi^{\alpha}(x) \partial_\mu \phi^{\alpha}(x)
-  \partial_\mu \bar \phi^{\alpha}(x) \phi^{\alpha}(x)) ,
\label{CPNLag1}
\end{eqnarray}
under the constraint 
\begin{eqnarray}
  \phi^\dagger(x) \phi(x) := \bar \phi^a(x) \phi^a(x)  =  1 .
 \label{constraint1}
\end{eqnarray}
It is not difficult to show that the above estimation
gives the correct order for the higher-order terms, e.g., (b) and (e) 
in Fig.~\ref{WilsonLoopDiagram}, by making use of the
relations (\ref{trT3}) and (\ref{trT4}).
For example,
$
g^3 \langle
{\cal V}_{\mu_1}^A(x_1) {\cal V}_{\mu_2}^B(x_2)
{\cal V}_{\mu_n}^C(x_n)
\rangle_{pYM_4} 
$
is proportional to $ig^4 f^{ABC}$,
and
$
 ig^4f^{ABC}{\rm tr}[T^A T^B T^C]/N^3 = - g^4 f^{ABC}f^{ABC}/(4N^3)
= O(g^4) ,
$
since
$
 f^{ABC}f^{ABD}= C_2(Adj) \delta^{CD}
$ 
and
$ 
C_2(Adj)=N.
$
\par
Another way to understand this result is based on the idea of the
{\it reduction of degrees of freedom} that are responsible to the Wilson
loop.  The flag space has dimension
${\rm dim}F_{N-1}=N(N-1)$, whereas $CP^{N-1}$ has dimension 
${\rm dim}CP^{N-1}=2(N-1)$.  Therefore, the number of relevant degrees of
freedom is reduced for the fundamental quark for large $N$, since
${\rm dim}CP^{N-1} \cong 2{\rm dim}F_{N-1}/N$ for large $N$.
Indeed, this result is expected from the NAST given by
(\ref{NAST2}),  
\begin{eqnarray}
 W^C[{\cal A}]  
 =    \int [d\mu(\xi)]_C  \exp \left( i g \oint_{C} a \right) .
\end{eqnarray}
The Abelian gauge field $a={\rm tr}({\cal H}{\cal A})$ has only two
physical degrees of freedom, while the non-Abelian gauge field ${\cal
A}={\cal A}^A T^A$ in the Wilson loop (\ref{Wldef2}) has
$2(N^2-1)$ components. 
Thus the large $N$ expansion is reduced to a perturbative expansion
in the coupling constant $g$.  In this sense, the large $N$ expansion
combined with the NAST justifies the identification of the
deformation part with the perturbative part.
\par
Then we find that the $f[C]$ is of order $O(\lambda/N)$. 
Therefore, 
to leading order in the large $N$ expansion, 
the static potential and the string tension are given by
\begin{eqnarray}
  V(R) &=& - \lim_{T \rightarrow \infty} {1 \over T} \ln \Biggr\langle  \exp \left[ i \oint_C \omega \right]
\Biggr\rangle_{CP^{N-1}_2} 
+ \lim_{T \rightarrow \infty} {f[C] \over T} 
+ O({\lambda^2\over N^2}) ,
\\
 \sigma &=& -\lim_{R,T \rightarrow \infty}{1\over RT}\left( \ln \Biggr\langle  \exp \left[ i \oint_C \omega \right]
\Biggr\rangle_{CP^{N-1}_2}
+
f[C]\right) 
 + O({\lambda^2\over N^2})  .
\end{eqnarray}
\par
Now we proceed to estimate the second term, $f[C]$.  
We will show that the second term gives at most the perimeter law, so
that the area law (if it exists) is provided by the first term. 
Because of the factor $\delta_{\mu\nu}$ in 
$\langle {\cal V}_\mu^{A}(x) {\cal V}_\nu^{B}(y) \rangle_{pYM_4}$,
only integration between parallel sides $dx$ and $dy$ gives a
contribution to $f[C]$.  
Thus $f[C]$ is reduced to
\begin{eqnarray}
 f[C] = {g^2 \over 4} 
\oint_C dx^\mu \oint_C dy^\mu D(x,y) G_C(x,y) ,
\label{2ndterm3}
\end{eqnarray}
where we have defined the correlation function for the composite
operators as
\begin{eqnarray}
G_C(x,y) &:=& 2 (T^A)_{ab} (T^A)_{cd} 
{\langle  P_{ab}(x) P_{cd}(y) \exp \left[ i \oint_C \omega
\right]\rangle_{CP^{N-1}_2} \over
\langle  \exp \left[ i \oint_C \omega \right]
\rangle_{CP^{N-1}_2} } 
\\
&\cong&  2 (T^A)_{ab} (T^A)_{cd} 
  \left( \delta_{ad} \delta_{bc} - {1 \over N}
 \delta_{ab} \delta_{cd} \right)  Q(x,y)   
\\
&=& (N^2-1) Q(x,y) .
\label{GC}
\end{eqnarray}
First, if we restrict our consideration to topologically trivial
configurations, i.e.,
\begin{eqnarray}
 n^A(x)n^A(y) \equiv P_{ab}(x) P_{cd}(y) (T^A)_{ab} (T^A)_{cd} 
\cong n^A(\infty)n^A(\infty) \equiv 
{1\over 2}(1-N^{-1}) ,\nonumber \\
\end{eqnarray}
then we obtain
$G_C(x,y) \cong 1+O(N^{-1})$ and
\begin{eqnarray}
 f[C] \cong  {\lambda \over 2N} (1+O(N^{-1})) h[C], \quad
h[C] := {1 \over 2}\oint_C dx^\mu \oint_C dy^\mu D(x,y) .
\end{eqnarray}
 By taking into account all the contributions from parallel sides
$dx$ and $dy$ (see Fig.\ref{loopR}), $h[C]$ is calculated as, 
for $T > R \gg 1$,
\begin{eqnarray}
 h[C] =   - {1 \over 4\pi}{T \over R} 
+ {1 \over 2\pi^2}{T+R \over \epsilon}
+ {1 \over 2\pi^2} \ln {R \over \epsilon}  ,
\label{frec}
\end{eqnarray}
where $\epsilon$ is the ultraviolet cutoff included to avoid the coincidence
of $x$ and $y$ (see Appendix of
Ref.\citen{KondoIV}). 
In  $h[C]$, the first term corresponds to the Coulomb
potential in four dimensions,
\begin{eqnarray}
 V_C(R) =    
- {g^2 \over 4\pi}{1 \over R} + {\rm const.}
 +  O(\lambda^2/N^2) ,
 \label{potential}
\end{eqnarray}
and the second term in $h[C]$ corresponds to the
self-energy of quark and anti-quark.
Furthermore, if we take into account
the  $O(g^4)$ correction, the coupling constant begins to
run and the bare coupling $g$ in (\ref{potential}) is replaced by the
running coupling constant $g=g(\mu)$ (see, e.g., Kogut
\cite{Kogut83}).\footnote{
The contribution up to
$O(\lambda^2)$ in the leading order diagrams (planar diagram) in the
large $N$ expansion leads to a running coupling that differsfrom that
in the usual perturbative calculation in the coupling constant
$g$.
}
In the topologically trivial case, therefore, the second term $f[C]$
cannot give a non-vanishing string tension.

\begin{figure}
\begin{center}
 \leavevmode
 \epsfxsize=60mm
 \epsfysize=60mm
 \epsfbox{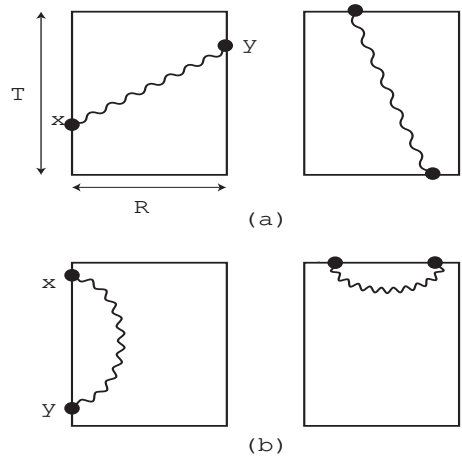}
\end{center} 
 \caption[]{A rectangular Wilson loop and the contribution to the Wilson
integral in which $x$ and $y$ run over (a) opposite sides, and (b)
same sides.}
 \label{loopR}
\end{figure}

\par
\begin{figure}
\begin{center}
 \leavevmode
 \epsfxsize=40mm
 \epsfysize=45mm
 \epsfbox{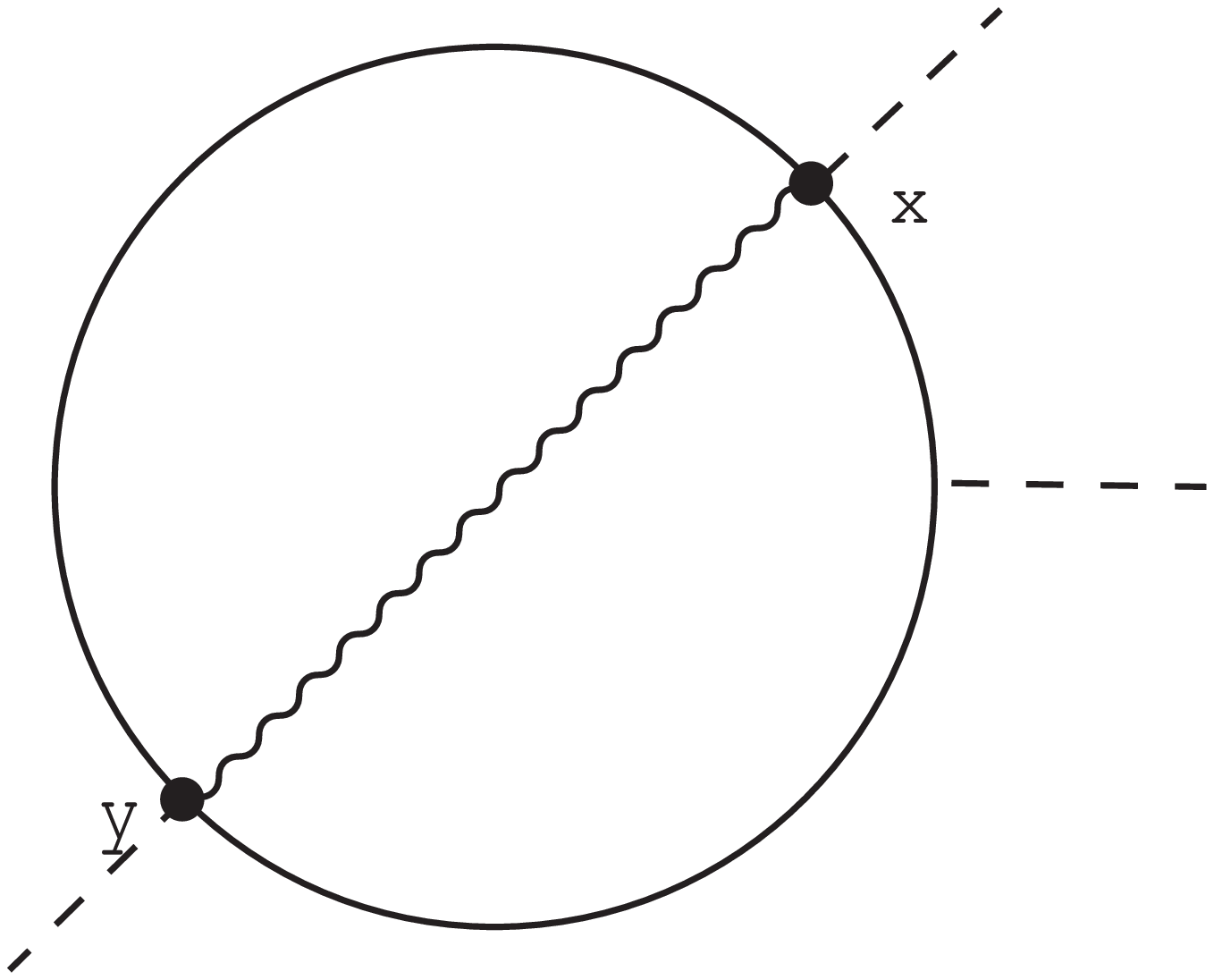}
\end{center} 
 \caption[]{A circular Wilson loop.}
 \label{loopC}
\end{figure}

\par
Next, we consider the topologically non-trivial case.
We begin to estimate (\ref{2ndterm3}) for a circular Wilson loop
$C$ with diameter $R$ (see Fig.\ref{loopC}). If
we avoid the coinciding case,
$x=y$,
$f[C]$ has a contribution only when $x$ and $y$ are at opposite ends of
a diameter, i.e., $|x-y|=R$. Therefore, $D(x,y)$ and $G(x,y)$ are
functions of
$R$, due to translational invariance. 
For any $x,y\in C$, $|x-y|=R$ and
\begin{eqnarray}
\oint_C dx^\mu \oint_C dy^\mu D(R)
 = {\pi^2 R^2 \over 4\pi^2 R^2} 
= {1 \over 4} .
\end{eqnarray}
Hence we obtain
\begin{eqnarray}
 f[C] =   {\lambda \over 8N} G_C(x,y)  . 
\end{eqnarray}
It is clear that $f[C]$ is not 
sufficient to give a non-vanishing string tension, since $G_C(x,y)$
exhibits exponential decay for large $R:=|x-y|$. 
Note that $P_{ab}(x)$ and $\oint_C \omega = \int_S d\omega$ are
U(1) gauge invariant quantities, so that $G_C(x,y)$ is also U(1)
gauge invariant. In the large
$N$ expansion, we can give a more precise estimate of the  second term (see Appendix E). 
\par
Finally, we consider the topologically nontrivial
case of a rectangular Wilson loop $C$ with side lengths
$R$ and $T$ (see Fig.\ref{loopR}).  In this case, we cannot give
a precise estimate of the second term, since we cannot perform the
integration exactly. 
To leading order in the $1/N$ expansion, it turns out that
\begin{eqnarray}
G_C(x,y) \cong \tilde G(x,y) := 2 (T^A)_{ab} (T^A)_{cd} 
\langle  P_{ab}(x) P_{cd}(y)  \rangle_{CP^{N-1}_2} ,
\label{GC2}
\end{eqnarray}
and that $\tilde G(x,y)$ decays exponentially for sufficiently large
$|x-y|$ (see Appendix E). 
Note that $x$ and $y$ are located on the opposite sides of the
rectangular Wilson loop.  Therefore, 
there exists an uniform upper bound, 
\begin{eqnarray}
   |\tilde G(x,y)| \le \tilde G(R) .
\end{eqnarray}
Hence, there exists an upper bound on $f[C]$  for a
sufficiently large Wilson loop such that $T > R \gg 1$:
\begin{eqnarray}
  |f[C]| \le {\lambda \over 2N}\tilde G(R) h[C] .
\end{eqnarray}
Here $h[C]$ is calculated in the same way as in (\ref{frec})  by
taking into account all the contributions from parallel sides
$dx$ and $dy$ (see Fig.\ref{loopR}).
Therefore the second term $f[C]$ cannot give non-vanishing string
tension in the topologically nontrivial case.
\par
Thus, within this reformulation, the area law of the non-Abelian
Wilson loop and the linear static potential in four-dimensional
$SU(N)$ Yang-Mills theory is realized if and only if the diagonal
Wilson loop
$
\Big\langle \exp \left( i \oint_C \omega \right)
\Big\rangle_{CP^{N-1}_2}
$
in the two-dimensional NLS model obeys the area law 
\begin{eqnarray}
\Big\langle \exp \left( i \oint_C \omega \right)
\Big\rangle_{CP^{N-1}_2}
\equiv \Big\langle \exp \left( i \int_{S} \Omega_K \right)
\Big\rangle_{CP^{N-1}_2}
\sim \exp (- \sigma_0 TR ) .
\label{2darealaw}
\end{eqnarray}
In other words, the area law or the linear potential between the
fundamental quark and anti-quark is obtained from the topological
$TQFT_4$ piece alone:
\begin{eqnarray}
  \Biggr\langle  \exp \left( i \oint_C \omega \right)
\Biggr\rangle_{TQFT_4}
= \Biggr\langle  \exp \left( i \oint_C \omega \right)
\Biggr\rangle_{CP^{N-1}_2}
\sim \exp (- \sigma_0 TR ) .
\end{eqnarray}
In any case, the derivation of the area law is reduced to a
two-dimensional problem.
\par
It should be remarked that only the total static potential,
\begin{eqnarray}
 V(R) =   \sigma_0 R
- {g^2(\mu) \over 4\pi}{1 \over R} + {\rm const.} ,
\end{eqnarray}
is gauge invariant.  Thus the linear potential piece alone is not gauge
invariant.  However, in the large $R$ limit, $R \rightarrow \infty$,
the linear potential is dominant in $V(R)$ so that the linear
potential piece becomes essentially gauge invariant.

\subsection{Area law to the leading order in the large $N$ expansion}

\par
By the rescaling of the field $\phi$ in the Lagrangian (\ref{CPNL}),
another form of the Lagrangian of the $CP^{N-1}$ model is obtained as
\begin{eqnarray}
  {\cal L}_{CP^{N-1}}    =  
\partial_\mu \bar \phi^{\alpha}(x) \partial_\mu \phi^{\alpha}(x)
+ {g_0^2 \over 4N} 
 (\bar \phi^{\alpha}(x) \partial_\mu \phi^{\alpha}(x)
- \partial_\mu \bar \phi^{\alpha}(x) \phi^{\alpha}(x))^2 ,
\label{CPNLag2}
\end{eqnarray}
with the constraint 
\begin{eqnarray}
  \phi^\dagger(x) \phi(x)  
:= \bar \phi^a(x) \phi^a(x)
 =  {N \over g_0^2} ,  \quad g_0^2 := {g_{YM}^2N \over \pi} .
 \label{constraint3}
\end{eqnarray}
It is useful to consider  the Schwinger
parameterization, \cite{Schwinger67} 
\begin{eqnarray}
 \phi = \pmatrix{\phi_1 \cr \vdots \cr \phi_N}
 = e^{i\varphi} \left( 
 1 + {g_0^2 \over 4N} {\bf u}^\dagger {\bf u} \right)^{-1}
  \pmatrix{ {\bf u} 
 \cr \sqrt{N \over g_0^2} ( 1-{g_0^2 \over 4N} {\bf u}^\dagger {\bf
u} ) }   ,
\end{eqnarray}
where
\begin{eqnarray}
{\bf u} = \pmatrix{u_1 \cr \vdots \cr u_{N-1}}, \quad
 {\bf u}^\dagger {\bf u} := \bar u_\alpha u_\alpha 
 \quad (\alpha = 1, \cdots, N-1) .
\end{eqnarray}
Note that there is no constraint for the variable ${\bf u}$, since
the Schwinger parameterization automatically satisfies the constraint
(\ref{constraint3}).
We can rewrite various quantities in terms of ${\bf u}$
without the constraint, e.g.,
\begin{eqnarray}
  \phi^\dagger d \phi - d \phi^\dagger \phi
 =  {{\bf u}^\dagger d {\bf u} - d {\bf u}^\dagger {\bf u}
  \over (1+{g^2 \over 4N} {\bf u}^\dagger {\bf u})^2} 
  + 2i {N \over g_0^2} d\varphi .
\end{eqnarray}
Then the $U(1)$ gauge field $V=V_\mu dx^\mu$ in the $CP^{N-1}$ model
is written as
\begin{eqnarray}
 V :=  {i \over 2} {g_0^2 \over N}
  (\phi^\dagger d \phi - d \phi^\dagger \phi)
 =   {i \over 2} {g_0^2 \over N}
 {{\bf u}^\dagger d {\bf u} - d {\bf u}^\dagger {\bf u}
  \over (1+{g_0^2 \over 4N} {\bf u}^\dagger {\bf u})^2} - d\varphi .
\end{eqnarray}
\par
We identify the complex coordinate $w$ in the K\"ahler manifold 
with the  Schwinger variable as
\begin{eqnarray}
 w_\alpha 
 = \sqrt{g_0^2 \over N} {u_\alpha \over 1-{g_0^2 \over 4N} {\bf
u}^\dagger {\bf u} } . \quad (\alpha = 1, \cdots, N-1)  
\end{eqnarray}
This leads to
\begin{eqnarray}
 \bar w_\alpha d w_\alpha - d \bar w_\alpha w_\alpha
 = {1 \over (1-{g_0^2 \over 4N} {\bf u}^\dagger {\bf u})^2}
 {g_0^2 \over N} (\bar u_\alpha d u_\alpha - d \bar u_\alpha
u_\alpha) ,
\end{eqnarray}
and
\begin{eqnarray}
 1 + \bar w_\alpha  w_\alpha  
 = {(1+{g_0^2 \over 4N} {\bf u}^\dagger {\bf u})^2 
 \over (1-{g_0^2 \over 4N} {\bf u}^\dagger {\bf u})^2} .
\end{eqnarray}
Then we find the following expression for $\omega$ in terms of $u$:
\begin{eqnarray}
\omega := {i \over 2}{
 \bar w_\alpha d w_\alpha - d \bar w_\alpha w_\alpha
 \over 1 + \bar w_\alpha  w_\alpha}
 =  {i \over 2} {g_0^2 \over N}
 {{\bf u}^\dagger d {\bf u} - d {\bf u}^\dagger {\bf u}
  \over (1+{g_0^2 \over 4N} {\bf u}^\dagger {\bf u})^2}  .
\end{eqnarray}
Thus  the connection one-form is given by
\begin{eqnarray}
V =  {i \over 2}{
 \bar w_\alpha d w_\alpha - d \bar w_\alpha w_\alpha
 \over 1 + \bar w_\alpha  w_\alpha} - d\varphi
 = \omega - d\varphi  ,
\end{eqnarray}
and the Abelian curvature two-form is equal to the K\"ahler
two-form:
\begin{eqnarray}
 dV =  d\omega = \Omega_K = i g_{\alpha\bar \beta} 
 dw_\alpha \wedge d\bar w_\beta, 
 \\
g_{\alpha\bar \beta} 
=  {\Delta \delta_{\alpha\beta} - \bar w_\alpha w_\beta
 \over \Delta},
 \quad \Delta := 1 + |||w|||^2 .
\end{eqnarray}

By way of the variable $u$, we have found that the connection
one-form
$\omega$  appearing in the NAST is equal to the gauge-invariant
part of
$V$. Therefore the diagonal Wilson loop for $CP^{N-1}$ model is equal
to 
\begin{equation}
 \Big\langle \exp \left( i \oint_{C} \omega \right)
\Big\rangle_{CP^{N-1}} 
= \Big\langle \exp \left( i \oint_{C} V \right)
\Big\rangle_{CP^{N-1}} .
\end{equation}
For the $CP^1$ model, this reduces to $V$ (6.51) in Ref.\citen{KondoII} for $G=SU(2)$.

\par
The expectation value
$
\langle \exp \left( i \oint_{C} V \right)\rangle_{CP^{N-1}} 
$
is calculated in Appendix D in the large $N$ expansion in a manner based on 
pioneering works.
\cite{DVL78,DVL79,Witten79}  The result
agrees with the result of Campostrini and Rossi.\cite{CR92} 
 To leading order in  the $1/N$ expansion, the Wilson loop
obeys the area law for all non-self-intersecting loops:
\begin{equation}
 \Big\langle \exp \left( i \oint_{C} V \right)
\Big\rangle_{CP^{N-1}}
= \exp \left[ - {6\pi \over N} m_\phi^2 |{\rm Area}(C)| \right] ,
\end{equation}
where
\begin{equation}
 m_\phi^2 = \mu^2 \exp \left[ -{2\pi^2 \over g^2(\mu)} \right] .
\end{equation}
Thus the string tension is obtained as
\begin{equation}
 \sigma_0 = {6\pi \over N} m_\phi^2 .
\end{equation}
Here $m_\phi$ is the mass of the $CP^{N-1}$ field $\phi$.
The $m_\phi^2$ is
equal to the vacuum expectation value 
$\langle \sigma(x) \rangle$
of the Lagrange multiplier
field $\sigma$ from the correspondence
$\sigma(x) \bar \phi(x) \phi(x) \rightarrow m_\phi^2  \bar \phi(x)
\phi(x)$. In the propagator of the vector field $V$, a massless pole
appears.  Hence, the auxiliary vector field $V$ becomes a dynamical
gauge field, giving rise to a linear confining potential between
$\phi$ and
$\bar \phi$.  On the other hand, the Lagrange multiplier field
$\sigma$ for the constraint (\ref{constraint3}) becomes massive, so
that it does not contribute to the confining potential between
$\phi$ and $\bar \phi$.  
Thus we have completed a proof of quark confinement in four-dimensional
$SU(N)$ Yang-Mills theory based on the Wilson criterion to
leading order in the large $N$ expansion {\it within}
our reformulation of the Yang-Mills theory.

\section{Remarks}
\setcounter{equation}{0}

Some remarks are in order to avoid confusion.

\subsection{Calculating gauge invariant quantity in the gauge non-invariant theory}

The Wilson loop operator $W_C[{\cal A}]$ is invariant under the gauge transformation denoted by $U(x)$, which is defined by the decomposition of the variable in (\ref{deco}) or (\ref{cov}),
since it is a gauge-invariant quantity. 
Therefore, the expression of the Wilson loop operator given by the non-Abelian Stokes theorem derived in this paper is also invariant under arbitrary (i.e., infinitesimal and finite) gauge transformation.  
Therefore, the variable $U(x)$ does not contribute to the Wilson loop operator, and we can write
\begin{equation}
 W_C[{\cal A}] = W_C[{\cal V}] ,
\end{equation}
which is a mathematical identity.  This is indeed the situation before taking the expectation value 
$\langle W_C[{\cal A}] \rangle_{YM}$ in terms of the Yang-Mills theory.  
However, this seems to contradict the claim of this paper; the area law can be derived from the contribution of the topologically non-trivial degrees of freedom expressed by the variable $U(x)$, which is described by the TQFT.
In fact, we have used the formula (\ref{expec}) to calculate the expectation value of the Wilson loop operator in $\S$ 8.
Therefore, if we set
\begin{equation}
 g({\cal V},U) = g({\cal V}) = W_C[{\cal V}], \quad h(U) = 1,
\label{gt0}
\end{equation}
in the expectation value $\langle \cdot \rangle_{YM}$, we would obtain 
the inconsistent result 
\begin{equation}
  \langle W_C[{\cal A}] \rangle_{YM} = \langle W_C[{\cal V}] \rangle_{pYM}
\langle 1 \rangle_{TQFT}
= \langle W_C[{\cal V}] \rangle_{pYM} ? 
\end{equation}
since this implies that there is no contribution from the topological part expressed by $U$ and that the Wilson loop can be calculated by the perturbative part only.%
\footnote{The author would like to thank Giovanni Prosperi and a referee for pointing out this issue.
}
\par
This apparent contradiction can be understood as follows. 
Elimination of the $U$-dependence is possible only when the gauge theory itself is formulated in a gauge invariant way, just as the lattice gauge theory can be written in a manifestly gauge invariant manner without gauge fixing.  In such a case, we can perform the gauge transformation so that a gauge invariant quantity, e.g., the Wilson loop, has no dependence on $U(x)$ also in the expectation value.
However, our strategy is totally different from this case. That is to say, we calculate the gauge-invariant quantity using the gauge fixed (i.e, gauge non-invariant) theory, where the off-diagonal components and the diagonal components are subject to different gauge fixing conditions.   
Therefore, the total action $S_{YM}^{tot}[{\cal A},\cdots]=S_{YM}[{\cal A}]+S_{GF+FP}[{\cal A},\cdots]$ obtained by adding the GF+FP term $S_{GF+FP}[{\cal A},\cdots]$ to the Yang-Mills action $S_{YM}[{\cal A}]$ does not have the same form as the gauge-transformed total action 
$S_{YM}^{tot}[{\cal V},B,C,\bar C]$ obtained by the gauge rotation $U$, although 
$S_{YM}[{\cal A}]=S_{YM}[{\cal V}]$. 
In this sense,
\begin{equation}
  \langle W_C[{\cal A}] \rangle_{YM} 
\not= \langle W_C[{\cal V}] \rangle_{YM},
\end{equation}
where
\begin{equation}
 \langle W_C[{\cal A}] \rangle_{YM} := Z_{YM}^{-1} \int {\cal D}{\cal A}_\mu  \cdots  \exp \{ i S_{YM}^{tot}[{\cal A},\cdots] \} W_C[{\cal A}] .
\end{equation}
Thus it should be remarked that {\it the formula (\ref{expec}) holds only when the field variable ${\cal A}$ is separated into ${\cal V}$ and $U$ and at the same time the respective component is identified with the topologically trivial (perturbative) and non-trivial (non-perturbative) components, respectively}, where the ${\cal V}$-dependent part is calculated using perturbative Yang-Mills (pYM) theory. This is an assumption of our approach. 
In this sense, Eq.~(\ref{expec}) holds under this assumption and it is not a mathematical identity.  In our approach, {\it an arbitrary gauge transformation is not allowed under the functional of the expectation value}, i.e., 
$\langle \cdot \rangle_{YM}$, once the respective component is identified with the relevant degrees of freedom in Yang-Mills theory.  
(Thus, if we wish to change the integration variable as in (\ref{gt0}), the total Yang-Mills theory should be modifed simultaneously.  This  leads to a more complicated theory.)  In this paper we have fixed the gauge degrees of freedom for the variable $U(x)$ using the  modified MA gauge.  
  In other words, we have chosen a specific field configuration of $\{ U(x)\}$ to calculate the physical quantity in such a way that the $U$-dependence is controlled by the MA gauge fixing term.  This is an essential point of our strategy.  In this sense, the physics enters in when we consider the expectation value.

\subsection{The Gribov problem and dimensional reduction}
 
Dimensional reduction of the Parisi-Sourlas type was applied to the random field Ising model \cite{RFIM} and scalar field theories with random external sources \cite{Cardy83,KP83,KLP84}  based on non-perturbative \cite{Cardy83,KP83} and rigorous methods.\cite{RFIM,KLP84}  In these models it has been recognized \cite{PS82,Zinn-Justin93} that the Parisi-Sourlas correspondence (between random systems in $d$ dimensions and the corresponding pure systems in $d-2$ dimensions, or supersymmetric theory in $d$ dimensions and the corresponding bosonic theory in $d-2$ dimensions) is exact only in the case of unique solutions to the classical equation of motion.  
In fact, it was rigorously shown \cite{RFIM} that the three-dimensional Ising model in a random magnetic field exhibits long-range order at zero temperature and small disorder.  This implies that the lower critical dimension $d_\ell$ for this model is 2 (ruling out $d_\ell=3$), where the lower-critical dimension is the dimension above which ($d>d_\ell$) long-range ferromagnetic order can exist.  Therefore, this result contradicts the naive prediction \cite{IM75} between the random field Ising model in $d$ dimensions and the pure Ising model in $d-2$ dimensions.
\par
In gauge theories, a similar problem can in principle arise due to the existence of Gribov copies, although much is not known on this issue.  
Indeed, it is known \cite{BHTW00} that Gribov copies exist for the naive MA gauge. On the other hand, it is not known whether the modified MA gauge with $OSp(D|2)$ symmetry possesses the Gribov copies or not.  
If Gribov copies exist, first of all, the naive BRST formulation breaks down.  Even if a modified BRST formulation can exist, the BRST symmetry can be spontaneously broken,\cite{Fujikawa83} at least in the topological sector (described by the modified MA gauge action), where the large (or finite) gauge transformation with non-trivial winding number plays a crucial role.  In this case, the vacuum in the topological sector is not annihilated by the BRST charge $Q_B$ (i.e., $Q_B|0\rangle_{TQFT}\not=0$), whereas the perturbative sector (in which there exist small quantum fluctuation around an arbitrary but fixed topological background) is characterized by the unbroken BRST charge, i.e., $Q_B|0\rangle_{pYM} =0$.   
Even in this case,  dimensional reduction should take place, since we have  used neither the field equations nor the above property of the BRST charge acting on the state.   
In other words, the dimensional reduction is an origin of the spontaneous breaking of the BRST symmetry (see Refs.\citen{KondoVI,KS00a}).
Moreover, the BRST symmetry can also be broken by other mechanisms, e.g., radiative corrections.
\par
In any case, the spontaneous breaking of the BRST symmetry in the topological sector implies that the expectation value of the gauge invariant operator depends on the gauge fixing parameter $\alpha$ for the topological sector.  In this case, we can not conclude the $\alpha$-independence of the physical (gauge-invariant) quantities. Rather, the value of $\alpha$ should be determined as a physical parameter by the theory \cite{Kondo00} or experiments.  This is reasonable if we believe that the real world is described by a unique quantum theory and that nature is realized at an appropriate gauge fixing, since the classical theory does not really exist in nature, and hence ambiguities arising in the quantization are absent from the beginning.
It is a challenge to give a definite answer to these questions.
The author would like to thank the referee for pointing out the Gribov problem as an obstacle to the dimensional reduction.

\section{Conclusion and discussion}
\setcounter{equation}{0}

We have given a new version of the non-Abelian Stokes theorem for
$G=SU(N)$ with $N\ge 2$ which reduces to the previous result
\cite{KondoIV} for
$SU(2)$.  
This version of the non-Abelian Wilson loop is very helpful to see the
role played by the magnetic monopole in the calculation of the
expectation value of the non-Abelian Wilson loop.
Combining this non-Abelian Stokes theorem
with the Abelian-projected effective gauge theory for
$SU(N)$, we have explained the Abelian dominance for the Wilson loop
in $SU(N)$ Yang-Mills gauge theory.
For $SU(N)$ with $N\ge 3$, we must distinguish the
maximal stability group $\tilde H$ and the residual gauge group $H$, 
which is taken to be the maximal torus group $H=U(1)^{N-1}$.
\par
In order to demonstrate the magnetic monopole dominance and the area law of the
Wilson loop, we have used a novel reformulation of the
Yang-Mills theory which has been proposed by one of the authors.\cite{KondoII}   This reformulation is based on the identification
of the Yang-Mills theory with the perturbative deformation of a
topological quantum field theory.
This framework deals with the gauge action $S_{YM}$ and the
gauge-fixing action $S_{GF}$ on an equal footing.  The gauge action is
characterized from the viewpoint of the geometry of connections.  On the
other hand, the gauge-fixing part is related to the topological
invariant (Euler characteristic) determined by a global topology. 
Therefore, the gauge-fixing part can have a geometric meaning from
a global viewpoint.  This point has not been emphasized in textbooks on quantum field theory. (See Ref.\citen{KondoVI} for more details.)
\par
 Our approach relies heavily on a specific gauge, the MA gauge.
In spite of the absence of an elementary scalar field in Yang-Mills
theory, the MA gauge allows the existence of the magnetic monopole.  The magnetic monopole is considered as a topological
soliton composed of gauge degrees of freedom, providing the
composite scalar field.
 At least in this gauge,  (Parisi-Sourlas)
dimensional reduction occurs due to the supersymmetry hidden
in the gauge fixing part in the MA gauge.    By making use of the
non-Abelian Stokes theorem within this reformulation of the
Yang-Mills theory, the derivation of the area law of the non-Abelian
Wilson loop in four-dimensional Yang-Mills theory has been reduced
to the two-dimensional problem of calculating the expectation value
of the Abelian Wilson loop in the coset
$G/H$ non-linear sigma model.
This is the main result of this article.
\par
In particular, in order to demonstrate confinement of the fundamental quark in
the four-dimensional $SU(N)$ Yang-Mills theory in the MA gauge, we
have only to consider the two-dimensional $CP^{N-1}$ model.
From the topological point of
view, the Abelian Wilson loop is equivalent to the area integral
(enclosed by the Wilson loop) of the instanton density in  the two-dimensional NLS model. This implies that the calculation of the magnetic
monopole contribution to the Wilson loop in four-dimensional
Yang-Mills theory was translated into that of the instanton contribution
in the two-dimensional NLS model  when the Wilson loop is contained in
the two-dimensional plane, i.e. when the Wilson loop is planar. 
In addition, the
two-dimensional instanton is considered as a subclass of the
four-dimensional Yang-Mills instanton (see
Ref.\citen{KondoII}). This suggests that the area law of the Wilson loop can be  derived by taking into account the contributions of a
restricted class of Yang-Mills instantons.
A Monte Carlo simulation on a lattice
will be efficient in confirming this dimensionally reduced
picture of quark confinement.\cite{Kondo99Lattice99} 
\par
Moreover,  naive instanton calculus (the dilute gas approximation) in
the coset NLS model leads to the area law of the Wilson loop of the
original four-dimensional Yang-Mills theory.  Improvements of the
dilute gas approximation are necessary to confirm the area law based
on the above picture.  However, this is rather difficult, as found more than
twenty years ago.\cite{BL79} 
An advantage of the extension of the above strategy to $SU(N)$ with
arbitrary $N$ is that the large $N$ systematic expansion can be used
in the calculation of the Wilson loop, and that the area law of the Wilson loop has been demonstrated.
These results confirm the area law of the Wilson
loop.   
It is known that the large $N$ Yang-Mills theory is related to the
string theory. The correspondence between instanton calculus and the
large $N$ expansion and a related issue in the large
$N$ expansion in Yang-Mills theory will also be
discussed in a subsequent article.\cite{Kondo00}  
 These considerations should shed more light on the
confining string picture.\cite{Polyakov96}

\appendix
\section{Normalization of the Coherent State 
}
\setcounter{equation}{0}
\par
In this appendix we derive the normalization facter $N$ of coherent state 
 $\left|\xi ,\Lambda \right>$ which parametrizes $G/\tilde{H}$,
\begin{eqnarray}\label{N}
&N& :=  \left<\Lambda \right|\exp {[\bar{\tau} _{\alpha}E_{\alpha}]}\exp {[\tau _{\beta}E_{-\beta}]}\left|\Lambda \right>\\
  &=&\sum_{K,L=0}^{\infty}\frac{1}{K!}\frac{1}{L!}
  \bar{\tau}_{\alpha^1}\ldots \bar{\tau}_{\alpha^K}\tau_{\beta^1}\ldots
  \tau_{\beta^L}\left<\Lambda \right|E_{\alpha^{1}}\ldots E_{\alpha^{K}}
E_{-\beta^{1}}\ldots E_{-\beta^{L}}\left|\Lambda \right>.
\end{eqnarray}

\subsection{$SU(3)$ coherent state}

We use the positive root $\alpha^{(i)}$ 
and Dynkin index 
$[m,n]$ in $SU(3)$, which are defined by Fig.\ref{root} and 
(\ref{Dynkin index}).
This definition may be rewritten as 
\begin{eqnarray}
\alpha^{(1)}\cdot \vec\Lambda =\frac{m}{2}
, \quad \ \alpha^{(3)}\cdot \vec\Lambda =\frac{n}{2}.
\end{eqnarray}
For $[m,n]=[m,0]$ (resp., $[0,n]$), $\alpha $ and $\beta$ run over $1,2$ (resp., $2,3$).
Then the corresponding coherent state parametrizes $CP^2$.
From the orthogonality of the states which span the representation space,
the terms contributing to $N$ in (\ref{N}) must satisfy the condition 
\begin{eqnarray}
\alpha^{1}+\ldots +\alpha^{K}-\beta^{1}-\ldots -\beta^{L}=0.
\end{eqnarray}
In the $[m,0]$ case, this condition implies that the number of 
$\alpha^{(1)} $ and $\alpha^{(2)}$ in $K$ positive roots is equal to that of 
$\alpha^{(1)} $ and $\alpha^{(2)}$ in $L$ negative roots. Since $[E_1 ,E_2]=0$,
 we have only to estimate the terms
\begin{eqnarray}\label{k,l}
N_{k,l}:= \left<\Lambda \right|\underbrace{E_{1}\ldots E_{1}}_{k}
\underbrace{E_{2}\ldots E_{2}}_{l}\underbrace{E_{-2}\ldots E_{-2}}_{l}
\underbrace{E_{-1}\ldots E_{-1}}_{k}\left|\Lambda \right>.
\end{eqnarray}
 We begin with the term 
\begin{eqnarray}
N_{0,l}&=&\left<\Lambda \right|\underbrace{E_{2}\ldots E_{2}}_{l}
\underbrace{E_{-2}\ldots E_{-2}}_{l}
\left|\Lambda \right>=N_{0,l}\left<\Lambda -l\alpha^{(2)} |\Lambda -l
\alpha^{(2)} \right>,
\end{eqnarray}
where we have used in the last equality the fact that $E_{-2}\ldots E_{-2}
\left|\Lambda \right>$ has a weight $\Lambda -l\alpha^{(2)}$ and is 
proportional to the state $\left|\Lambda -l\alpha^{(2)} \right>$,
which is normalized as
\begin{equation}
\left <\Lambda -l\alpha^{(2)} 
\Big|\Lambda -l\alpha^{(2)} \right>=1 .
\end{equation} 
 Exchanging the rightmost $E_{2}$ with $E_{-2}$ and using 
\begin{eqnarray}
[E_2 ,E_{-2}] &=&
\alpha^{(2)}\cdot H ,
\\
 \alpha^{(2)}\cdot H \left|\Lambda -j\alpha^{(2)} \right>
&=&(\frac{m}{2}-j)
\left|\Lambda -j\alpha^{(2)} \right>, \quad (0<j<m) 
\end{eqnarray}
we obtain the recursion relation for $N_{0,l}$,
\begin{eqnarray}
N_{0,l} &=& \left\{\frac{m}{2}-(l-1) \right\} N_{0,l-1}
+\left<\Lambda \right|\underbrace{E_{2}\ldots E_{2}}_{l-1}E_{-2}E_{2}
\underbrace{E_{-2}\ldots E_{-2}}_{l-1}\left|\Lambda \right> \nonumber \\
 &=& \left(\frac{m}{2}-(l-1)+\frac{m}{2}-(l-2)+\cdots +\frac{m}{2} \right)
 N_{0,l-1}\nonumber \\
 &=&\frac{l(m-l+1)}{2}N_{0,l-1},
\end{eqnarray}
where we have used $E_{2}\left|\Lambda\right>=0
$. From this relation and from $N_{0,0}= \left<\Lambda|\Lambda\right>=1$,
we obtain
\begin{eqnarray}\label{0,l}
N_{0,l}=\frac{(l!)^{2} {m \choose l}}{2^{l}}.
\end{eqnarray}
Similarly, 
\begin{eqnarray}\label{j,0}
N_{k,0}=\frac{(k!)^{2} {m \choose k}}{2^{k}}.
\end{eqnarray}
 For the general terms (\ref{k,l}), we have
\begin{eqnarray}
N_{k,l}&=&N_{k,0}\left<\Lambda -k\alpha^{(1)}\right|\underbrace{E_{2}\ldots
 E_{2}}_{l}\underbrace{E_{-2}\ldots E_{-2}}_{l}\left|\Lambda -k\alpha^{(1)}
  \right> \nonumber \\ 
 &=&N_{k,0}\frac{k(m-k-l+1)}{2}\left<\Lambda -k\alpha^{(1)}\right|
 \underbrace{E_{2}\ldots E_{2}}_{l-1}\underbrace{E_{-2}\ldots E_{-2}}_{l-1}\left|\Lambda -k\alpha^{(1)} \right> \nonumber \\
      &=&\frac{(k!)^{2} {m \choose k}}{2^{k}}\frac{(l!)^{2}
{m-k \choose l}}{2^{l}}.
\end{eqnarray}
Finally, the normalization factor $N$ is given by
\begin{eqnarray}
N&=&\sum_{k=0}^{\infty}\sum_{l=0}^{\infty}\frac{1}{(k+l)!^2}{k+l \choose l}
  \underbrace{\bar{\tau}_{2}\ldots \bar{\tau}_{2}}_{l}\underbrace{\bar{\tau}_
  {1}\ldots \bar{\tau}_{1}}_{k}\underbrace{
  \tau_{2}\ldots\tau_{2}}_{l}\underbrace{\tau_{1}\ldots
  \tau_{1}}_{k}N_{k,l} \nonumber \\
  &=&\sum_{k=0}^{n}\sum_{l=0}^{n-k}(\frac{\bar{\tau}_2\tau _2}{2})
^{l}(\frac{\bar{\tau}_1\tau _1}{2})^{k}{m \choose k}{m-k \choose l} \nonumber \\
  &=&\left( 1+(\frac{\bar{\tau}_1\tau _1}{2})+(\frac{\bar{\tau}_2 \tau _2}{2})
\right) ^m \nonumber \\
  &=&\left( 1+(\bar{w}_1 w _1)+(\bar{w}_2 w_2)\right) ^m=\exp{K_{CP^2}(\bar{w},w)},
\end{eqnarray}
where $w_1 := \tau_{1}/\sqrt{2} \ w_2 := \tau_2/\sqrt{2}$, and
 $K_{CP^2}(w)$ is the K\"ahler potential of $CP^2$.

For $F_2$, i.e., $mn\ne 0$, $\alpha$ and $\beta $ in (\ref{N}) run over $1,2,3$.  Here we apply almost the same steps as in the case of $CP^2$. The result is  
\begin{eqnarray}
N &:=&  \left<\Lambda \right|\exp {\sum_{\alpha =1}^{3}[\bar{\tau} _{\alpha}
E_{\alpha}]}\exp {\sum_{\beta =1}^{3}[\tau _{\beta}E_{-\beta}]}\left|\Lambda \right>\nonumber \\
 &=&\left( 1+|\frac{\tau _1}{\sqrt2}|^2+|\frac{\tau_2}{\sqrt{2}}
+\frac{\tau _1\tau_3}{4}|^2
\right) ^m\left( 1+|\frac{\tau _3}{\sqrt2}|^2+|\frac{\tau_2}{\sqrt{2}}
-\frac{\tau _1\tau_3}{4}|^2\right) ^n \nonumber \\
 &=&\left( 1+|w _1|^2+|w_2|^2)
\right) ^m\left( 1+|w_3|^2+|w_2
-w_1 w_3|^2\right) ^n \nonumber \\
&=&\exp{K_{F_2}(\bar{w},w)},
\end{eqnarray}
where $w_1:= \tau_1/\sqrt{2} ,\ w_2:= \tau_2/\sqrt{2}+
\tau _{1}\tau_{3}/4 ,\ w_3:=\tau_3/\sqrt{2}$.

\subsection{$SU(N)$ coherent state}
In $SU(N)$, the simple root  $\alpha^i$ and the Dynkin index 
$[m_1,\cdots m_{N-1}]$ are defined by (\ref{simple root}) and 
$\alpha^i\cdot\Lambda:=m_i/2$. 
In general, the roots $E_{\pm\alpha}$ belonging to the coset group 
$G/\tilde {H}$ are not orthogonal to the highest weight $\Lambda.$\footnote{
From the definition of the coherent state (\ref{cohrentdef}), we  have
$E_{\pm\alpha}\left|\Lambda\right>\ne 0$ and 
$[E_{\alpha},E_{-\alpha}]\left|\Lambda\right>=\alpha\cdot\Lambda
\left|\Lambda\right>\ne 0 $.} 
We restrict our consideration to only the case of $[m,0,\cdots ,0]$, 
in which case only
$\alpha^1$ is not orthogonal to $\Lambda$. 
Any positive root $\tilde{\alpha}$ is given by a linear combination of simple roots $\alpha^i$.
The positive roots $\tilde{\alpha}_i$ in (\ref{N}) are not orthogonal to $\Lambda$, so they must include $\alpha^1$.
 It turns out that such roots are given by $\tilde{\alpha}_i:=\alpha^1+\cdots+\alpha^i \ 
(i=1,\cdots N-1 )$ , which are realized by 
$\left(E_{\tilde{\alpha}_i}\right)_{kl}=\delta_{1,k}
\delta_{i+1,l}/\sqrt{2} $ in the ${\bf N}$ representation.
The corresponding coherent state parametrizes $CP^{N-1}$. 
We can easily extend the $CP^2$ case to that of $CP^{N-1}$ using the fact that 
$[E_{\tilde{\alpha}_i} ,E_{\tilde{\alpha}_j}]=0$ and $\tilde{\alpha}_i\cdot\vec
\Lambda
=m/2$. (This situation is same as in the $SU(3)$ case, i.e., $[E_1,E_2]=0$, and $\alpha^{(1)}\cdot\vec\Lambda=\alpha^{(2)}\cdot\vec\Lambda=m/2$.)
Thus we obtain
\begin{eqnarray}
N &:=&  \left<\Lambda \right|\exp {\sum_{i=1}^{N-1}[\bar{\tau} _{i}
E_{\tilde{\alpha}_i}]}
\exp {\sum_{j=1}^{N-1}[\tau _{j}E_{-\tilde{\alpha}_j}]}
\left|\Lambda \right>\nonumber \\
&=&\left( 1+\sum_{i=1}^{N-1}(\frac{\bar{\tau}_i\tau _i}{2})
 \right) ^m =:\left( 1+\sum_{i=1}^{N-1}(\bar{w}_{i}w_{i})\right) ^m \nonumber\\
 &=&\exp{K_{CP^{N-1}}(\bar{w},w)}.
\end{eqnarray}


\section{From $CP^1$ to $CP^2$}
\setcounter{equation}{0}

For an arbitrary element $(\phi_1,\phi_2,\phi_3)$ of $W$,
\begin{equation}
  W = {\bf C}^3 -(0,0,0) ,
\end{equation}
the entire set of ratios $\phi_1:\phi_2:\phi_3$ is called the
complex projective plane and is denoted by $P^2({\bf C})$ or 
${\bf C}P^2$:
\begin{equation}
  (\alpha \phi_1:\alpha \phi_2:\alpha \phi_3) =
(\phi_1:\phi_2:\phi_3) , \quad \alpha \in {\bf C} , \quad
\alpha \not= 0 . 
\end{equation}
Defining the subset $U_a (a=1,2,3)$ of ${\bf C}P^2$ as
\begin{equation}
  U_a = \{ (\phi_1:\phi_2:\phi_3) \in {\bf C}P^2; \phi_a \not= 0 \}
\subset  {\bf C}P^2 ,
\end{equation}
we observe that
\begin{equation}
  (\phi_1:\phi_2:\phi_3) 
= \left(1:{\phi_2 \over \phi_1}:{\phi_3 \over \phi_1}\right) \in U_1 .
\end{equation}
The mapping $\varphi_1$ from $U_1$ to
${\bf C}^2$,
\begin{equation}
  \varphi_1:  (\phi_1:\phi_2:\phi_3) \in U_1 \rightarrow  
 \left({\phi_2 \over \phi_1}:{\phi_3 \over \phi_1}\right) \in {\bf C}^2,
\end{equation}
is a bijection, i.e., a surjection (onto-mapping) and
injection (one-to-one mapping).  The inverse mapping is given by
\begin{equation}
  \varphi_1^{-1}:  (x,y) \in {\bf C}^2 \rightarrow (1:x:y) \in U_1 .
\end{equation}
Similarly, the following maps $\varphi_2$ and $\varphi_3$ are also 
bijections from $U_2$ and $U_3$ to ${\bf C}^2$:
\begin{equation}
  \varphi_2:  (\phi_1:\phi_2:\phi_3) \in U_2 \rightarrow  
 \left({\phi_1 \over \phi_2}:{\phi_3 \over \phi_2}\right) \in {\bf C}^2,
\end{equation}
\begin{equation}
  \varphi_3:  (\phi_1:\phi_2:\phi_3) \in U_3 \rightarrow  
 \left({\phi_1 \over \phi_3}:{\phi_2 \over \phi_3}\right) \in {\bf C}^2 .
\end{equation}
Since 
\begin{equation}
 {\bf C}P^2 = U_1 \cup \{ (0:\phi_2:\phi_3)  \},
\quad (\phi_2:\phi_3) \not= 0 ,
\end{equation}
$(\phi_2:\phi_3)$ determines a point in ${\bf C}P^2$.  Conversely,
for a point $(b_1:b_2)$ in ${\bf C}P^1$, $(0:b_1:b_2)$ defines a
point in
${\bf C}P^2 - U_1$.  Thus the map
\begin{equation}
  (0:\phi_2:\phi_3) \in {\bf C}P^2-U_1 \rightarrow (\phi_2:\phi_3)
\in {\bf C}P^1
\end{equation}
is   one-to-one and onto.  Then we can identify ${\bf
C}P^2-U_1$ with
${\bf C}P^1$ as ${\bf C}P^2 - U_1 \cong {\bf C}P^1$.
\par
On the other hand, the identification of $U_1$ and ${\bf C}^2$, $U_1 \cong {\bf C}^2$, by the
mapping $\varphi_1$ leads to 
\begin{equation}
  {\bf C}P^2 = U_1 \cup {\bf C}P^1 \cong {\bf C}^2 \cup {\bf C}P^1
.
\end{equation}
Using 
${\bf C}P^1={\bf C}^1 \cup \{ (0:1) \}$,
we can write
$
 {\bf C}P^2 = {\bf C}^2 \cup {\bf C}^1 \cup \{ (0:1) \} \equiv {\bf
C}^2
\cup {\bf C}^1 \cup {\bf C}^0 .
$
${\bf C}P^2-U_1$ is called the ``line at infinity'' and is denoted by
$\ell_\infty$,
${\bf C}P^2-U_1 = \ell_\infty$.
Thus we can also write 
\begin{equation}
 {\bf C}P^2 = U_1 \cup \ell_\infty, \quad U_1 \cong C^2, 
\quad \ell_\infty \cong {\bf C}P^1 ,
\end{equation}
where ${\bf C}^2$ is called the ``complex affine plane''.   The
homogeneous coordinates 
$(\phi_1:\phi_2:\phi_3)$
are related to an element $(x,y)$ in the affine plane as
\begin{equation}
 x={\phi_2 \over \phi_1}, \quad y={\phi_3 \over \phi_1}, \quad
\ell_\infty \cong \{ \phi_1 = 0 \} .
\end{equation}
\par
${\bf C}P^2$ is obtained as 
${\bf C}P^2={\cal U}\cup {\cal V}\cup {\cal W}$ by gluing the three
affine planes 
${\cal U}, {\cal V}$ and ${\cal W}$ using the relations
\begin{equation}
  u_1 = {1 \over v_1},  \quad u_2 = {v_2 \over v_1}, \quad
 v_1 = {w_1 \over w_2},  \quad v_2 = {1 \over w_2}, \quad
 w_1 = {1 \over u_2}, \quad w_2 = {u_1 \over u_2} ,
\end{equation}
where $(u_1,u_2), (v_1,v_2)$ and $(w_1,w_2)$ are the coordinates of ${\cal U},
{\cal V}$ and ${\cal W}$ defined in terms of homogeneous coordinates 
of ${\bf C}P^2$ as
\begin{equation}
  (u_1,u_2) = \left({\phi_2 \over \phi_1}, {\phi_3 \over \phi_1}\right), \quad
  (v_1,v_2) = \left({\phi_1 \over \phi_2}, {\phi_3 \over \phi_2}\right), \quad
  (w_1,w_2) = \left({\phi_1 \over \phi_3}, {\phi_2 \over \phi_3}\right) .
\end{equation}
Note that the method of gluing affine planes is not unique.  In fact,
${\bf C}P^2$ can be obtained from the following gluing:
\begin{equation}
  u_1 = {w_2 \over w_1}, \quad u_2 = {1 \over w_1}, \quad
 v_1 = {1 \over u_1}, \quad v_2 = {u_2 \over u_1}, \quad
 w_1 = {v_1 \over v_2}, \quad w_2 = {1 \over v_2} .
\end{equation}

\section{Nonlinear Sigma Model of the Flag Space}
\setcounter{equation}{0}

\subsection{Correspondence between $SU(N)/T$ and  $SL(N,C)/B$}
\par
When an element $\xi$ in $F_{N-1}$ is expressed by the complex
coordinate, 
\begin{eqnarray}
  \xi = 
 \pmatrix{
 1 & w_1 & w_2     & \cdots & \cdots &w_n \cr
 0 &  1  & w_{n+1} & \cdots & \cdots & w_{2n-1} \cr
 0 & 0   & 1 & w_{2n}  & \cdots & w_{3n-3} \cr
   &   &  &  &   &   \cr
 \vdots & \vdots &\vdots &\vdots & \vdots & \vdots \cr
   &   &  &  &   &   \cr
 0 & 0   & \cdots & 0 &  1 & w_{n(n+1)/2} \cr 
 0 & 0   & \cdots & \cdots & 0 & 1 \cr 
 }^T  \in F_n ,
\end{eqnarray}  
we find $\det \xi = 1$. Hence $\xi$ is an element of $SL(N,C)$.
There is the isomorphism 
$SU(N)/T \cong SL(N,C)/B$.
However, $\xi$ in this form is not necessarily unitary.  The
corresponding unitary matrix $V \in SU(N)$ is obtained as follows. 
First,  $\xi$ as an element of $SL(N,C)$ is expressed in terms of the
column vectors:
\begin{eqnarray}
  \xi = (E_1, E_2, \cdots, E_N) \in SL(N,C) = SU(N)^C .
\end{eqnarray}
By applying the Gramm-Schmidt orthogonalization, we can obtain a set
of mutually orthogonal vectors $(E_1', E_2', \cdots, E_N')$ from 
$(E_1, E_2, \cdots, E_N)$ as
\begin{eqnarray}
  E_1' &:=& E_1,
  \nonumber\\
  E_2' &:=& E_2 - {(E_2,E_1') \over (E_1', E_1')} E_1' ,
  \nonumber\\
  \vdots
  \nonumber\\
  E_N' &:=& E_N - {(E_N,E_{N-1}') \over (E_{N-1}',E_{N-1}')} E_{N-1}'
  - \cdots - {(E_N,E_1') \over (E_1',E_1')} E_1' ,
\end{eqnarray} 
where the inner product is defined by
$
 (E_i,E_j) := E_i^T \cdot \bar E_j .
$
Using the normalized vectors $ e_j:=E_j'/||E_j'||$, we obtain
an element in $SU(N)$,
\begin{eqnarray}
  V = (e_1, e_2, \cdots, e_N) \in SU(N).
\end{eqnarray}
In fact, $\det V=1$ and $V$ is unitary, since
\begin{eqnarray}
V^\dagger V = \pmatrix{ \bar e_1^T \cr \vdots \cr \bar e_N^T} 
(e_1, e_2, \cdots, e_N),
\quad
  (V^\dagger V)_{ij} = \bar e_i^T \cdot e_j 
  = (e_i, e_j) = \delta_{ij} .
\end{eqnarray} 
In general, the unitary matrix $V$ is related to its complexfication
$V^C$ by 
\begin{eqnarray}
  V = V^C B, \quad V \in SU(N), \quad V^C \in SL(N,C), 
\end{eqnarray} 
where $B$ is an upper triangular matrix. 
This is nothing but the Iwasawa decomposition.%
\footnote{
Any element $g_c \in G^C$ may be factorized as
\begin{eqnarray}
 g_c = g b, \quad g \in G, \quad b \in B 
\end{eqnarray} 
in a unique fashion, up to torus elements that are common to $G$
and $B$.
}
Since the upper triangular
matrices form a group, we have
\begin{eqnarray}
 V^C = V B^{-1} = V B' ,
\end{eqnarray} 
where $B'=B^{-1}$ is also upper triangular.  This implies
$\xi=V B$ by the above construction.
Therefore, $V$ is indeed the element of $SU(N)$ corresponding to
$\xi$.
Note that the multiplication by the matrix $B$ leaves the
highest-weight state $|\Lambda \rangle$ invariant, so that
\begin{eqnarray}
 \xi |\Lambda \rangle = VB |\Lambda \rangle= V|\Lambda \rangle .
\end{eqnarray} 
For example, $e_1=E_1'/||E_1'||=E_1/||E_1||$, i.e.,
\begin{eqnarray}
   e_1 = {1 \over \Delta^{1/2}} 
   \pmatrix{1 \cr -w_1 \cr \vdots \cr -w_N} , 
   \quad \Delta := 1+ ||w||^2 = 1 + \sum_{a=1}^{N} |w_a|^2 .
   \label{e1}
\end{eqnarray}
\par
The Mauer-Cartan form is 
\begin{eqnarray}
   V^{-1} dV = V^\dagger dV
  = \pmatrix{\bar e_1^T \cr \vdots \cr \bar e_N^T} 
  (de_1, \cdots, de_N)
  = \pmatrix{
\bar e_1^T de_1 & \bar e_1^T de_2 & \cdots & \bar e_1^T de_N \cr
\bar e_2^T de_1 & \bar e_2^T de_2 & \cdots & \bar e_2^T de_N \cr
\cdots  & \cdots & \cdots & \cdots \cr
\bar e_N^T de_1 & \bar e_N^T de_2 & \cdots & \bar e_N^T de_N } , 
\nonumber\\
\end{eqnarray}
which can be decomposed into diagonal and off-diagonal parts as
\begin{eqnarray}
   V^{-1} dV = \sum_{i=1}^{N} \bar e_i^T de_i I_{ii} 
   + \sum_{a\not=b} \bar e_a^T de_b E_{ab} .
\end{eqnarray}
In the fundamental representation, we have
\begin{eqnarray}
   \omega = \langle \Lambda | i V^{-1} dV | \Lambda \rangle
   = i (V^{-1} dV)_{11} = i \bar e_1^T de_1 ,
   \label{omega1}
\end{eqnarray}
\begin{eqnarray}
   n^A = \langle \Lambda |  V^{-1} T^A V | \Lambda \rangle
  = \bar e_1^T (T^A) e_1  .
\end{eqnarray}
Hence, substituting (\ref{e1}) into (\ref{omega1}), we obtain
\begin{eqnarray}
   \omega =  {i \over 2} {\bar w_a dw_a - d\bar w_a w_a \over
\Delta} .
\end{eqnarray}
The Lagrangian density of the coset $G/H$, i.e. the flag NLS model, 
\begin{eqnarray}
  {\cal L}_{NLSM} = {\beta_g \over 2}{\rm tr}_{G/H}
  (i V^{-1} \partial_\mu V iV^{-1} \partial_\mu V ) ,
\end{eqnarray}
can be  written in terms of the off-diagonal elements as
\begin{eqnarray}
  {\cal L}_{NLSM}   =  \beta_g\sum_{a,b:a<b}  
(\Omega_\mu)_{ab}(\Omega_\mu)_{ab} 
=  \beta_g \sum_{a,b:a<b}   
(e_a, \partial_\mu e_b) (e_a,\partial_\mu e_b) ,
 \label{CPN}
\end{eqnarray}
where we have used
$
 (e_i,e_j) := \bar e_i^T \cdot e_j = \delta_{ij} .
$
\par
Especially, the Lagrangian of the $CP^{N-1}$ model is obtained as a
special case of (\ref{CPN}) as follows.
Using the definition (\ref{CPrep2}), 
\begin{eqnarray}
 n^A = (UT^A U^\dagger)_{11} = U_{1a}(T^A)_{ab}\bar U_{1b}
 = U_{1a}(T^A)_{ab}U^\dagger_{b1} ,
\end{eqnarray}
and $UU^\dagger =1$,
we find 
\begin{eqnarray}
  \partial_\mu n^A \partial_\mu n^A
  &=& \partial_\mu U_{1a} \partial_\mu U^\dagger_{a1}
  U_{1b} U^\dagger_{b1}
  + U_{1a} \partial U^\dagger_{a1} U_{1b} \partial_\mu U^\dagger_{b1}
  \nonumber\\
  &=& \sum_{b=2}^{N} (iU\partial_\mu U^\dagger)_{1b}
  (iU\partial_\mu U^\dagger)_{1b} 
 = \sum_{b=2}^{N}   (e_1, \partial_\mu e_b)^2.
\end{eqnarray}
If we use (\ref{CPrep})($\phi_a = \bar U_{1a} = U^\dagger_{a1}$), we obtain another expression,  
\begin{eqnarray}
  \partial_\mu n^A \partial_\mu n^A
   =   \partial_\mu \bar \phi \cdot \partial_\mu \phi
   + (\bar \phi \cdot \partial_\mu \phi)(\bar \phi \cdot
\partial_\mu \phi) ,
\end{eqnarray}
where we have used $\phi^\dagger  \cdot \phi =1$.  
Thus the Lagrangian of $CP^{N-1}$ model is obtained as
\begin{eqnarray}
  {\cal L}_{CP^{N-1}} 
    &=& {\beta_g \over 2}  
 \partial_\mu {\bf n} \cdot \partial_\mu {\bf n} 
 \quad (\mu=1, \cdots, d)
 \\
  &=& {\beta_g \over 2}  
  \sum_{b=2}^{N}   (e_1, \partial_\mu e_b)^2
  \\
  &=& {\beta_g \over 2}  
g_{\alpha\bar \beta}(\phi)
  {\partial \phi^{\alpha} \over \partial x_\mu}
  {\partial \bar \phi^{\beta} \over \partial x_\mu} ,
\label{CPNLag}
\end{eqnarray}
where
\begin{eqnarray}
 g_{\alpha\bar \beta}(\phi) :=  \delta_{\alpha\beta}- \bar
\phi_\alpha \phi_\beta  .
\end{eqnarray}
This agrees with the Lagrangian obtained from the K\"ahler potential,
\begin{eqnarray}
 {\cal L}_{CP^{N-1}} = {\beta_g \over 2}
   g_{\alpha\bar \beta}(w) \partial_\mu w^\alpha
\partial_\mu \bar w^\beta ,
\end{eqnarray}
with
\begin{equation}
 g_{\alpha\bar \beta}(w) 
 = {(1+|||w|||^2)\delta_{\alpha\beta} - \bar w_\alpha w_\beta \over
 (1+|||w|||^2)^2} .
\end{equation}

\par
The explicit construction of $V$ is given in the following section.

\subsection{SU(2)}
For $G=SU(2)$, we have
\begin{eqnarray}
 \xi = \pmatrix{1 & 0 \cr -w & 1} = (E_1, E_2) ,
\quad
 E_1 = \pmatrix{1 \cr -w } , \quad E_2 = \pmatrix{0 \cr 1} .
\end{eqnarray}
It is easy to see  that
\begin{eqnarray}
 (E_1,E_1) = 1+w\bar w ,  \quad (E_2,E_1) = -\bar w .
\end{eqnarray}
Hence we obtain
\begin{eqnarray}
e_1 = \Delta^{-1/2} \pmatrix{1 \cr -w }, \quad
e_2 = \Delta^{-1/2} \pmatrix{\bar w \cr 1},
\quad 
\Delta := 1+|w|^2 ,
\end{eqnarray}
and
\begin{eqnarray}
V = (e_1, e_2) = \Delta^{-1/2} \pmatrix{1 & \bar w \cr -w & 1} .
\end{eqnarray}
The elements of the one-form $V^{-1} dV$ are
\begin{eqnarray}
 e_1^T d \bar e_1 &=&  {1 \over 2}\Delta^{-1} (wd\bar w - \bar w dw) ,
\\
 e_2^T d \bar e_1 &=& - \Delta^{-1} d\bar w  . 
\end{eqnarray}
The Lagrangian of the $F_1=CP^1$ model reads
\begin{eqnarray}
  {\cal L}_{NLSM} 
  = - \beta_g  \Delta^{-2} \partial_\mu w \partial_\mu \bar w 
  = - \beta_g {1 \over \left(1+|w|^2\right)^2} \partial_\mu w \partial_\mu \bar w
.  
\end{eqnarray}

\subsection{SU(3)}
For $G=SU(3)$, we have
\begin{eqnarray}
 \xi = \pmatrix{1 & 0 & 0 \cr -w_1 & 1 & 0 \cr -w_2 & w_3 & 1} ,
\end{eqnarray}
or,
\begin{eqnarray}
 E_1 = \pmatrix{ 1 \cr -w_1 \cr -w_2 } ,
\quad E_2 = \pmatrix{ 0 \cr 1 \cr w_3 } ,
\quad E_3 = \pmatrix{ 0 \cr 0 \cr 1 } .
\end{eqnarray}
Then we have
\begin{eqnarray}
 (E_1,E_1) = 1+w_1\bar w_1+w_2\bar w_2, 
 \quad (E_2,E_1) = -\bar w_1 - w_3 \bar w_2 , \cdots .
\end{eqnarray}
A straightforward calculation leads to
$V = (e_1, e_2, e_3) \in SU(3)$, with
\begin{eqnarray}
e_1 &=& (\Delta_1)^{-1/2} \pmatrix{1 \cr -w_1 \cr -w_2 },
\nonumber\\
e_2 &=& (\Delta_1 \Delta_2)^{-1/2} \pmatrix{\bar w_1+w_3 \bar w_2 
\cr 1+|w_2|^2-w_1 \bar w_2 w_3 
\cr -w_2 \bar w_1 + w_3 +w_3 |w_1|^2},
\nonumber\\
e_3 &=& (\Delta_2)^{-1/2}
\pmatrix{\bar w_2 - \bar w_1 \bar w_3 \cr -w_3 \cr 1},
\end{eqnarray}
where
\begin{eqnarray}
\Delta_1 := 1+|w_1|^2+|w_2|^2 , \quad
\Delta_2 := 1+|w_2-w_1 w_3|^2 + |w_3|^2 .
\end{eqnarray}
The off-diagonal elements of the one-form $V^{-1} dV$ are
\begin{eqnarray}
 e_2^T d \bar e_1 &=& (\Delta_1)^{-1} (\Delta_2)^{-1/2} 
 [(1+|w_2|^2-w_1\bar w_2 w_3) d\bar w_1
 + (-w_2 \bar w_1 + w_3 + w_3|w_1|^2) d\bar w_2] ,
\nonumber\\
 e_3^T d \bar e_1 &=& (\Delta_1)^{-1/2} (\Delta_2)^{-1/2}
 [d\bar w_2 - \bar w_3 d\bar w_1]  , 
\nonumber\\
 e_3^T d \bar e_2 &=& (\Delta_1)^{-1/2} (\Delta_2)^{-1} 
 [(w_1+\bar w_3 w_2)(d\bar w_2 - \bar w_3 d\bar w_1) - \Delta_1
d\bar w_3] . 
\end{eqnarray}
\par
For $CP^2$, we have
\begin{eqnarray}
e_1 &=& (\Delta_1)^{-1/2} \pmatrix{1 \cr -w_1 \cr -w_2 },
\\
e_2 &=& (\Delta_1 \Delta_2)^{-1/2} \pmatrix{\bar w_1   
\cr 1+|w_2|^2 
\cr -w_2 \bar w_1  },
\\
e_3 &=& (\Delta_2)^{-1/2}
\pmatrix{\bar w_2  \cr 0 \cr 1},
\end{eqnarray}
where
\begin{eqnarray}
\Delta_1 := 1+|w_1|^2+|w_2|^2 = 1+ |||w|||^2 , \quad
\Delta_2 := 1+|w_2|^2  .
\end{eqnarray}
The off-diagonal elements of the one-form $V^{-1} dV$ are
\begin{eqnarray}
 e_2^T d \bar e_1 &=& (\Delta_1)^{-1} (\Delta_2)^{-1/2} 
 [(1+|w_2|^2) d\bar w_1
  -w_2 \bar w_1  d\bar w_2] ,
\nonumber\\
 e_3^T d \bar e_1 &=& (\Delta_1)^{-1/2} (\Delta_2)^{-1/2}
 [d\bar w_2]  , 
\nonumber\\
 e_3^T d \bar e_2 &=& (\Delta_1)^{-1/2} (\Delta_2)^{-1} 
 [ w_1 d\bar w_2 ] . 
\end{eqnarray}
The Lagrangian of $CP^2$ model is given by
\begin{eqnarray}
  {\cal L}_{CP^2}
    &=& {\beta_g \over 2}  
  \sum_{b=2}^{3}   |(e_1, \partial_\mu e_b)|^2
  \\
  &=& {\beta_g \over 2}  
  {1  \over (1+|||w|||^2)^2}
  [  (1+|w_2|^2) \partial_\mu w_1 \partial_\mu \bar w_1
  + (1+|w_1|^2) \partial_\mu w_2 \partial_\mu \bar w_2
  \nonumber\\&&  \quad \quad \quad \quad  \quad \quad \quad \quad
  - \bar w_1 w_2 \partial_\mu w_1 \partial_\mu \bar w_2
  - w_1 \bar w_2 \partial_\mu w_2 \partial_\mu \bar w_1
  ] .
\end{eqnarray}

\section{Large N Expansion of the $CP^{N-1}$ Model}
\setcounter{equation}{0}

The generating function of the $CP^{N-1}$ model is defined by
\begin{eqnarray}
  Z[J,\bar J,J_\mu] &:=& \int {\cal D}\phi {\cal D}\bar \phi 
\prod_{x} \delta \left(|\phi(x)|^2-{N \over g_0^2}\right) 
\nonumber\\ 
&\times& \exp  \left\{ - S + \int d^2x [\bar J \cdot \phi 
+ \bar \phi \cdot J + J_\mu V_\mu] \right\} ,
\end{eqnarray}
where $S$ is the action of the $CP^{N-1}$ model,
\begin{eqnarray}
  S &:=& \int d^2x \left[ 
\partial_\mu \bar \phi \cdot \partial_\mu \phi + {g_0^2 \over 4N}
(\bar \phi \cdot \stackrel{\leftrightarrow}{\partial_\mu} \phi) 
(\bar \phi \cdot \stackrel{\leftrightarrow}{\partial_\mu}
\phi) \right] 
\\
&=&  \int d^2x \left[ 
\partial_\mu \bar \phi \cdot \partial_\mu \phi 
-  {N \over g_0^2} V_\mu V_\mu \right],
\end{eqnarray}
and the auxiliary vector field $V_\mu$ is defined by
\begin{eqnarray}
  V_\mu(x) := {g_0^2 \over 2N} i  
(\bar \phi(x) \cdot \stackrel{\leftrightarrow}{\partial_\mu} \phi(x))
= {g_0^2 \over 2N} i  
(\bar \phi(x) \cdot \partial_\mu \phi(x) - 
\partial_\mu \bar \phi(x) \cdot \phi(x)) .
\end{eqnarray}
Introducing the Lagrange multiplier fields $\sigma(x)$ and
$A_\mu(x)$, we can rewrite as
\begin{eqnarray}
 && \prod_{x} \delta \left( |\phi(x)|^2-{N \over g_0^2} \right) 
 \exp \left\{ \int d^2x \left[  {N \over g_0^2} V_\mu(x)V_\mu(x) +
J_\mu(x) V_\mu(x) \right] \right\} ,
\nonumber\\
&=& \int {\cal D}\sigma \int {\cal D}A_\mu \exp \Biggr[ \int d^2x
\Biggr\{ {i \over \sqrt{N}} \sigma 
\left( |\phi|^2-{N \over g_0^2} \right)
- {1 \over N} A_\mu A_\mu |\phi|^2 
- m^2 |\phi|^2
\nonumber\\&&
+ {1 \over \sqrt{N}} A_\mu [i (\bar \phi \cdot
\stackrel{\leftrightarrow}{\partial_\mu} \phi) + J_\mu] 
- {g_0^2 \over 4N} J_\mu J_\mu \Biggr\} \Biggr] ,
\end{eqnarray}
where we have inserted the mass term $m^2 |\phi|^2$ for later
convenience and have chosen a specific normalization for the field
$\sigma$. Then we obtain
\begin{eqnarray}
  Z[J,\bar J,J_\mu] &:=& \int {\cal D}\phi {\cal D}\bar \phi 
\int {\cal D}\sigma \int {\cal D}A_\mu 
\exp \Biggr\{ - \int d^2x \left[ \bar \phi \cdot \Delta_B \phi 
+ {i \sqrt{N} \over g_0^2} \sigma \right] 
\nonumber\\
&+& \int d^2x  \left[\bar J \cdot \phi + \bar \phi \cdot J +
{1 \over \sqrt{N}} A_\mu J_\mu - {g_0^2 \over 4N} J_\mu J_\mu \right]
 \Biggr\}  ,
\end{eqnarray}
where
\begin{eqnarray}
 \Delta_B := - D_\mu D_\mu + m^2 - {i \over \sqrt{N}} \sigma(x) ,
\quad D_\mu := \partial_\mu + {i \over \sqrt{N}}A_\mu(x) .
\end{eqnarray}
This theory has global $SU(N)$ invariance corresponding  to
rotations of $\phi_a$.
Moreover, it has local $U(1)$ gauge invariance
under the transformation 
\begin{eqnarray}
 \phi_a'(x) &=& e^{i\Lambda(x)} \phi_a(x),  \ (a=1, \cdots, N)  
\nonumber\\
 A_\mu'(x) &=& A_\mu(x) - \sqrt{N} \partial_\mu \Lambda(x) ,  
\nonumber\\
 \sigma'(x) &=& \sigma(x) .
\end{eqnarray}

\par
We can perform the integration over $\phi$ and $\bar \phi$ to obtain
\begin{eqnarray}
  Z[J,\bar J,J_\mu] &:=&  
\int {\cal D}\sigma \int {\cal D}A_\mu 
\exp \Biggr\{ - S_{eff} 
\nonumber\\
&+& \int d^2x  \left[\bar J \Delta_B^{-1} J +
{1 \over \sqrt{N}} A_\mu J_\mu - {g_0^2 \over 4N} J_\mu J_\mu \right]
 \Biggr\}  ,
\end{eqnarray}
where
\begin{eqnarray}
 S_{eff} &:=&  N {\rm Tr} \ln \Delta_B  
+ {i \sqrt{N} \over g_0^2} \int d^2x \sigma(x) .
\end{eqnarray}
The effective action can be expanded in a power series of $1/N$:
\begin{eqnarray}
S_{eff} =  \sum_{n=1}^{\infty} N^{1-n/2} S^{(n)} 
= \sqrt{N} S^{(1)} + N^0 S^{(2)} + N^{-1/2} S^{(3)} + \cdots .
\end{eqnarray}
The diagramatic representation of this expansion is given in Fig.~\ref{CPnDiagram}
using the rule given in Fig.~\ref{CPnlargeNrule}.

\par
\begin{figure}[htbp]
\begin{center}
 \leavevmode
 \epsfxsize=100mm
 \epsfysize=100mm
\epsfbox{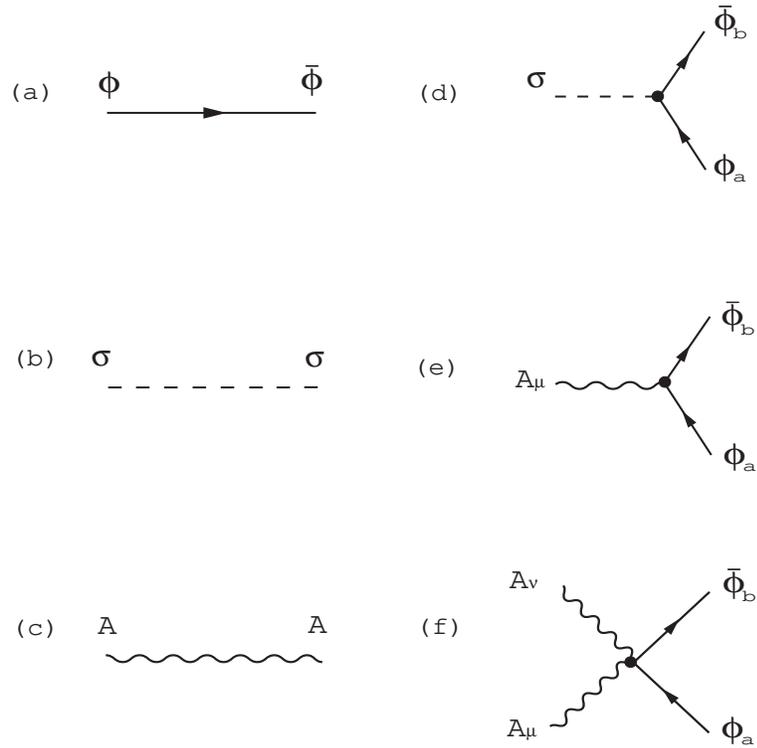}
\end{center} 
 \caption[]{Graphical representation of the large $N$ expansion  in the 
$CP^{N-1}$ model. 
 The propagators are as follows:
(a) $\phi$ propagtor, $\delta_{ab} (p^2+m^2)^{-1}$;
(b) $\sigma$ propagator, $\Gamma(p)^{-1}$;
(c) $A_\mu$ propagator, 
$\left( \delta_{\mu\nu} - {p_\mu p_\nu \over p^2} \right) 
 \left[ (p^2+4m^2)\tilde \Gamma(p)-{1 \over \pi} \right]^{-1}$.
The vertices are as follows:  
(d) $\sigma \phi_a \bar \phi_b$ vertex
 (${i \over \sqrt{N}} \delta_{ab}$); 
(e) $A_\mu \phi_a \bar \phi_b$ vertex
(${-1 \over \sqrt{N}} \delta_{ab}(p_\mu+p_\mu ')$); 
(f) $A_\mu A_\nu \phi_a \bar \phi_b$ vertex 
(${-1 \over N} \delta_{ab}\delta_{\mu\nu}$). }
 \label{CPnlargeNrule}
\end{figure}

\par
\begin{figure}
\begin{center}
 \leavevmode
 \epsfxsize=100mm
 \epsfbox{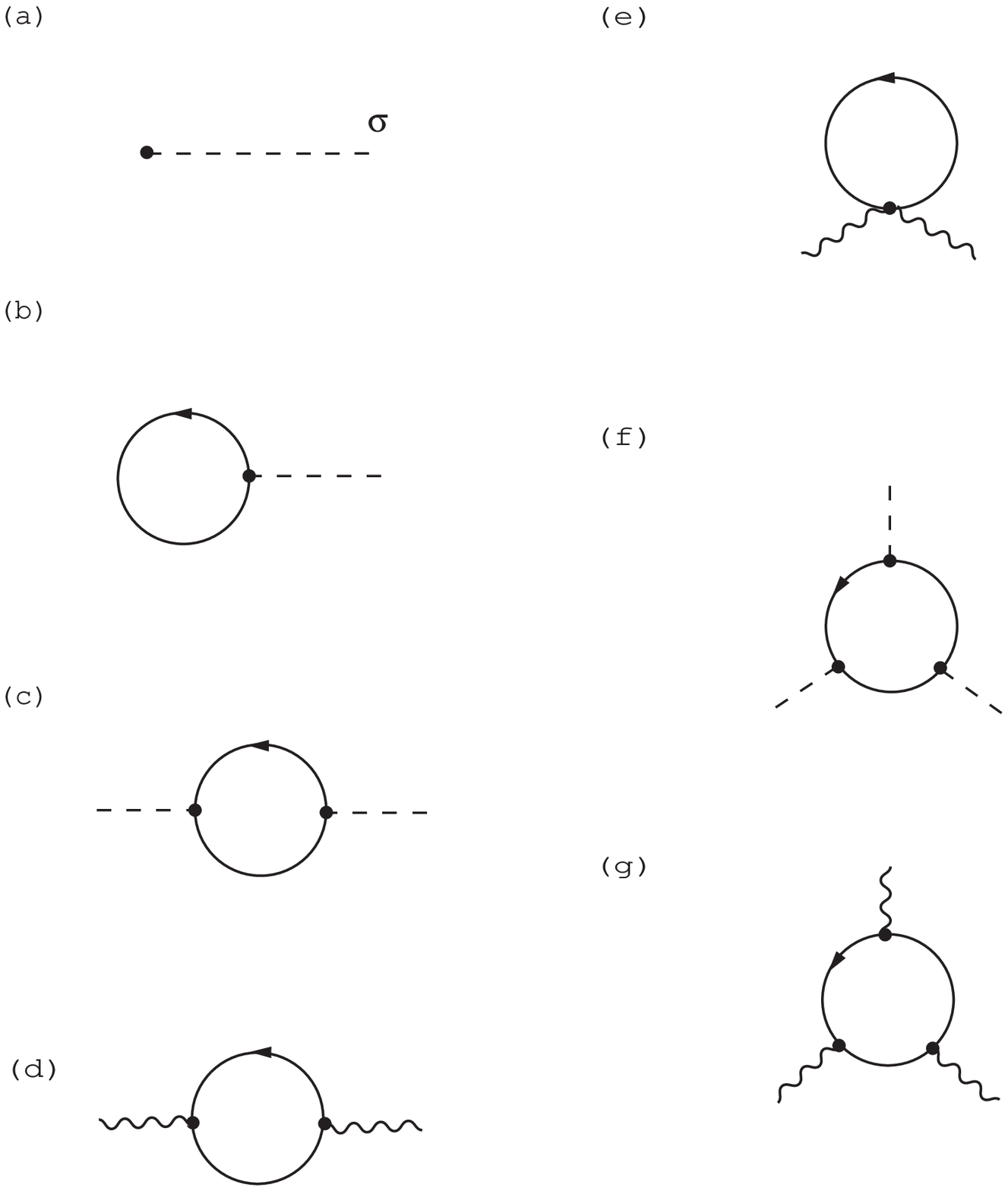}
\end{center} 
 \caption[]{Examples of Feynmann diagrams. 
(a) and (b) are tadpole diagrams of order $N^{1/2}$.
(c),(d) and (e) are vacuum polarization diagrams of order $N^0$.
(f) and (g) are order $N^{-1/2}$ diagrams.
}
 \label{CPnDiagram}
\end{figure}

\par
 First, the order $N^{1/2}$  term corresponds to the diagrams (a) and (b)
in Fig.~\ref{CPnDiagram}:
\begin{eqnarray}
  S^{(1)} = {i \over g_0^2} \int d^2x \sigma(x) 
- i {\rm Tr}[(-\partial^2 +m^2)^{-1} \sigma] 
\\
= i \tilde \sigma(0) \left[ {1 \over g_0^2} 
- \int {d^2q \over (2\pi)^2} {1 \over q^2+m^2}  \right]  ,
\label{integral}
\end{eqnarray}
where we have used the Fourier transformation 
\begin{eqnarray}
   \tilde \sigma(p)  =  \int d^2x e^{-i px} \sigma(x) . 
\end{eqnarray}
The integral in (\ref{integral}) is ultraviolet divergent.  It can be
regularized by introducing the cutoff $\Lambda$.  The saddle point
condition $S^{(1)}=0$ requires the bare coupling constant $g_0$ to vary
with the cutoff $\Lambda$ according to
\begin{eqnarray}
  {1 \over g_0^2(\Lambda)} = {1 \over 4\pi}\ln {\Lambda^2 \over m^2}
.
\end{eqnarray}
In other words, if the bare coupling $g_0$ varies with respect to the
cutoff according to
\begin{eqnarray}
  {1 \over g_0^2(\Lambda)} - {1 \over g_R^2(\mu)}
 = {1 \over 4\pi}\ln {\Lambda^2 \over \mu^2} ,
\end{eqnarray}
the divergences cancel in $S^{(1)}$.  This implies the asymptotic
freedom of the $CP^{N-1}$ model.
 Imposing the condition 
$S^{(1)}=0$, therefore, we obtain
\begin{eqnarray}
  m^2 = \mu^2 \exp \left[ - {4\pi \over g_R^2(\mu)} \right]  .
\end{eqnarray}
\par
Next, the order $N^{0}$ term corresponds to the diagrams (c),(d) and (e)
in Fig.~\ref{CPnDiagram}:
\begin{eqnarray}
  S^{(2)} = {1 \over 2} \int d^2x \int d^2y 
 [\sigma(x) \Gamma(x,y) \sigma(y) 
+ A_\mu(x) \Gamma_{\mu\nu}(x,y) A_\nu(y) ] ,
\end{eqnarray}
where the Fourier transformation of $\Gamma(x,y)$ and
$\Gamma_{\mu\nu}(x,y)$ are respectively given by (see
Fig.~\ref{CPnDiagram})
\begin{eqnarray}
  \tilde \Gamma(p) &=&  \int  {d^2q \over (2\pi)^2}
{1 \over (p^2+m^2)((p+q)^2+m^2)} 
\nonumber\\
&=& {1 \over 2\pi} {1 \over \sqrt{p^2(p^2+4m^2)}}
\ln {\sqrt{p^2+4m^2}+\sqrt{p^2} \over \sqrt{p^2+4m^2}-\sqrt{p^2}} ,
\end{eqnarray}
and
\begin{eqnarray}
  \tilde \Gamma_{\mu\nu}(p) &=&  2\delta_{\mu\nu}
\int  {d^2q \over (2\pi)^2}{1 \over q^2+m^2}
-  \int  {d^2q \over (2\pi)^2}{(p_\mu+2q_\mu)(p_\nu+2q_\nu)
 \over (p^2+m^2)((p+q)^2+m^2)} 
\nonumber \\
 &=& \left( \delta_{\mu\nu} - {p_\mu p_\nu \over p^2} \right) 
 \left[ (p^2+4m^2)\tilde \Gamma(p)-{1 \over \pi} \right] .
\end{eqnarray}
In the neighbourhood of $p^2=0$, we have
\begin{eqnarray}
 \tilde \Gamma(p) = {1 \over 24\pi m^4}(6m^2-p^2) +O(p^4) ,  \quad
  \tilde \Gamma_{\mu\nu}(p) 
  =  \left[ {p^2 \over 12\pi m^2}+O(p^4) \right]
  \left( \delta_{\mu\nu} - {p_\mu p_\nu \over p^2} \right) .
\nonumber\\
\end{eqnarray}
Thus we obtain the low-energy effective action,
\begin{eqnarray}
  S^{(2)} \cong   \int d^2x      {1 \over 24\pi m^4}
  \sigma(x) \left(  \partial^2 + 6m^2  \right) \sigma(x) 
+ \int d^2x {1 \over 48\pi m^2}
[\partial_\mu A_\nu(x) - \partial_\nu A_\mu(x)]^2   .
\nonumber\\
\end{eqnarray}
It is important to note that the kinetic terms for the Lagrangian
multiplier fields are generated. The field $\sigma$ becomes massive,
while the field
$A_\mu (=\sqrt{N}V_\mu)$ is massless.
\par
The line integral in the Wilson loop can be rewritten as
\begin{eqnarray}
   \oint_C d\xi_\mu V_\mu(\xi) = \int d^2z V_\mu(z) J_\mu(z) 
=:  (V_\mu, J_\mu) ,
\end{eqnarray}
where
\begin{eqnarray}
  J_\mu(z) = \oint_C d\xi_\mu \delta^2(z-\xi)
  = \epsilon_{\mu\nu} \partial_\nu \Phi(z),
  \quad \Phi(z) = \cases{1 &($z \in S_C$) \cr 0 &($z \notin S_C$)}  .
\end{eqnarray}
Here $S_C$ is the area bounded by the loop $C$.
Up to  leading order, we can perform the Gaussian integration to
obtain
\begin{eqnarray}
  \Big\langle \exp \left[ i \oint_C d\xi_\mu V_\mu(\xi) \right] \Big\rangle_{CP^{N-1}} 
  &\cong& Z^{-1}[0,0,0] \int {\cal D}\sigma {\cal D}A_\mu 
  e^{-S^{(2)}} \exp \left( {i\over \sqrt{N}} (A_\mu, J_\mu) \right) 
  \nonumber\\
  &=& {\rm const }\exp \left[ -{1 \over 2}{12\pi m^2 \over N}
 (J_\mu, \Delta^{-1} J_\mu) 
  \right] .
\end{eqnarray}
Thus we obtain the area law,
\begin{eqnarray}
  \Big\langle \exp \left[ i \oint_C d\xi_\mu V_\mu(\xi) \right] \Big\rangle_{CP^{N-1}} 
  = {\rm const }\exp \left[ -{6\pi m^2 \over N} |S_C|  \right] ,
\end{eqnarray}
since $J={}^*d \Phi$ and  
$\Delta:=d\delta + \delta d$, and hence
\begin{eqnarray}
  (J_\mu, \Delta^{-1} J_\mu) 
:= \int d^2x \int d^2 y J_\mu(x) \Delta^{-1} (x,y) J_\mu(y) =  (\Phi,
\Phi) = |S_C| .
\end{eqnarray}
The dynamically generated gauge field $A_\mu$ produces a long-range
force with a linear potential that confines the $\phi$s.
Both global $SU(N)$ and local $U(1)$ symmetries are unbroken in two dimensions due to the Coleman theorem.  
In dimensions $D>2$, it has been shown \cite{AA80} that there is a critical point $g_c$ such that for
$g<g_c$ $SU(N)$ and $U(1)$ symmetries are broken,
$\langle \phi_a \rangle \not=0$, while 
$\langle \sigma \rangle=0$.  In this phase $\phi$ is
regarded as the Nambu-Goldstone particle, since $m_\phi=0$. 
A massless vector pole exists in the propagator
$\langle A_\mu(x) A_\nu(0) \rangle$.\cite{KT81}  
For $g>g_c$, on ther other hand, the $SU(N)$ and $U(1)$ symmetries
are exact, implying $\langle \phi_a \rangle =0$, and 
$\langle \sigma \rangle \not=0$.  In this phase, $\phi$ is massive,
$m_\phi = \langle \sigma \rangle$.
For $D=2$, $g_c=0$.
For more details on  large
$N$ results, see
Refs.\citen{Stone79,HHE80,Macfarlane79,BLS76}.

\section{Large N Estimation}
\setcounter{equation}{0}

\par
If we write $n^A(x)$ in terms of the $CP^{N-1}$ variable $\phi_a(x)$ as
\begin{equation}
 n^A(x) = \bar \phi_a(x) (T^A)_{ab} \phi_b(x) ,
\end{equation}
we obtain
\begin{eqnarray}
 n^A(x)n^A(y) &=& 
 \bar \phi_a(x) (T^A)_{ab} \phi_b(x) 
 \bar \phi_c(y) (T^A)_{cd} \phi_d(y)
 \nonumber\\
 &=& P_{ab}(x) P_{cd}(y) (T^A)_{ab} (T^A)_{cd} 
 \nonumber\\
 &=& {1 \over 2} \left[ (\bar \phi(x) \cdot \phi(y))
 (\phi(x) \cdot \bar \phi(y)) - {1 \over N} 
 (\bar \phi(x) \cdot \phi(x))(\bar \phi(y) \cdot \phi(y))
 \right] ,\nonumber\\
\end{eqnarray}
where we have used
\begin{eqnarray}
 \sum_{A=1}^{N^2-1} (T^A)_{ab} (T^A)_{cd} 
 = {1 \over 2} \left( \delta_{ad} \delta_{bc} - {1 \over N}
 \delta_{ab} \delta_{cd} \right) .
\end{eqnarray}
Hence we obtain
\begin{eqnarray}
  {\bf n}(x) \cdot {\bf n}(x) 
=  n^A(x) n^A(x) = {1 \over 2} \left[ 1 - {1 \over N} \right] 
 (\bar \phi(x) \cdot \phi(x))^2 .
\end{eqnarray}
The contraint 
$
\bar \phi(x) \cdot \phi(x)=1
$ 
leads to
$
{\bf n}(x) \cdot {\bf n}(x)
={1 \over 2} \left[ 1 - {1 \over N} \right] .
$
The expectation value reads
\begin{eqnarray}
 2 \langle n^A(x)n^A(y) \rangle
 &=&  \langle  (\bar \phi(x) \cdot \phi(y))
 (\phi(x) \cdot \bar \phi(y)) \rangle
 - {1 \over N}  \langle 
 | \phi(x) |^2 | \phi(y) |^2  \rangle .
\end{eqnarray}
The factorization in the large $N$ expansion leads to
\begin{eqnarray}
 2 \langle n^A(x)n^A(y) \rangle
 &\cong&   
 \langle \bar \phi(x) \cdot \phi(y) \rangle
 \langle \phi(x) \cdot \bar \phi(y) \rangle
 - {1 \over N}  \langle 
 | \phi(x) |^2 \rangle
 \langle | \phi(y) |^2  \rangle ,
\nonumber\\&=&   
 |\langle \bar \phi(x) \cdot \phi(y) \rangle |^2
 - {1 \over N}  \langle 
 | \phi(x) |^2 \rangle
 \langle | \phi(y) |^2  \rangle .
\end{eqnarray}
The large $N$ expansion shows that the field $\phi$ becomes massive,
so that the two-point function
$\langle \bar \phi(x) \cdot \phi(y) \rangle$
exhibits exponential decay.
\par

To leading order in the large $N$ expansion, the
correlation function $G_C(x,y)$ defined by (\ref{GC}) in the $CP^{N-1}_2$
model reads
\begin{eqnarray}
G_C(x,y) \cong \tilde G(x,y) 
:= \left( \delta_{ad} \delta_{bc} - {1 \over N}
 \delta_{ab} \delta_{cd} \right)
\langle  P_{ab}(x) P_{ba}(y) \rangle_{CP^{N-1}_2}   .
\end{eqnarray}
Note that $P$ is a projection operator with the properties 
\begin{eqnarray}
 P^2(x)  = P(x), \quad {\rm tr}P(x) = 1 . 
\end{eqnarray}
The composite operator $P$ is $U(1)$ gauge invariant.
$SU(N)$ invariance implies that
\begin{eqnarray}
 \langle P_{ab}(x) \rangle 
= \langle \bar \phi_a(x) \phi_b(x) \rangle 
= {1 \over N} \delta_{ab} . 
\end{eqnarray}
It is easy to show that
\begin{eqnarray}
\tilde G(x,y) &=&  \langle \bar \phi_a(x) \phi_b(x) \bar \phi_b(y)
\phi_a(y)
\rangle  - {1 \over N}
\nonumber\\
&=&  \langle D^2(x,y) \rangle
+ {1 \over N} \langle D(x,x) D(y,y) \rangle
- {1 \over N} 
\nonumber\\
&=& \langle D^2(x,y) \rangle
+ {1 \over N} \langle D(x,x) D(y,y) \rangle_{conn}  ,
\end{eqnarray}
where $\langle D(x,x) \rangle = 1$.

\par

\section*{Acknowledgments}
The authors would like to thank Minoru Hirayama for sending valuable comments on the non-Abelian Stokes theorem derived in this paper after submitting this article for publication.
This work is supported in part by
a Grant-in-Aid for Scientific Research from the Ministry of
Education, Science and Culture (No.10640249).
 
\end{document}